\documentclass[iop,apj,useAMS,usenatbib]{emulateapj-rtx4}

\usepackage{graphicx,epstopdf,epsfig}
\usepackage{lscape}
\usepackage{amsmath}
\usepackage{longtable,natbib}
\usepackage[usenames,dvipsnames]{color}
\usepackage{gensymb}
\usepackage{multirow}
\usepackage{capt-of}
\usepackage{float}
\usepackage{amssymb}
\usepackage[utf8]{inputenc}
\usepackage{booktabs}
\usepackage{wasysym}
\usepackage{rotating}
\usepackage{afterpage}
\usepackage{hyperref,url}
\hypersetup{colorlinks=true, citecolor=blue, filecolor=magenta, urlcolor=cyan, linkcolor=blue}

\begin{document}


\title{Stacked Average Far-Infrared Spectrum of Dusty Star-Forming Galaxies from the {\it Herschel}\altaffilmark{*}/SPIRE Fourier Transform Spectrometer}
\author{Derek Wilson\altaffilmark{1}, Asantha Cooray\altaffilmark{1}, Hooshang Nayyeri\altaffilmark{1}, 
Matteo Bonato\altaffilmark{2,3}, 
Charles M. Bradford\altaffilmark{4},
David L. Clements\altaffilmark{5},
Gianfranco De Zotti\altaffilmark{2},
Tanio D{\'{\i}}az-Santos\altaffilmark{6}, Duncan Farrah\altaffilmark{7},  
Georgios Magdis\altaffilmark{8},  
Micha{\l} J.~Micha{\l}owski\altaffilmark{9},  
Chris Pearson\altaffilmark{10,11,12},   
Dimitra Rigopoulou\altaffilmark{12}, 
Ivan Valtchanov\altaffilmark{13}, 
Lingyu Wang\altaffilmark{14,15},  
Julie Wardlow\altaffilmark{16}
}
\altaffiltext{*}{{\it Herschel} is an ESA space observatory with science instruments provided by European-led Principal Investigator consortia and with important participation from NASA.}
\altaffiltext{1}{Department of Physics \& Astronomy, University of California, Irvine, CA 92697, USA}
\altaffiltext{2}{INAF, Osservatorio Astronomico di Padova, Vicolo Osservatorio 5, I-35122 Padova, Italy}
\altaffiltext{3}{SISSA, Via Bonomea 265, I-34136 Trieste, Italy}
\altaffiltext{4}{California Institute of Technology, 1200 E. California Blvd., Pasadena, CA 91125, USA}
\altaffiltext{5}{Astrophysics Group, Imperial College London, Blackett Laboratory, Prince Consort Road, London SW7 2AZ, UK}
\altaffiltext{6}{Nucleo de Astronomia de la Facultad de Ingenieria, Universidad Diego Portales, Av. Ejercito Libertador 441, Santiago, Chile}
\altaffiltext{7}{Department of Physics, Virginia Tech, Blacksburg, VA 24061, USA}
\altaffiltext{8}{Dark Cosmology Centre, Niels Bohr Institute, University of Copenhagen, Juliane Mariesvej 30, DK-2100 Copenhagen, Denmark}
\altaffiltext{9}{Institute for Astronomy, University of Edinburgh, Royal Observatory, Edinburgh EH9 3HJ, UK}
\altaffiltext{10}{RAL Space, CCLRC Rutherford Appleton Laboratory, Chilton, Didcot, Oxfordshire OX11 0QX, United Kingdom}
\altaffiltext{11}{Department of Physical Sciences, The Open University, Milton Keynes, MK7 6AA, UK}
\altaffiltext{12}{Oxford Astrophysics, Denys Wilkinson Building, University of Oxford, Keble Rd, Oxford OX1 3RH, UK}
\altaffiltext{13}{{\it Herschel} Science Centre, European Space Astronomy Centre, ESA, E-28691 Villanueva de la Ca\~nada, Spain}
\altaffiltext{14}{SRON Netherlands Institute for Space Research, Landleven 12, 9747 AD, Groningen, The Netherlands}
\altaffiltext{15}{Kapteyn Astronomical Institute, University of Groningen, Postbus 800, 9700 AV, Groningen, The Netherlands}
\altaffiltext{16}{Centre for Extragalactic Astronomy, Department of Physics, Durham University, South Road, Durham, DH1 3LE, UK}


\begin{abstract}
We present stacked average far-infrared spectra of a sample of 197 dusty, star-forming galaxies (DSFGs) at $0.005 < z < 4$ using about 90\% of the {\it Herschel} Space Observatory SPIRE Fourier Transform Spectrometer (FTS) extragalactic data archive based on 3.5 years of science operations. These spectra explore an observed-frame 447 GHz\,-\,1568 GHz frequency range allowing us to observe the main atomic and molecular lines emitted by gas in the interstellar medium. The sample is sub-divided into redshift bins, and a subset of the bins are stacked by infrared luminosity as well. These stacked spectra are used to determine the average gas density and radiation field strength in the photodissociation regions (PDRs) of dusty, star-forming galaxies. For the low-redshift sample, we present the average spectral line energy distributions (SLED) of CO and H$_2$O rotational transitions and consider PDR conditions based on observed [C\,I]\,370\,$\mu$m and 609\,$\mu$m, and CO (7-6) lines. For the high-$z$ ($0.8<z<4$) sample, PDR models suggest a molecular gas distribution in the presence of a radiation field that is at least a factor of 10$^3$ larger than the Milky-Way and with a neutral gas density of roughly 10$^{4.5}$ to 10$^{5.5}$ cm$^{-3}$. The corresponding PDR models for the low-$z$ sample suggest a UV radiation field and gas density comparable to those at high-$z$. Given the challenges in obtaining adequate far-infrared observations, the stacked average spectra we present here will remain the highest signal-to-noise measurements for at least a decade and a half until the launch of the next far-infrared facility.
\end{abstract}



\section{Introduction}

Our understanding of galaxy formation and evolution is directly linked to understanding the physical properties of the interstellar medium (ISM) of galaxies \citep{Kennicutt1998, Leroy2008, Hopkins2012, Magdis2012, Scoville2016}. Dusty star-forming galaxies (DSFGs), 
with star-formation rates in excess of 100\,M$_{\odot}$\,yr$^{-1}$, are an important contributor to the star-formation rate density of the Universe \citep{Chary2001, Elbaz2011}. 
However, our knowledge of the interstellar medium within these galaxies is severely limited due to high dust extinction with typical optical attenuations of $A_V \sim  6–10$\,mag \citep{Caseycoorayreview}. 
Instead of observations of rest-frame UV and optical lines, crucial diagnostics of the ISM in DSFGs can be obtained with spectroscopy at mid- and far-infrared wavelengths \citep{Spinoglio1992}. 

In particular, at far-infrared wavelengths, the 
general ISM is best studied through atomic fine-structure line transitions, such as the [C\,II]\,158 $\mu$m line transition. Such studies complement
rotational transitions of molecular gas tracers, such as CO, at mm-wavelengths that are  effective at
tracing the proto-stellar and dense star-forming cores of DSFGs (e.g. \citealp{Carilli2013}). 
Relative to the total infrared luminosities,  certain atomic fine-structure emission lines can have line luminosities that are the level of a few
tenths of a percent \citep{Stacey1989, Carilli2013, Riechers2014, Aravena2016, Spilker2016, Hemmati2017}.  Far-infrared fine-structure lines are capable of probing the ISM over the whole
range of physical conditions, from those that are found in the neutral to ionized  gas in photodissociation regions (PDRs; \citealt{Tielens1985, Hollenbach1997, Hollenbach1999, Wolfire1993, Spaans1994, Kaufman1999})  to X-ray dominated regions (XDRs; \citealt{Lepp1988, Bakes1994, Maloney1996, Meijerink2005}), such as those associated with an AGN, or shocks \citep{Flower2010}. Different star-formation modes
and the effects of feedback are mainly visible in terms of differences in
the ratios of fine-structure lines and the ratio of fine-structure line to the total IR luminosity \citep{Sturm2011, Kirkpatrick2012, Fernandez2016}.  Through PDR modeling and under assumptions such as local thermodynamic equilibrium (LTE), line ratios can then be used as a probe of the gas density, temperature, and the strength of the radiation field
that is ionizing the ISM gas. An example is [C\,II]/[O\,I] vs. [O\,III]/[O\,I] ratios that are used to separate starbursts from AGNs (e.g. \citealt{Spinoglio2000,Fischer1999}).

\begin{figure*}[th]
\centering
\includegraphics[trim=2cm 1cm 0cm 0cm, scale=0.9]{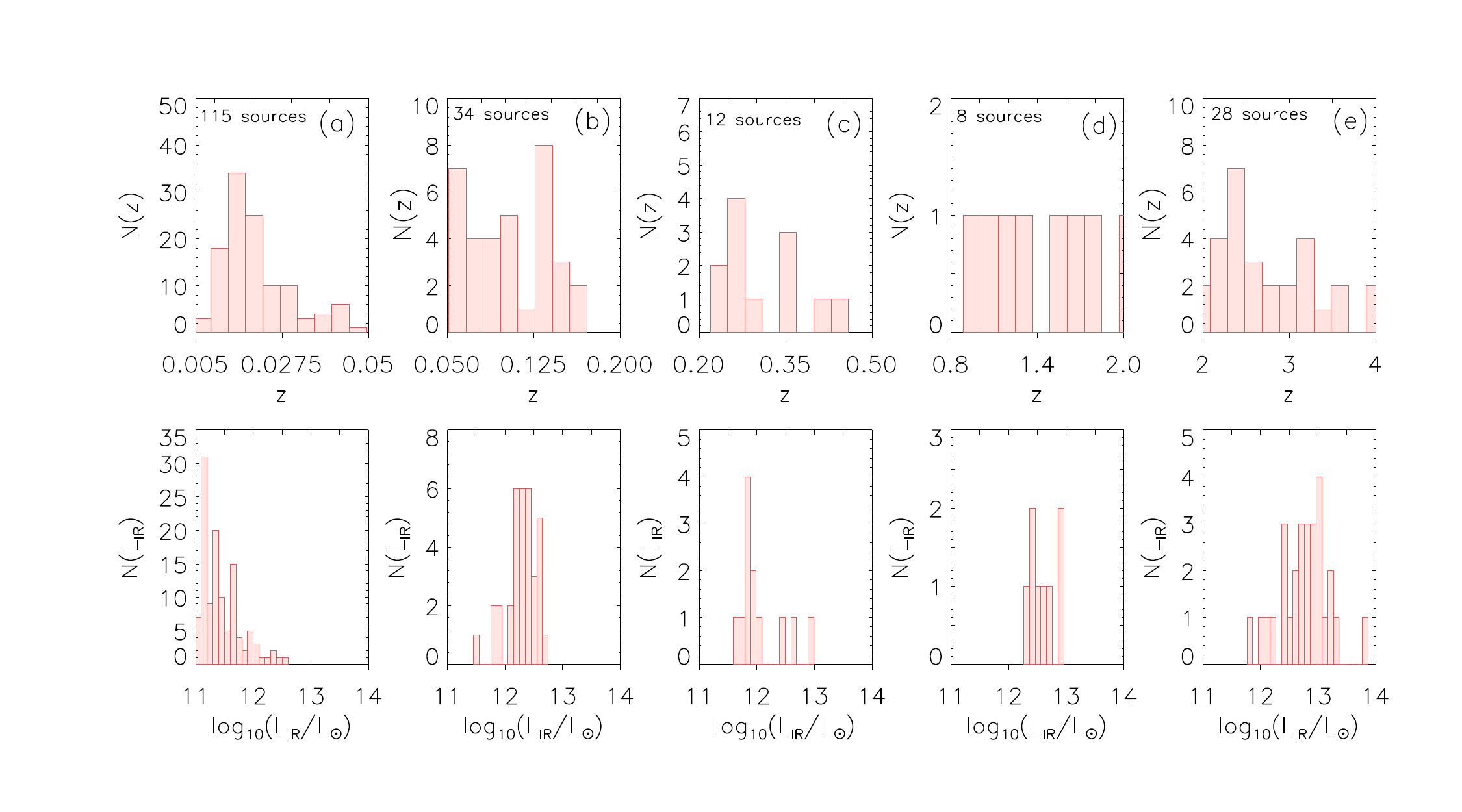}
\caption{\textit{Top:} Distribution of redshifts for sources included in each of the five redshift bins: (a) 115 sources with $0.005 < z < 0.05$, (b) 34 sources with $0.05 < z < 0.2$, (c) 12 sources with $0.2 < z < 0.5$, (d) 8 sources with $0.8 < z < 2$, and (e) 28 sources with $2 < z < 4$. The low number of sources in the two intermediate redshift bins of $0.2 < z < 0.5$ and $0.8 < z < 2$ is due to lack of observations. \textit{Bottom:} Total infrared luminosities (rest-frame $8-1000\,\mu$m) for sources included in each of the five redshift bins above with a median luminosity of log$_{10}$(L$_{\rm IR}$/L$_{\odot}$) = 11.35, 12.33, 11.89, 12.53, and 12.84, respectively. For lensed sources in the  $2 < z < 4$ range, we have made a  magnification correction using best-determined lensing models published in the literature (See Section 2).}
\label{fig:lumhist}
\end{figure*}

In comparison to the study presented here using {\it Herschel} SPIRE/FTS (\citealt{Pilbratt2010, Griffin2010}) data, we highlight a similar recent study by \citet{Wardlow2017}
on the average rest-frame mid-IR spectral line properties
using all of their archival high-redshift data from the {\it Herschel}/PACS instrument \citep{Poglitsch2010}. While the sample observed by SPIRE/FTS
is somewhat similar, the study with SPIRE extends the wavelength range to rest-frame far-IR lines from the mostly rest-frame mid-IR lines detected with PACS.
In a future publication, we aim to present a joint analysis of the overlap sample between SPIRE/FTS and PACS, but here we  mainly concentrate on the
analysis of FTS data and the average stacked spectra as measured from the SPIRE/FTS data. We also present a general analysis with interpretation based on PDR models
and comparisons to results in the literature on ISM properties of both low- and high-$z$ DSFGs.

The paper is organized as follows. In Sections 2 and 3, we describe the archival data set and the method by which the data were stacked, respectively. Section 4 presents the stacked spectra.
In Section 5, the average emission from detected spectral lines is used to model the average conditions in PDRs
of dusty, star-forming galaxies. In addition, the fluxes derived from the stacked spectra are compared to various measurements from the literature.
We discuss our results and conclude with a summary.
A flat-$\Lambda$CDM cosmology of $\Omega_{m_0}$ = 0.27, $\Omega_{\Lambda_0}$ = 0.73, and $H_0$ = 70 $\rm{km} ~ \rm{s^{-1}} ~  \rm{Mpc^{-1}}$ is assumed. 
With {\it Herschel}\ operations now completed, mid- and far-IR spectroscopy of DSFGs will not be feasible until the launch of next far-IR mission,
expected in the 2030s, such as SPICA \citep{SPICA2010} or the Origins Space Telescope \citep{Meixner2016}. The average spectra we present here will remain the standard in the field 
and will provide crucial input for the planning of the next mission.


\begin{figure*}
\centering
\includegraphics[trim=0cm 0cm 0cm 0cm, scale=0.7]{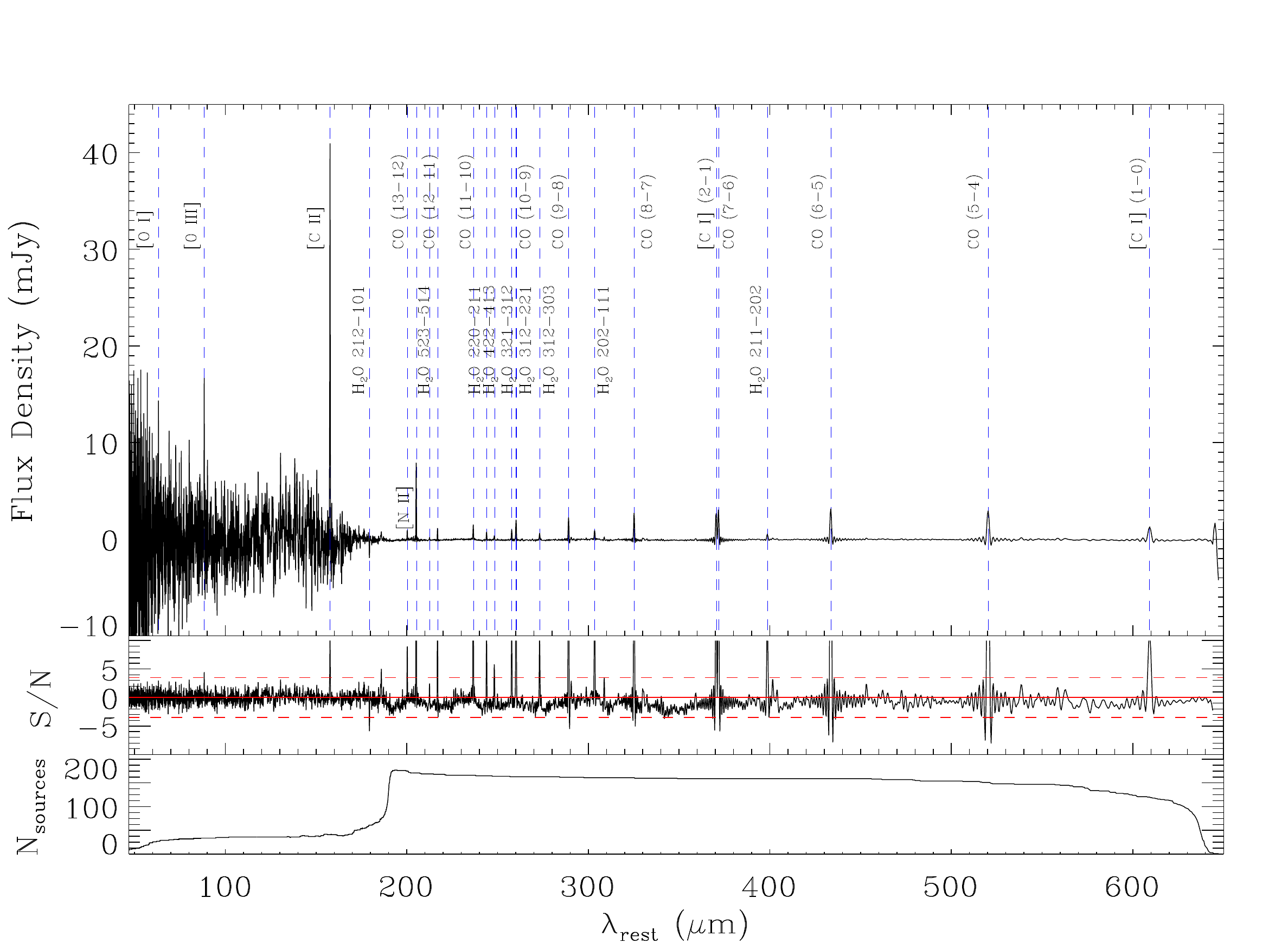}
\caption{\textit{Top:} Average far-infrared stacked spectrum containing all data. Sources range in redshift from $0.005<z<4$. This stack serves as a qualitative representation of the average spectrum of all of the \textit{Herschel} spectra. For the purposes of analysis and interpretation, the dataset is split into redshift and luminosity bins for the remainder of this paper. Dashed blue vertical lines indicate the locations of main molecular emission lines. We detect the fine-structure lines [C\,II], [O\,I], and [O\,III] as well as the CO emission line ladder from $J = 13-12$ to $J = 5-4$. Also detected are the two lowest [C\,I] emissions at 492\,GHz (609 $\mu$m) and 809\,GHz (370 $\mu$m), [N\,II] at 1461\,GHz (205\,$\mu$m) and the water lines within the frequency (wavelength) range covered in this stack from 50\,$\mu$m to 652\,$\mu$m). \textit{Middle}: Signal-to-noise ratio. The horizontal dashed line indicates $\rm S/N = 3.5$, and the solid red line represents $\rm S/N = 0$. \textit{Bottom}: The number of sources that contribute to the stack at each wavelength.}
\label{fig:stack_all}
\end{figure*}

\begin{figure*}
\centering
\includegraphics[trim=0cm 0cm 0cm 0cm, scale=0.7]{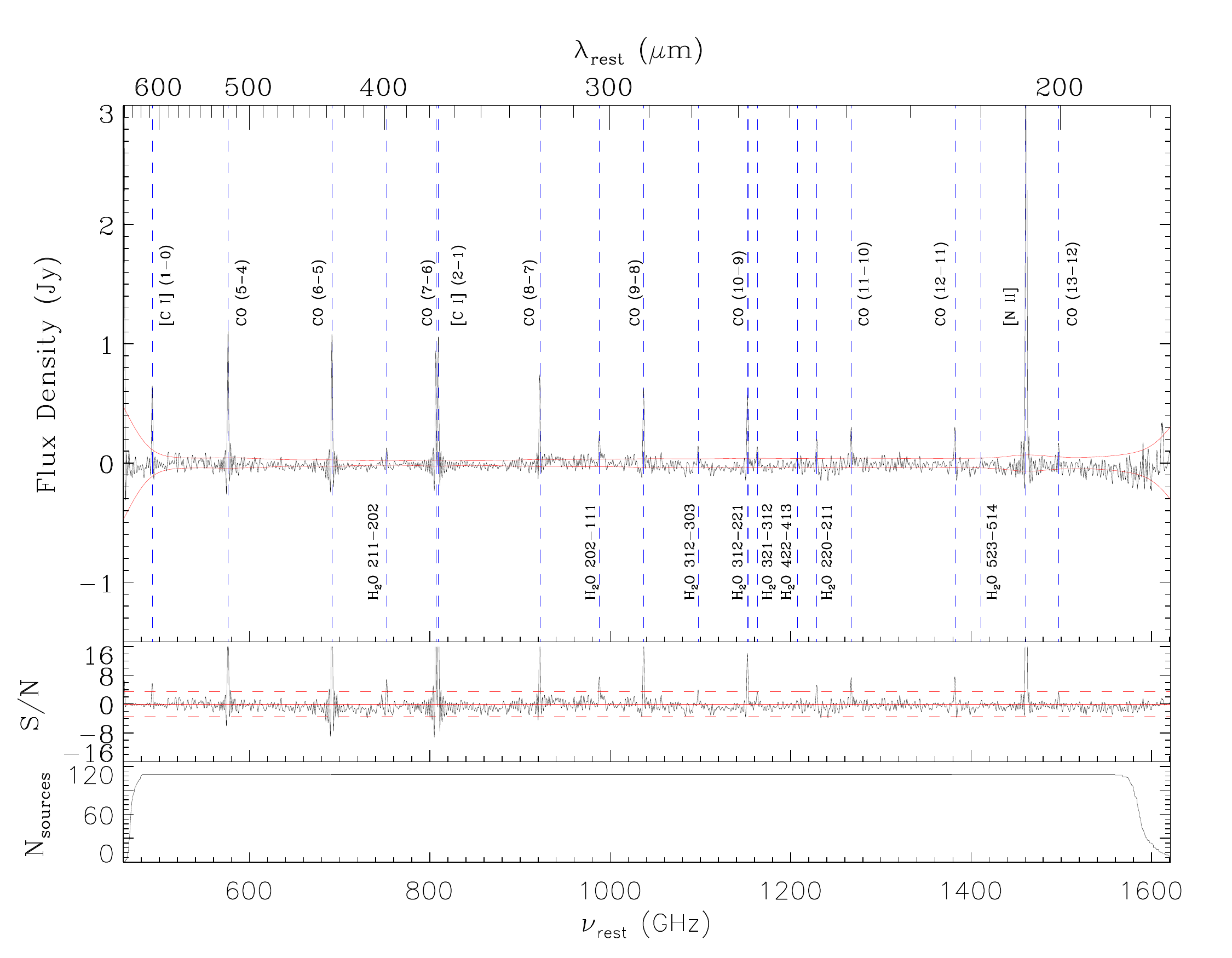}
\caption{\textit{Top}: Stacked SPIRE/FTS spectrum of archival sources with $0.005 < z < 0.05$. Overlaid is the $1\sigma$ jackknifed noise level in red and dashed vertical lines showing the locations of main molecular emission lines. We detect the CO emission line ladder from $J = 13-12$ to $J = 5-4$, as well as the two lowest [C\,I] emissions at 492\,GHz (609 $\mu$m) and 809\,GHz (370 $\mu$m), [N\,II] at 1461\,GHz (205\,$\mu$m) and the water lines within the rest frequencies (wavelengths) covered in this stack from 460\,GHz to 1620\,GHz (185\,$\mu$m to 652\,$\mu$m). \textit{Middle}: Signal-to-noise ratio. The horizontal dashed line indicates $\rm S/N = 3.5$, and the solid red line indicates $\rm S/N = 0$. Lines with $\rm S/N > 3.5$ were considered detected. \textit{Bottom}: The number of sources that contribute to the stack at each frequency. }
\label{fig:z0-005}
\end{figure*}

\begin{figure*}
\centering
\includegraphics[trim=0cm 0cm 0cm 0cm, scale=0.7]{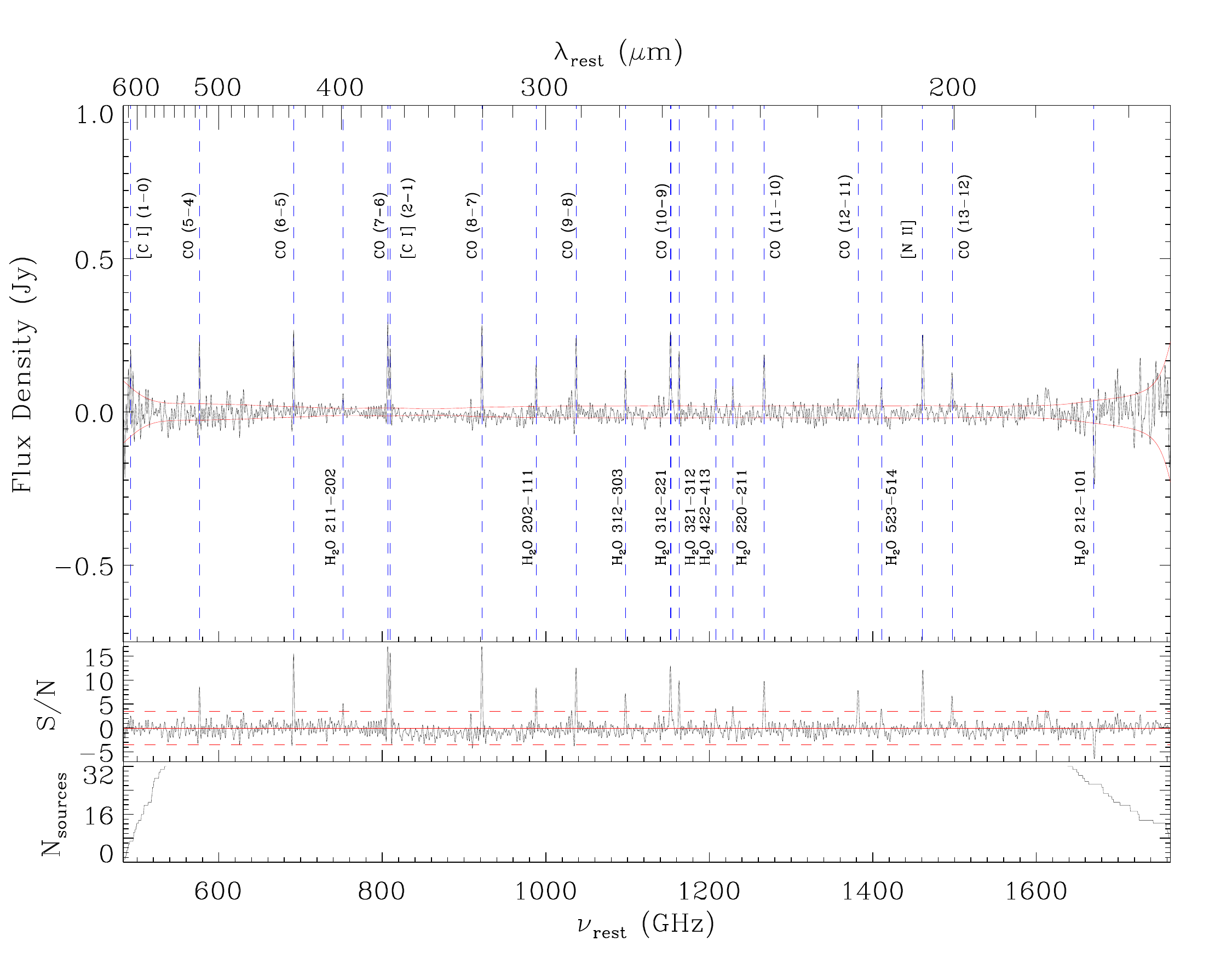}
\caption{Same as Figure \ref{fig:z0-005}, but for the redshift range $0.05 < z < 0.2$. We detect all the CO emission line ladder within the frequency (wavelength) covered by the stack from 480\,GHz to 1760\,GHz (170\,$\mu$m to 625 $\mu$m). The stacked spectrum also shows 3.5$\sigma$ detection for $\rm [C\,I](2-1)$ at 809 GHz ($370 \,\mu$m), [N\,II] at 1461\,GHz (205 $\mu$m), and water lines.}
\label{fig:z005-02}
\end{figure*}

\begin{figure*}
\centering
\includegraphics[trim=0cm 0cm 0cm 0cm, scale=0.7]{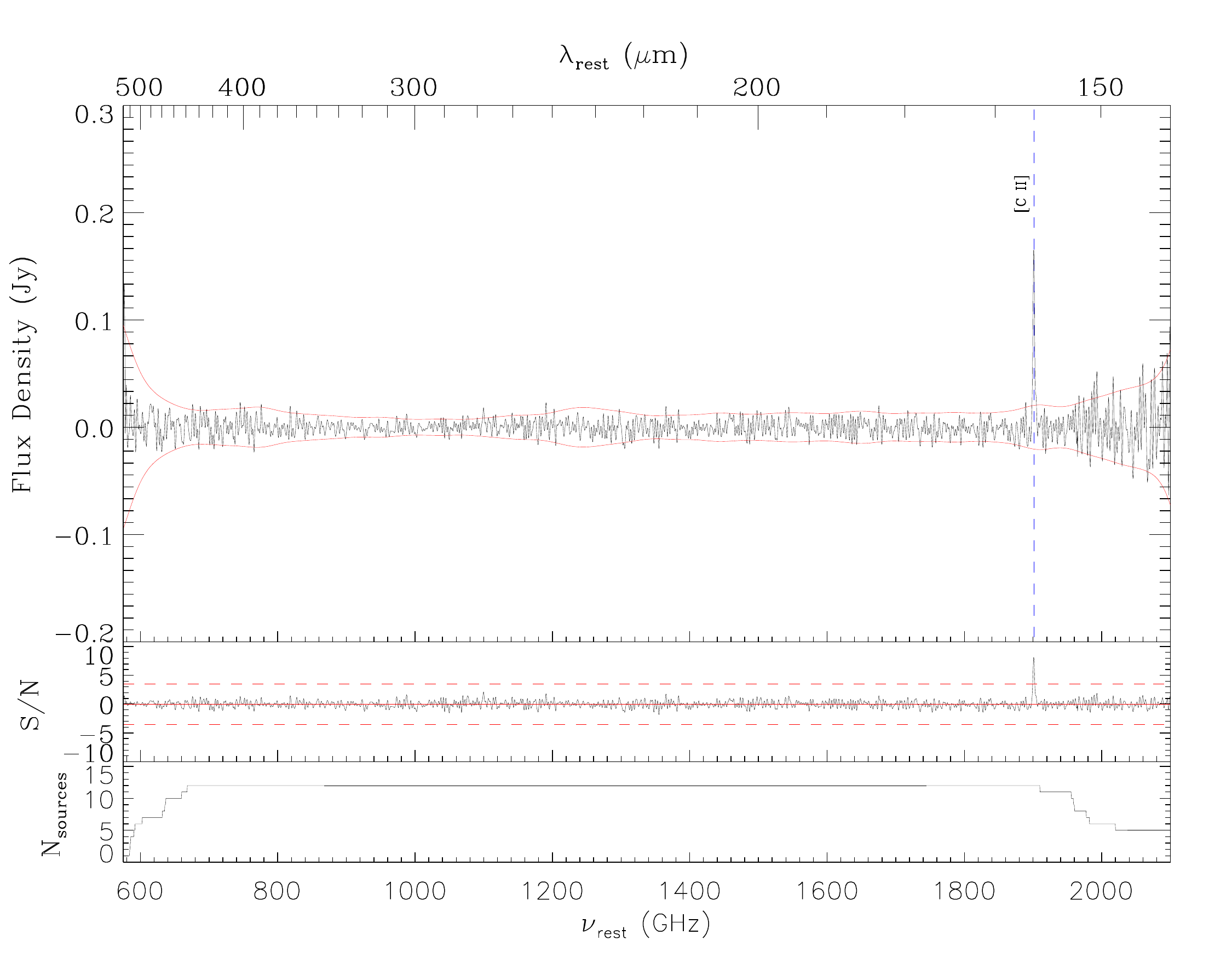}
\caption{Same as Figure \ref{fig:z0-005}, but for the redshift range $0.2 < z < 0.5$. We only detect the [C\,II] at 1901\,GHz (158\,$\mu$m) line in this stack with frequency (wavelength) coverage 580\,GHz to 2100\,GHz (143\,$\mu$m to 517\,$\mu$m).}
\label{fig:z02-5}
\end{figure*}

\begin{figure*}
\centering
\includegraphics[trim=0cm 0cm 0cm 0cm, scale=0.7]{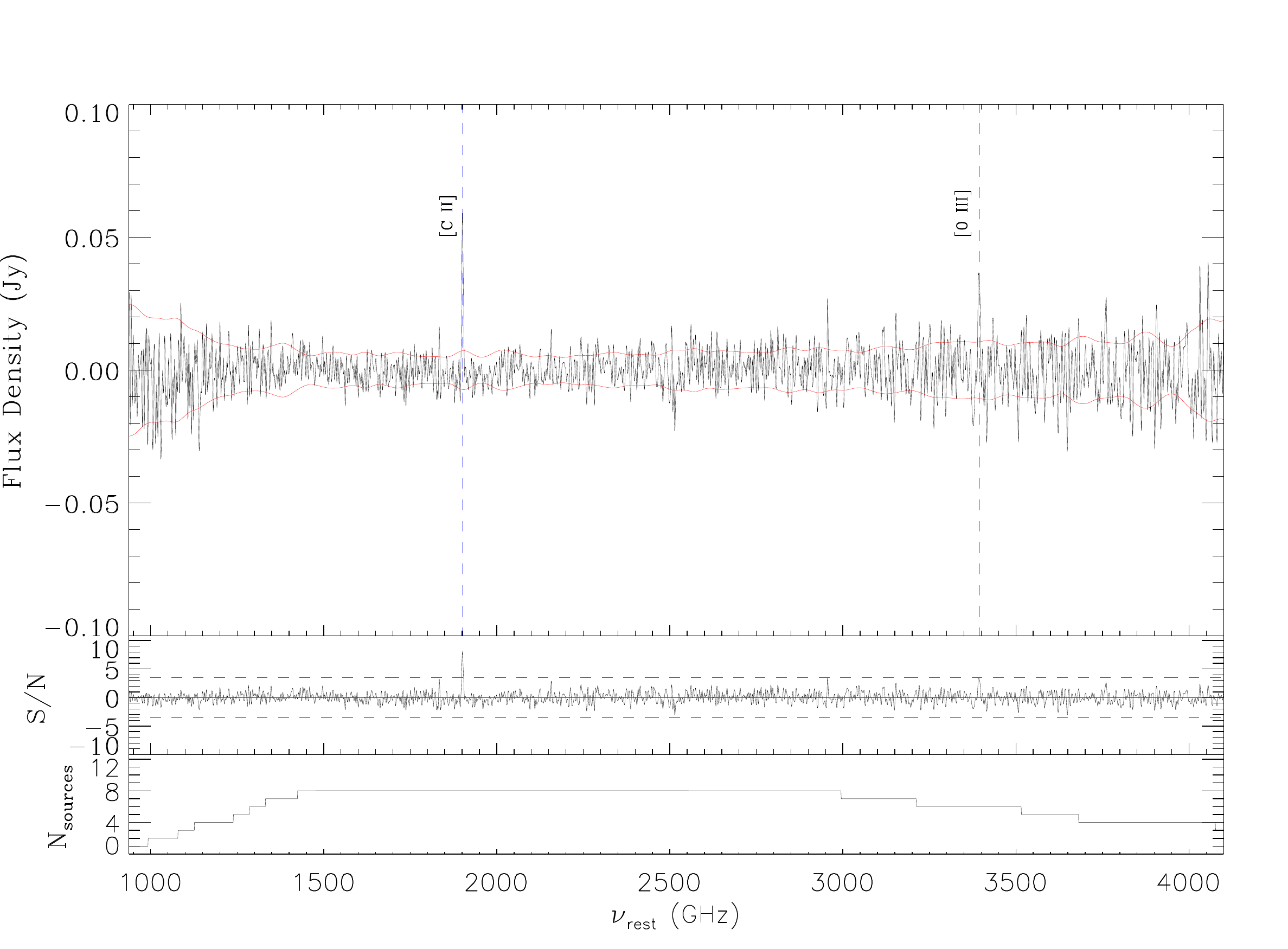}
\caption{Same as Figure \ref{fig:z0-005}, but for the redshift range $0.8 < z < 2$. We detect [N\,II] at 1461\,GHz (205\,$\mu$m), [C\,II] at 1901\,GHz (158\,$\mu$m) and [O\,III] at 3391\,GHz (88\,$\mu$m) in the frequency (wavelength) range of 950\,GHz to 4100\,GHz (70\,$\mu$m to 316\,$\mu$m) covered by the stack.}
\label{fig:z08-2}
\end{figure*}

\begin{figure*}
\centering
\includegraphics[trim=0cm 0cm 0cm 0cm, scale=0.7]{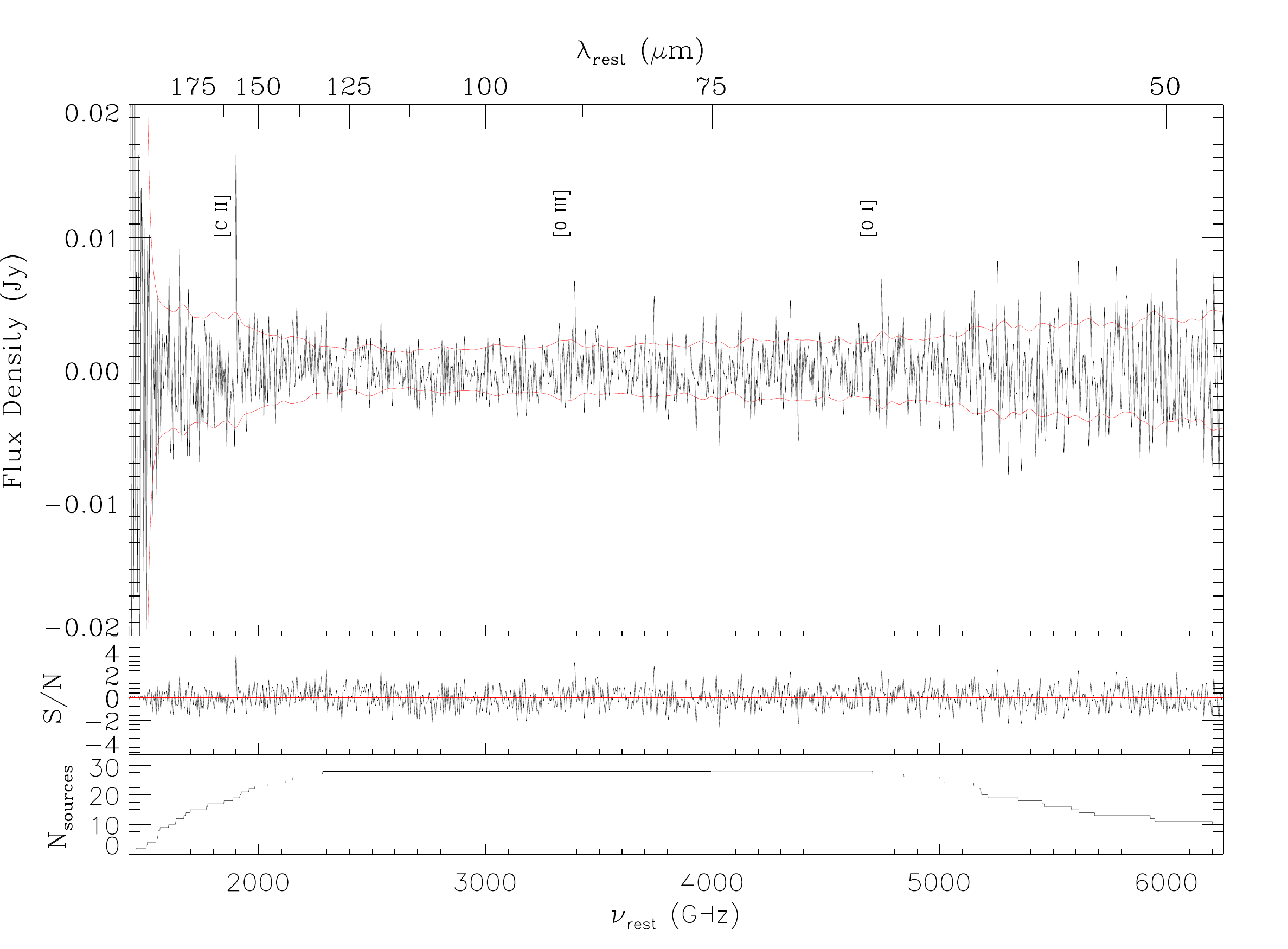}
\caption{Same as Figure \ref{fig:z0-005}, but for the redshift range $2 < z < 4$. We detect [C\,II] at 1901\,GHz (158\,$\mu$m) and [O\,III] at 3391\,GHz (88\,$\mu$m) in the frequency (wavelength) range of 1400\,GHz to 6200\,GHz (48\,$\mu$m to 214\,$\mu$m).}
\label{fig:z2-4}
\end{figure*}

\begin{figure*}
\centering
\includegraphics[trim=0cm 0cm 0cm 0cm, scale=0.7]{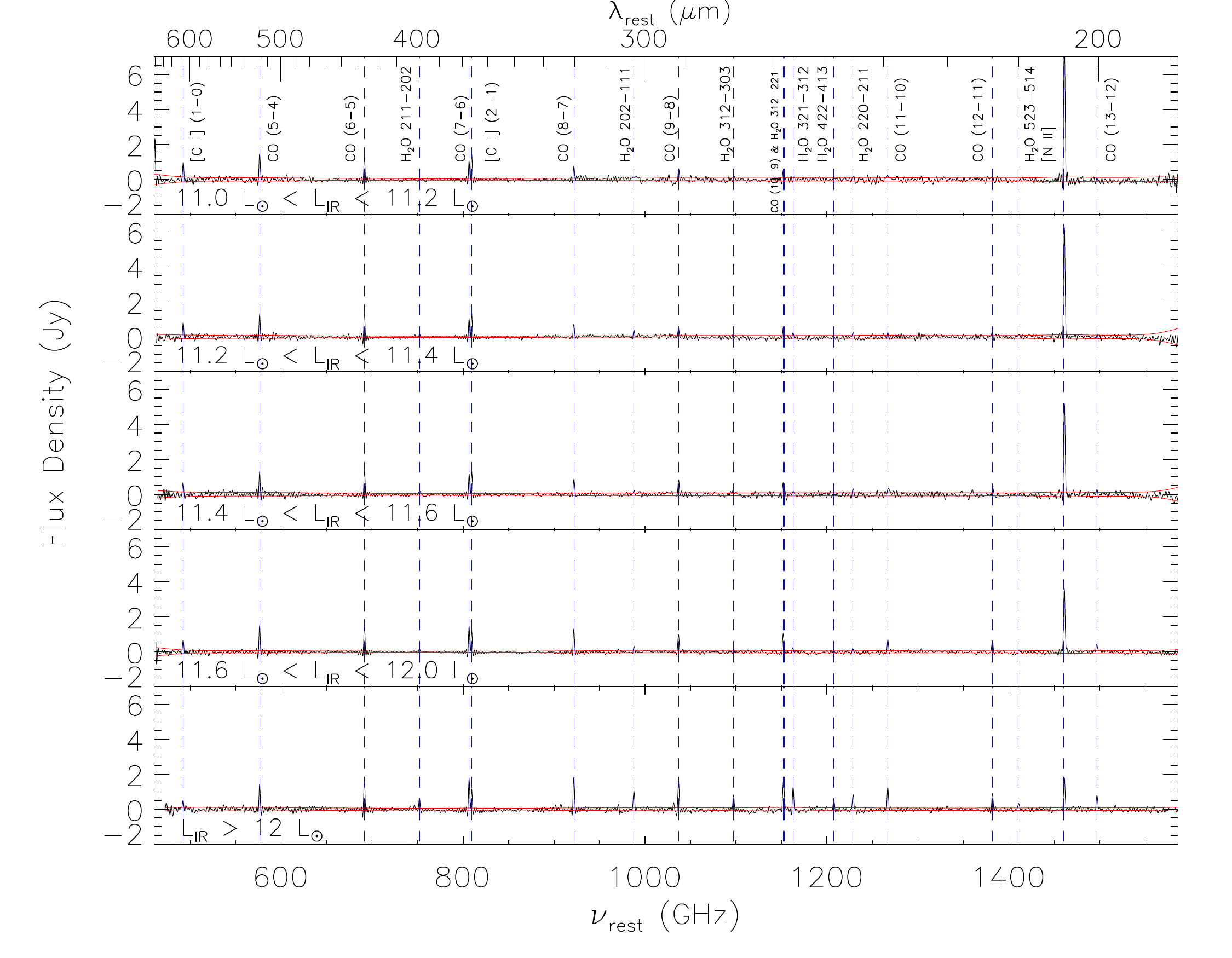}
\caption{The lowest redshift bin ($0.005 < z < 0.05$) is stacked using a straight mean (without inverse-variance weighting) in five luminosity bins as outlined in each panel. From top to bottom, the median luminosities in each bin are $10^{11.12}$ L$_{\odot}$, $10^{11.32}$ L$_{\odot}$, $10^{11.49}$ L$_{\odot}$, $10^{11.69}$ L$_{\odot}$, and $10^{12.21}$ L$_{\odot}$. The mean redshifts in each bin are 0.015, 0.018, 0.021, 0.027, and 0.038. The number of sources contributing to each bin are 37, 28, 17, 24, 9. and The CO molecular line excitations, [C\,I] atomic emissions, and [N\,II] at 205\,$\mu$m are detected in all five luminosity bins.
}
\label{fig:lowz_lirbins}
\end{figure*}


\section{Data}

Despite the potential applications of mid- and far-IR spectral lines, the limited wavelength coverage and sensitivity of far-IR facilities have restricted the vast majority of observations to galaxies in the nearby universe. A significant leap came from the {\it Herschel} Space Observatory \citep{Pilbratt2010}, thanks to the spectroscopic capabilities of the Fourier Transform Spectrometer (FTS; \citealt{Naylor2010, Swinyard2014}) of the SPIRE instrument \citep{Griffin2010}. SPIRE covered the wavelength range of $194\,\rm{\mu m} - 671 \,\rm{\mu m}$, making it useful in the detection of ISM fine structure cooling lines, such as [C\,II] 158\,$\rm{\mu m}$, [O\,III] 88\,$\rm{\mu m}$, [N\,II] 205\,$\rm{\mu m}$, and [O\,I] 63\,$\rm{\mu m}$, in high-redshift galaxies and carbon monoxide (CO) and water lines (H$_2$O) from the ISM of nearby galaxies. The {\it Herschel} data archive contains SPIRE/FTS data for a total of 231 galaxies, with 197 known to be in the redshift interval $0.005 <  z< 4.0$, completed through multiple programs either in guaranteed-time or open-time programs. While most of the galaxies at $0.5 < z < 4$ are intrinsically ultra-luminous IR galaxies (ULIRGS; \citealt{Sanders1996}), with luminosities greater than $10^{12}\,$L$_{\odot}$, archival observations at $z > 2$ are mainly limited to the brightest dusty starbursts with apparent L $ > 10^{13}\,$L$_{\odot}$ or hyper-luminous IR galaxies (HyLIRGs). Many of these cases, however, are gravitationally lensed DSFGs and their intrinsic luminosities are generally consistent with ULIRGS. At the lowest redshifts, especially in the range $0.005< z< 0.05$, many of the targets have L $ < 10^{12}\,$L$_{\odot}$ or are luminous IR galaxies (LIRGs). While fine-structure lines are easily detected for such sources, most individual archival observations of brighter ULIRGs and HyLIRGs at $z > 1$ do not reveal clear detections of far-infrared fine-structure lines despite their high intrinsic luminosities \citep{Georgethesis}, except in a few very extreme cases such as the Cloverleaf quasar host galaxy \citep{Uzgil2016}. Thus, instead of individual spectra, we study the averaged stacked spectra of DSFGs, making use of the full SPIRE/FTS archive of {\it Herschel}.

Given the wavelength range of SPIRE and the redshifts of observed galaxies, to ease stacking, we subdivide the full sample of 197 galaxies into five redshift bins (Figure \ref{fig:lumhist}),
namely, low-redshift galaxies at $0.005 < z  <  0.05$ and $0.05 < z  <  0.2$,  intermediate redshifts $0.2<  z  <  0.5$, and 
high-redshift galaxies at $0.8 < z < 2$ and $2 < z  <  4$.  Unfortunately, due to lack of published redshifts, we exclude observations of 24 targets or roughly 10\% of the total archival sample (231 sources) from
our stacking analysis expected to be mainly at $z > 1$ based on the sample
selection and flux densities. This is due to the fact that redshifts are crucial to shift spectra to a common redshift, usually taken to be the mean of the redshift distribution in each of our bins.
For these 24 cases we also did not detect strong individual lines, which would allow us to establish a redshift conclusively with the SPIRE/FTS data. Most of these sources are
likely to be at $z > 1$ and we highlight this subsample in the Appendix to encourage follow-up observations. We also note that the SPIRE/FTS archive does not contain any observations of galaxies in the redshift interval of 0.5 to 0.8 and even in the range of $0.8 < z < 2$, observations are simply limited to 8 galaxies,
compared to attempted observations of at least 28 galaxies, and possibly as high as 48 galaxies when including the subsample without redshifts, at $z > 2$.

The data used in our analysis consist of 197 publicly-available {\it Herschel} \, SPIRE/FTS spectra, as part of
various Guaranteed Time (GT) and Open-Time (OT) {\it Herschel} programs summarized in the Appendix (Table \ref{table:obsids}). Detailed properties of the sample are also presented in the Appendix (Table \ref{table:all_targets}) for both low and high redshifts where the dividing line is at $z=0.8$, with 161 and 36 objects respectively. Table \ref{table:all_targets} also lists 34 sources at the end with existing FTS observations but which were not used in the analysis. The majority of unused sources have unknown or uncertain spectroscopic redshifts. This includes MACS J2043-2144 for which a single reliable redshift is currently uncertain as there is evidence for three galaxies with $z=2.040$, $z=3.25$, and $z=4.68$ within the SPIRE beam \citep{Zavala2015}. The sources SPT 0551-50 and SPT 0512-59 have known redshifts but do not have magnification factors. The low-redshift sample is restricted to DSFGs with $z > 0.005$ only. This limits the bias in our stacked low-$z$ spectrum from bright near-by galaxies such as M81 and NGC 1068. Our selection does include bright sources such as Arp 220 and Mrk 231 in the stack, but we study their impact by breaking the lowest redshift sample into luminosity bins, including a ULIRG bin with L$_{\rm IR} > 10^{12}\,$L$_{\odot}$.

The {\it Herschel} sample of dusty, star-forming galaxies is composed of LIRGS with $10^{11}$ L$_{\odot} \, < $ L $ < 10^{12}\,$L$_{\odot}$ and ULIRGS with L $> 10^{12}\,$L$_{\odot}$. The sample is heterogeneous, consisting of AGN, starbursts, QSOs, LINERs, and Seyfert types 1 and 2. The low-redshift SPIRE/FTS spectra were taken as part of the HerCULES program (\citealp{Rosenberg2015}; PI van der Werf), HERUS program (\citealp{Pearson2016}; PI Farrah), 
and the Great Observatory All-Sky LIRG Survey (GOALS; \citealp{Armus2009}, \citealp{Lu2017}, PI: N. Lu)
along with supplementary targets from the $\rm{KPGT\textunderscore wilso01\textunderscore 1}$ (PI: C. Wilson) and $\rm{OT2\textunderscore drigopou\textunderscore 3}$ (PI: D. Rigopoulou) programs.  At $0.2 < z < 0.5$, the SPIRE/FTS sample of 11 galaxies is limited to \citet{Magdis2014}, 
apart from one source, IRAS 00397-1312, from \citet{Helouwalker1988} and \citet{Farrah2007}. Note that the \citet{Magdis2014} sample contained two galaxies initially identified
to be at $z < 0.5$, but later found to be background $z > 2$ galaxies that were lensed by the $z < 0.5$ foreground galaxy. Those data are included in our high-redshift sample.

The high-redshift sample at $z > 0.8$ primarily comes from open-time programs that followed-up lensed galaxies
from HerMES \citep{Oliver2012} and {\it H}-ATLAS \citep{Eales2010}, and discussed in \citet{Georgethesis}. Despite the boosting from lensing,
only a few known cases of individual detections exist in the literature: NB.v1.43 at $z=1.68$ \citep{George2013, Timmons2016}, 
showing a clear signature of [C\,II] that led to a redshift determination for the first-time
with a far-IR line, SMMJ2135-0102 (Cosmic eyelash; \citealp{Ivison2010}), ID.81 and ID.9 \citep{Negrello2014}.
With lens models for {\it Herschel}\,-selected lensed sources now in the literature  (e.g., \citealp{Bussmann2013,Calanog2014}), 
the lensing magnification factors are now known with reasonable enough accuracy that the intrinsic luminosities of many of these
high-redshift objects can be established.  The $z > 0.8$ sample is composed of
30 high-redshift, gravitationally-lensed galaxies (e.g., OT1\textunderscore rivison\textunderscore 1, OT2\textunderscore rivison\textunderscore 2) and 6 un-lensed galaxies  (OT1\textunderscore apope\textunderscore 2 and one each from OT1\textunderscore rivison\textunderscore 1 and OT2\textunderscore drigopou\textunderscore 3). 

The distribution of redshifts can be found in Figure \ref{fig:lumhist}, where
we have subdivided the total distribution into five redshift bins: $0.005 < z < 0.05$, $0.05 < z < 0.2$, $0.2 < z < 0.5$, $0.8 < z < 2$, and $2 < z < 4$. The mean redshifts in the five redshift bins are $z = 0.02$, $z = 0.1$, and $z = 0.3$, $z = 1.4$, and $z = 2.8$, respectively. For reference, in Figure \ref{fig:lumhist}, we also show the $8-1000\,\mu$m  luminosity distribution in the five redshift bins. The distribution spans mostly from LIRGS at low-redshifts to ULIRGS at $0.05 < z  < 0.2$ and above. In the highest redshift bins we find ULIRGS again, despite increase in redshift,
due to the fact that most of these are lensed sources; with magnification included, the observed sources will have apparent luminosities consistent with HyLIRGS. Unfortunately, there is a lack of data between redshifts of $z \sim 0.2$ and $z \sim 1$, with the \citet{Magdis2014} sample and the spectrum of IRAS 00397-1312 from HERUS (\citealp{Pearson2016}) being the only SPIRE/FTS observed spectra in this range.

In general, SPIRE/FTS observations we analyze here were taken in high resolution mode, with a spectral resolving power of $300-1000$ through a
 resolution of 1.2\,GHz and frequency span of $\rm 447\,GHz-1568\,GHz$.
The data come from two bolometer arrays: the spectrometer short wavelength (SSW) array, covering $\rm 194\,\mu m-318\,\mu m$ ($\rm 944\,GHz–1568\,GHz$) and the spectrometer long wavelength (SLW) array, covering $\rm 294\,\mu m-671\,\mu m$ ($\rm 447\,GHz–1018\,GHz$).
The two arrays have different responses on the sky with the full-width half-maximum (FWHM) of the SSW beam at 18$^{\prime\prime}$ and
the SLW beam varying from 30$^{\prime\prime}$ to 42$^{\prime\prime}$ with frequency \citep{Swinyard2014}. The SPIRE/FTS data typically involves $\sim90-100$ scans of the faint, high-redshift sources 
and about half as many scans for the lower-redshift sources. Total integration times for each source are presented in Table \ref{table:obsids}. Typical total integration times of order 5000 seconds achieve unresolved spectral line sensitivities down to $\sim 10^{-18}\,{\rm W\,m^{-2}} (3\sigma)$.

\begin{figure*}[!th]
\centering
\includegraphics[trim=0cm 0cm 0cm 0cm, scale=0.75]{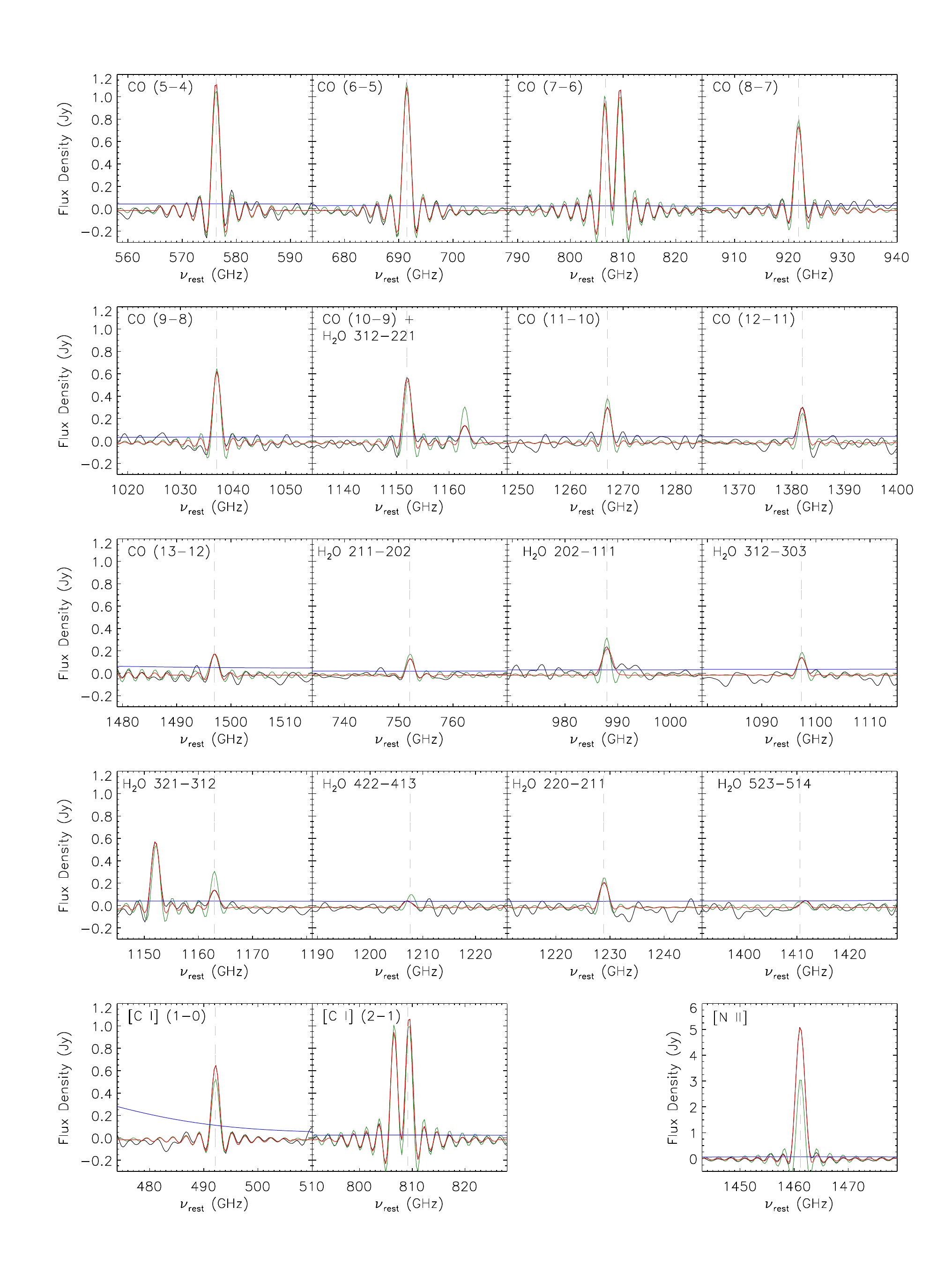}
\caption{Sinc-Gauss and sinc fits to the detected atomic and molecular lines in the low-redshift stack at $0.005<z<0.05$. The spectrum itself is shown in black. The green curve shows a sinc fit, red shows sinc-Gauss fit, and the blue curve is the 1$\sigma$ jackknife noise level. The sinc fit is often too thin to capture the full width of the spectral lines. The lines are shifted to the rest-frame based on the public spectroscopic redshifts reported in the literature. Fluxes are measured from the best-fit models. The fluxes of the lines are reported in Table \ref{table:linefluxes1}.  }
\label{fig:z0-005_post}
\end{figure*}

\begin{figure*}[!th]
\centering
\includegraphics[trim=0cm 0cm 0cm 0cm, scale=0.75]{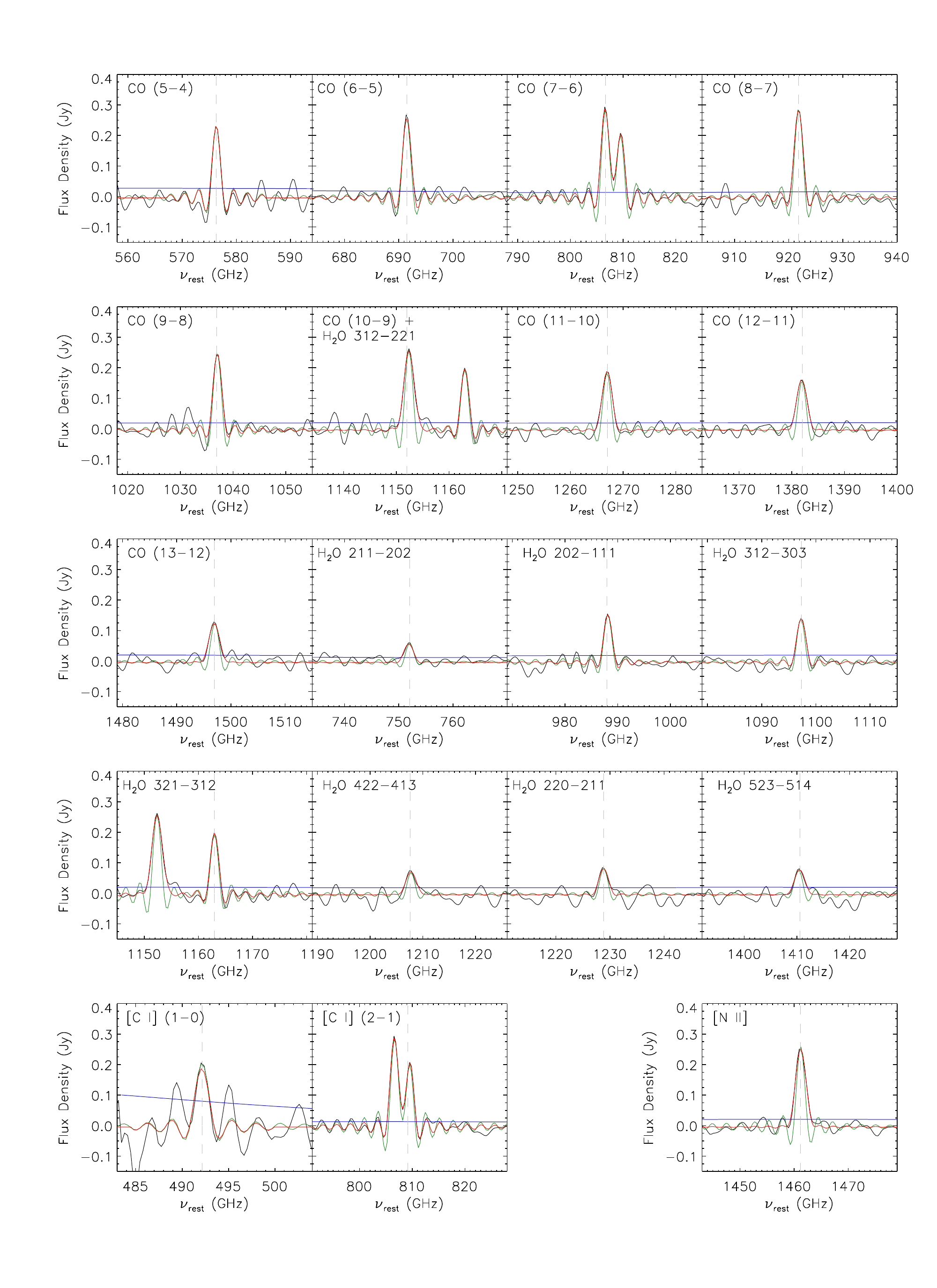}
\caption{Sinc-Gauss (red) and sinc (green) fits to the detected atomic and molecular lines in the stack at $0.05<z<0.2$, with the spectrum itself in black. We detect all the lines same as the low redshift stack (Figure \ref{fig:z0-005_post}) albeit with a different detection significance. In particular [C\,I] (1-0) is marginally detected in this redshift bin as fewer than ten sources contribute to the stack at this frequency, leading to a higher jackknife noise level. Fluxes of lines detected in this stack are also reported in Table \ref{table:linefluxes1}.}
\label{fig:z005-02_post}
\end{figure*}

\begin{figure}[!th]
\centering
\includegraphics[trim=0cm 0cm 0cm 0cm, scale=0.75]{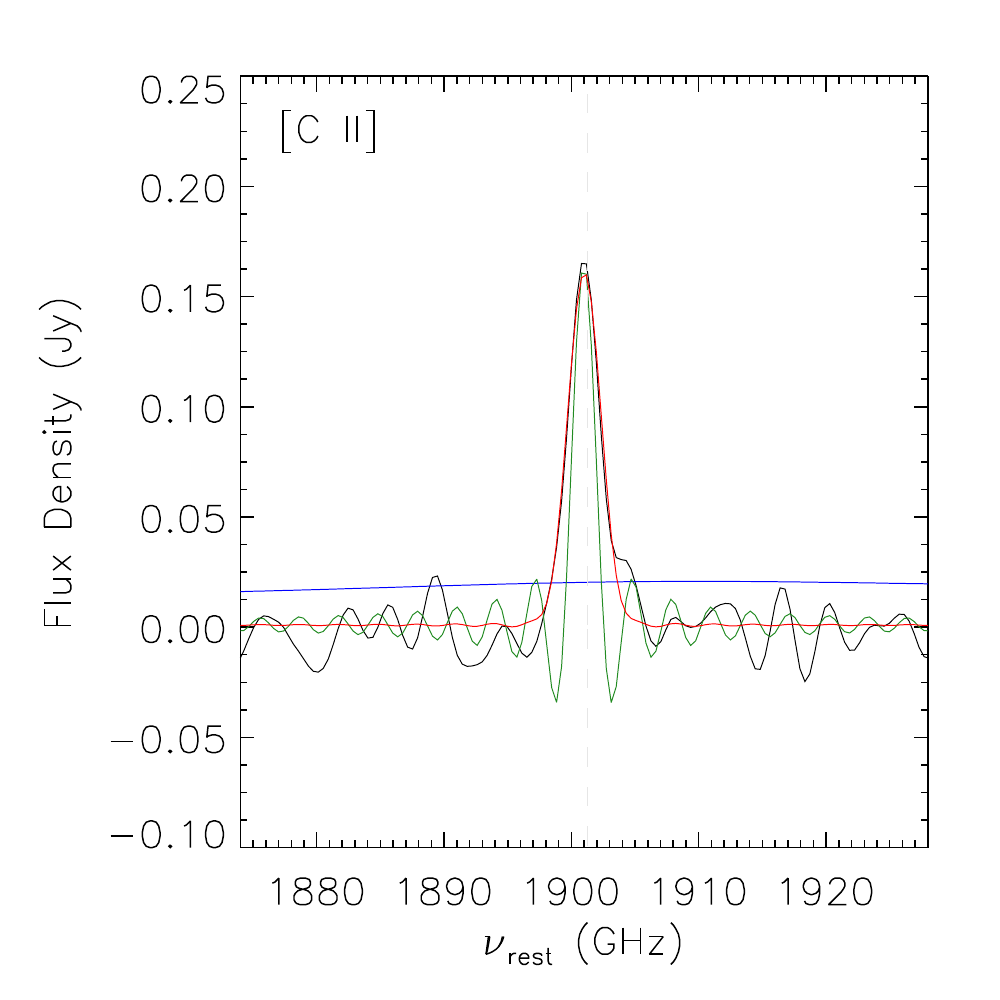}
\caption{Sinc-Gauss (red) and sinc (green) fits to the [C\,II] line in the $0.2<z<0.5$ stack. The spectrum itself is shown in black with the 1$\sigma$ noise level in blue.}
\label{fig:int_z_post}
\end{figure}

\begin{figure*}[!th]
\centering
\includegraphics[trim=0cm 0cm 0cm 0cm, scale=0.7]{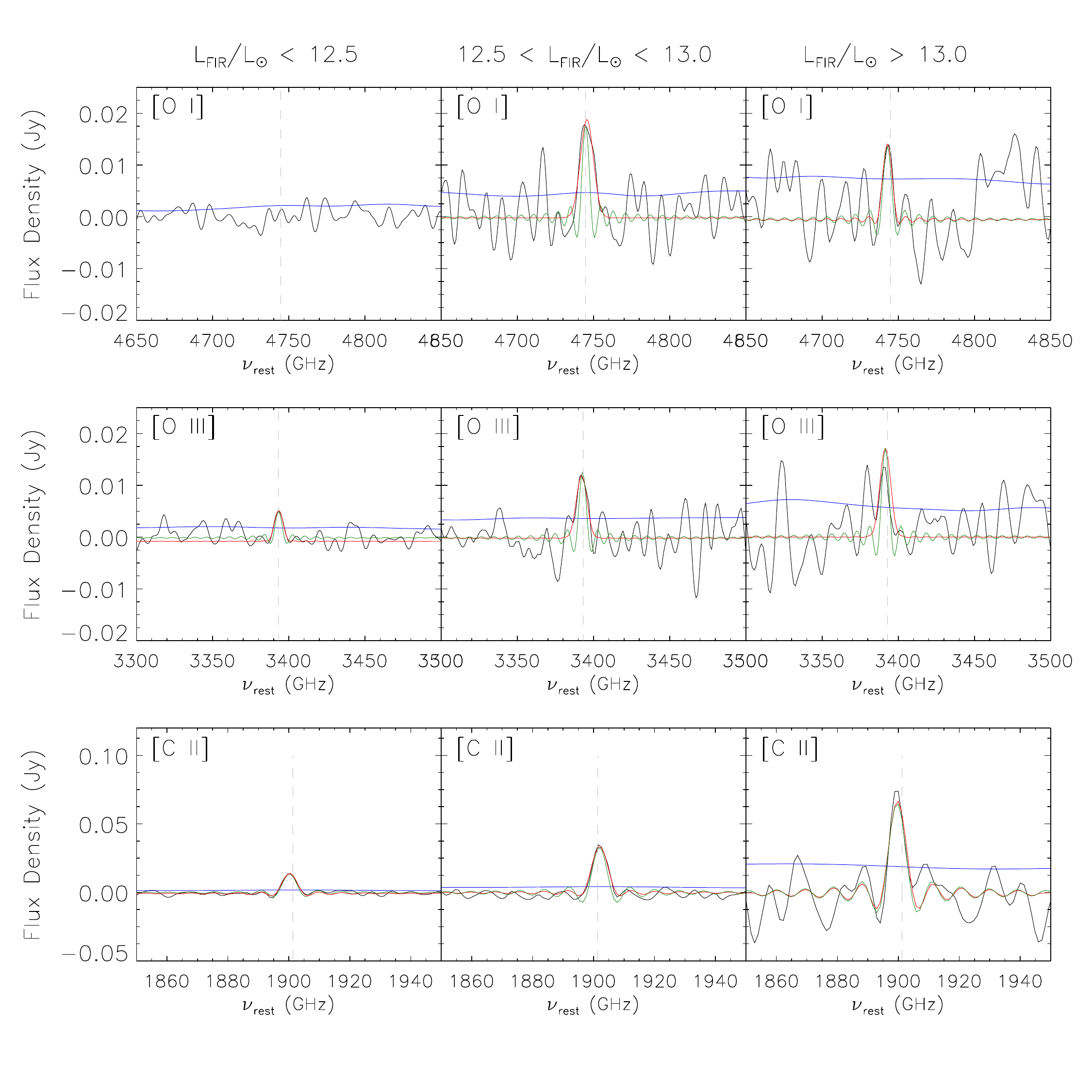}
\caption{Fits to lines for the three luminosity bins of the high-redshift sources. The sinc-Gauss fit is shown in red, and the sinc-only fit is shown in green. The spectrum itself in black, and the 1$\sigma$ jackknife noise level is in blue.}
\label{fig:high_z_post}
\end{figure*}


\begin{figure}[!th]
\centering
\includegraphics[trim=1cm 0cm 0cm 0cm, scale=0.75]{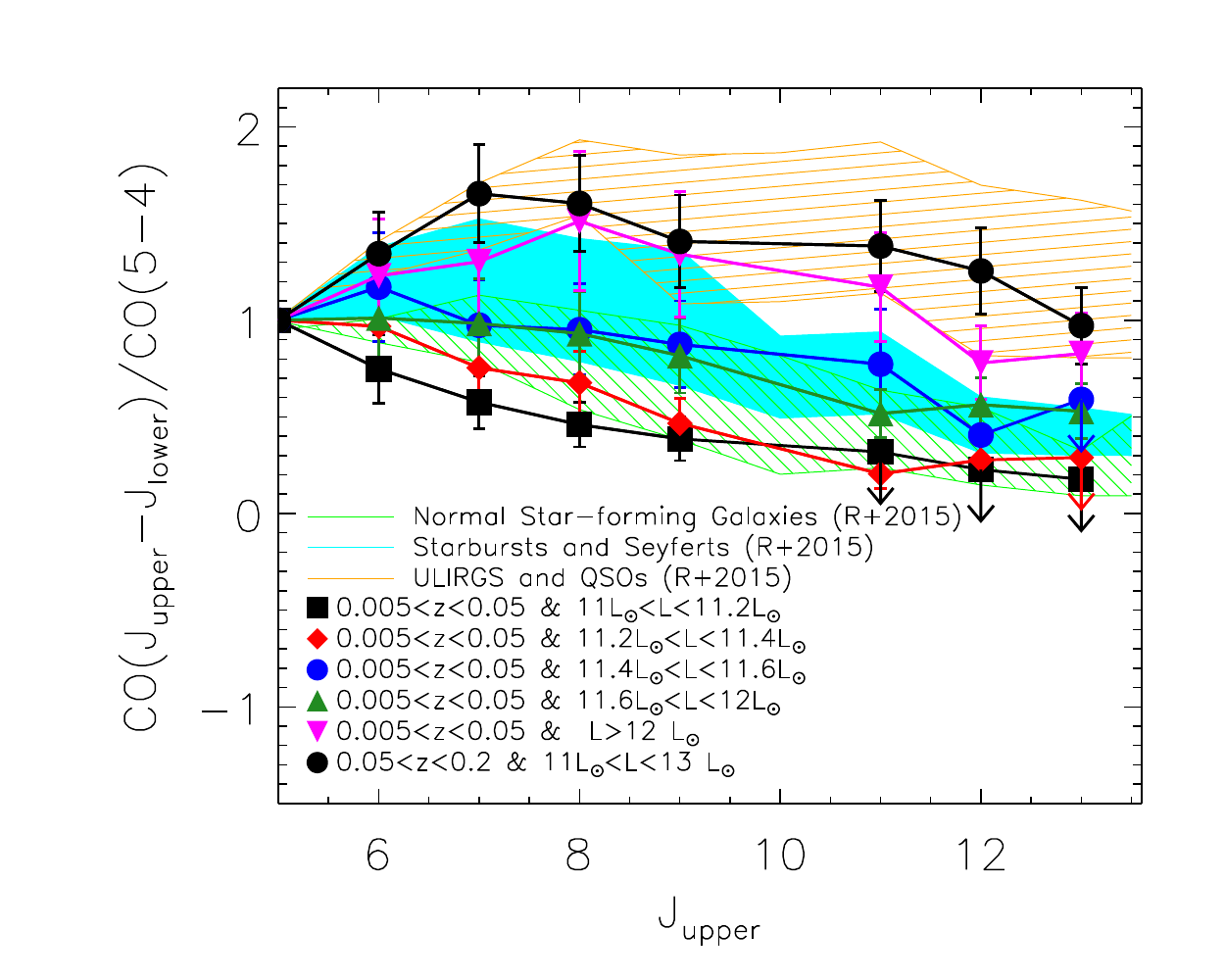}
\includegraphics[trim=1.5cm 0cm 0cm 0.6cm, scale=0.65]{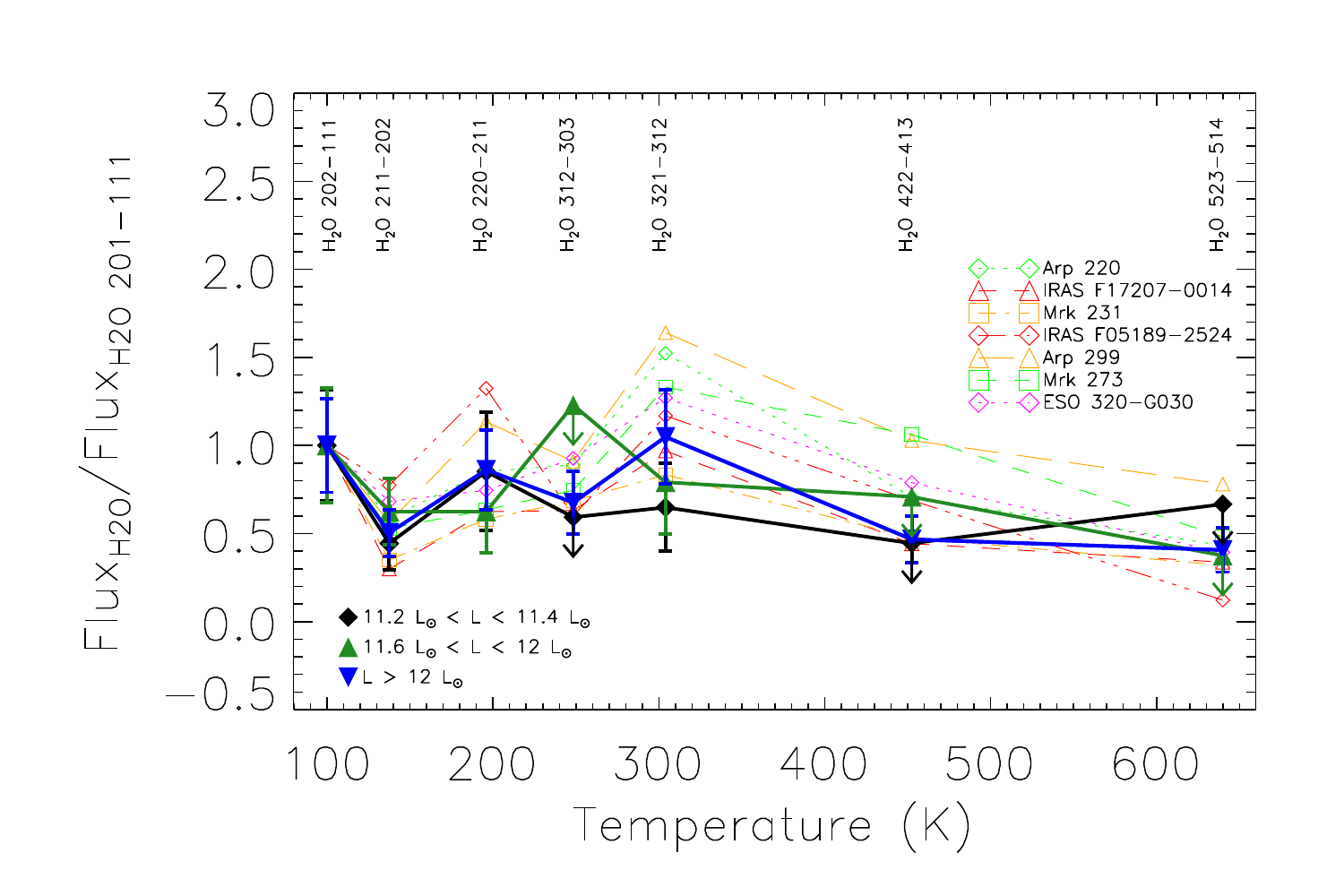}
\caption{{\it Top:} The carbon monoxide spectral line energy distribution for $0.005 < z < 0.05$ in five luminosity bins as presented in Figure \ref{fig:lowz_lirbins}. The filled regions are taken from \citet{Rosenberg2015} (see also \citealt{RobertsBorsani2017}), and they correspond to the range of CO flux ratios in normal star-forming galaxies (green stripes), starbursts and Seyferts (solid cyan), and ULIRGs and QSOs (orange stripes). {\it Bottom:} Spectral line energy distribution for transitions in water as a function of excitation temperature as in \citet{Yang2013} at $0.005 < z < 0.05$ in the luminosity bins in which water lines were strongly detected. These detections are compared to the water spectral line energy distribution for individual sources fit using sinc-Gauss profiles.}
\label{fig:co_lumbins}
\end{figure}

\begin{figure*}[t]
\centering
\includegraphics[trim=2cm 0cm 0cm 0cm, scale=0.85]{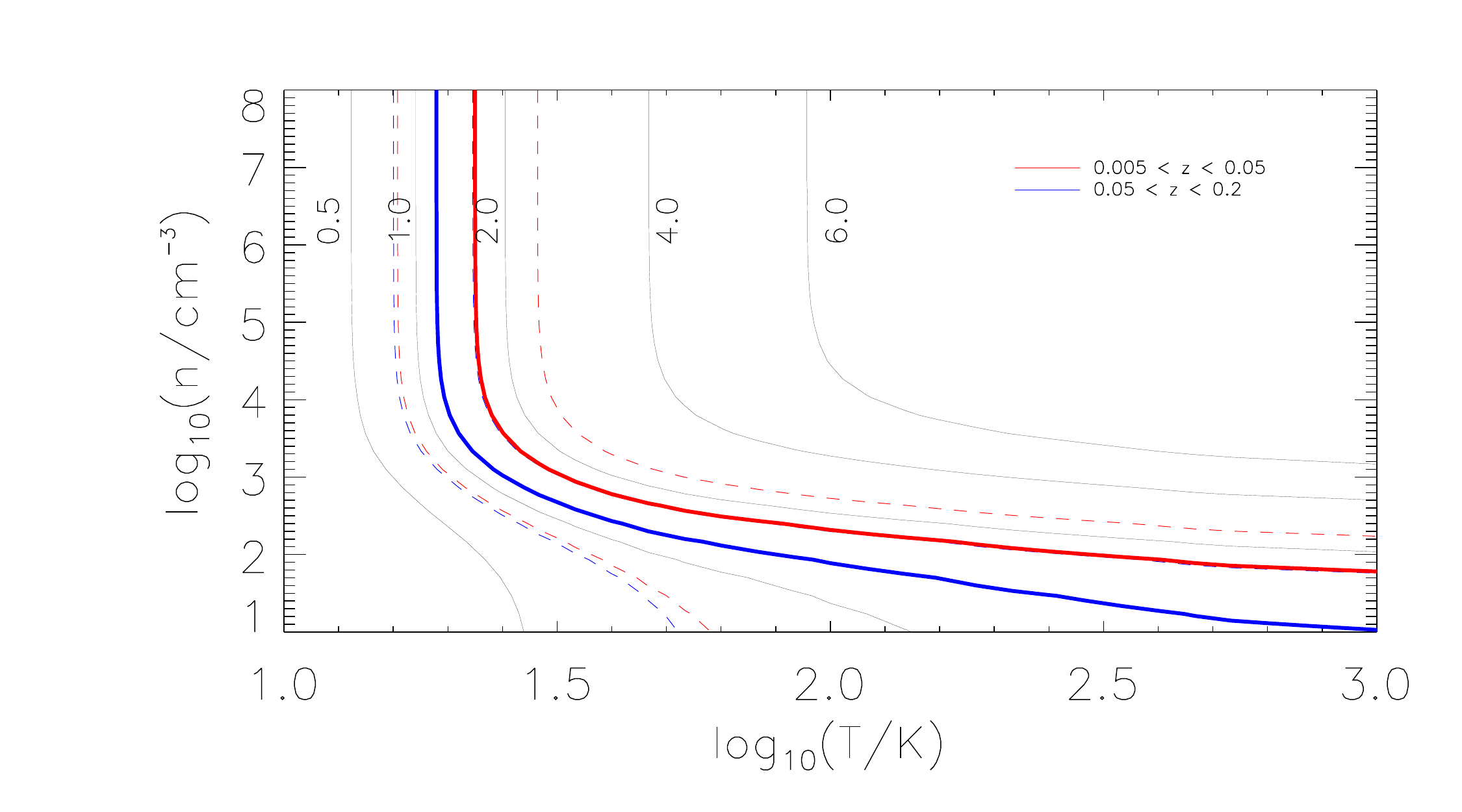}
\caption{Conditions in the ISM as probed by neutral [C\,I] (2-1)/[C\,I] (1-0) line ratio for $0.005 < z < 0.05$ and $0.05 < z < 0.2$ redshift bins. RADEX contours for an array of theoretical [C\,I] (2-1)/[C\,I] (1-0) ratios are shown in black. The dashed lines represent the 1$\sigma$ uncertainty.}
\label{temp_ci_lumbins}
\end{figure*}


\section{Stacking Analysis}

The Level-2 FTS spectral data are procured from the {\it Herschel} Science Archive (HSA) where they have already been reduced using version SPGv14.1.0 of the \textit{Herschel} Interactive Processing Environment (HIPE, \citet{Ott2010}) SPIRE spectrometer single pointing pipeline \citep{Fulton2016} with calibration tree \textsc{SPIRE\_CAL\_14\_2}. We use the point-source calibrated spectra. Additional steps are required to further reduce the data. An important step is the background subtraction. While {\it Herschel}/SPIRE-FTS observations include blank sky dark observations taken on or around the same observing day as the source observations are taken, they do not necessarily provide the best subtraction of the background \citep{Pearson2016}. The same study also showed that attempts to use a super-dark by combining many dark-sky observations into an average background do not always yield an acceptable removal of the background from science observations. Instead, the off-axis detectors present in each of the SPIRE arrays are used to construct a ``dark'' spectrum \citep{Polehampton2015}. These off-axis detectors provide multiple measurements of the sky and telescope spectra simultaneous with the science observations and are more effective at correcting the central spectrum. The background is constructed by taking the average of the off-axis detector spectra, but only after visually checking the spectra via HIPE's background subtraction script \citep{Polehampton2015} to ensure that the background detectors do not contain source emission. If any outliers are detected, they are removed from the analysis. Such outliers are mainly due to science observations that contain either an extended source or a random source that falls within the arrays. We use the average from all acceptable off-axis detectors from each science observation as the background to subtract from the central one. In a few unusual cases, a continuum bump from residual telescope emission in some spectra was better subtracted using a blank sky dark observation rather than an off-axis subtraction. In these cases, background subtraction was performed using the blank sky dark observation.

As part of the reduction, and similar to past analysis (e.g., \citealt{Rosenberg2015, Pearson2016}), we found a sizable fraction of the sources to show a clear discontinuity in flux between the continuum levels of the central SLW and SSW detectors in the overlap frequency interval between 944\,GHz and 1018\,GHz. If this discontinuity is still visible after the background subtraction (off-axis detector background or blank sky observation background) as discussed above, then we considered this offset to be an indication of extended source emission. For extended sources, we subtract a blank sky dark (and not an off-axis dark, as off-axis detectors may contain source emission) and correct for the source's size with HIPE's semiExtendedCorrector tool (SECT, \citealt{Wu2013}), following the \citet{Rosenberg2015} method of modeling the source as a Gaussian and normalizing the spectra for a Gaussian reference beam of 42$^{\prime\prime}$.

There are two other sources of discontinuity between the SLW and SSW detectors, one from a flux droop in the central SLW detector due to the recycling of the SPIRE cooler \citep{Pearson2016} and another due to potential pointing offsets \citep{Valtchanov2014}. Due to the differences in the size of the SLW and SSW SPIRE beams, a pointing offset can cause a larger loss of flux in the SSW beam than in the SLW beam. If an extended source correction was not able to fix the discontinuity between the SLW and SSW detectors, the discontinuity may likely be coming from the cooler recycling or from a pointing offset. We assume that these two effects are negligible, as we remove any continuum remaining after the application of SECT from the central SLW and SSW detectors by subtracting a second-order polynomial fit to the continuum.

Once the corrected individual spectra are obtained, the high-redshift lensed sample was corrected for lensing magnification. The magnification factors come
from lens models based on Sub-millimeter Array (SMA) and Keck/NIRC2-LGS adaptive optics observations \citep{Bussmann2013, Calanog2014}. Though these are mm-wave and optical magnifications while the present study involves far-IR observations, we ignore any effects of differential magnification \citep{Serjeant2014}. We simply make use of the best determined magnification factor, mainly from SMA analysis \citep{Bussmann2013}. For the overlapping lensed source sample with PACS spectroscopy, the lensing magnification factor used here is consistent with values used in \citet{Wardlow2017}. Sources with PACS spectroscopy that appear in \citet{Wardlow2017} are marked in Table \ref{table:all_targets}.


\begin{figure}[!th]
\centering
\includegraphics[trim=0.5cm 0cm 0cm 0cm, scale=0.65]{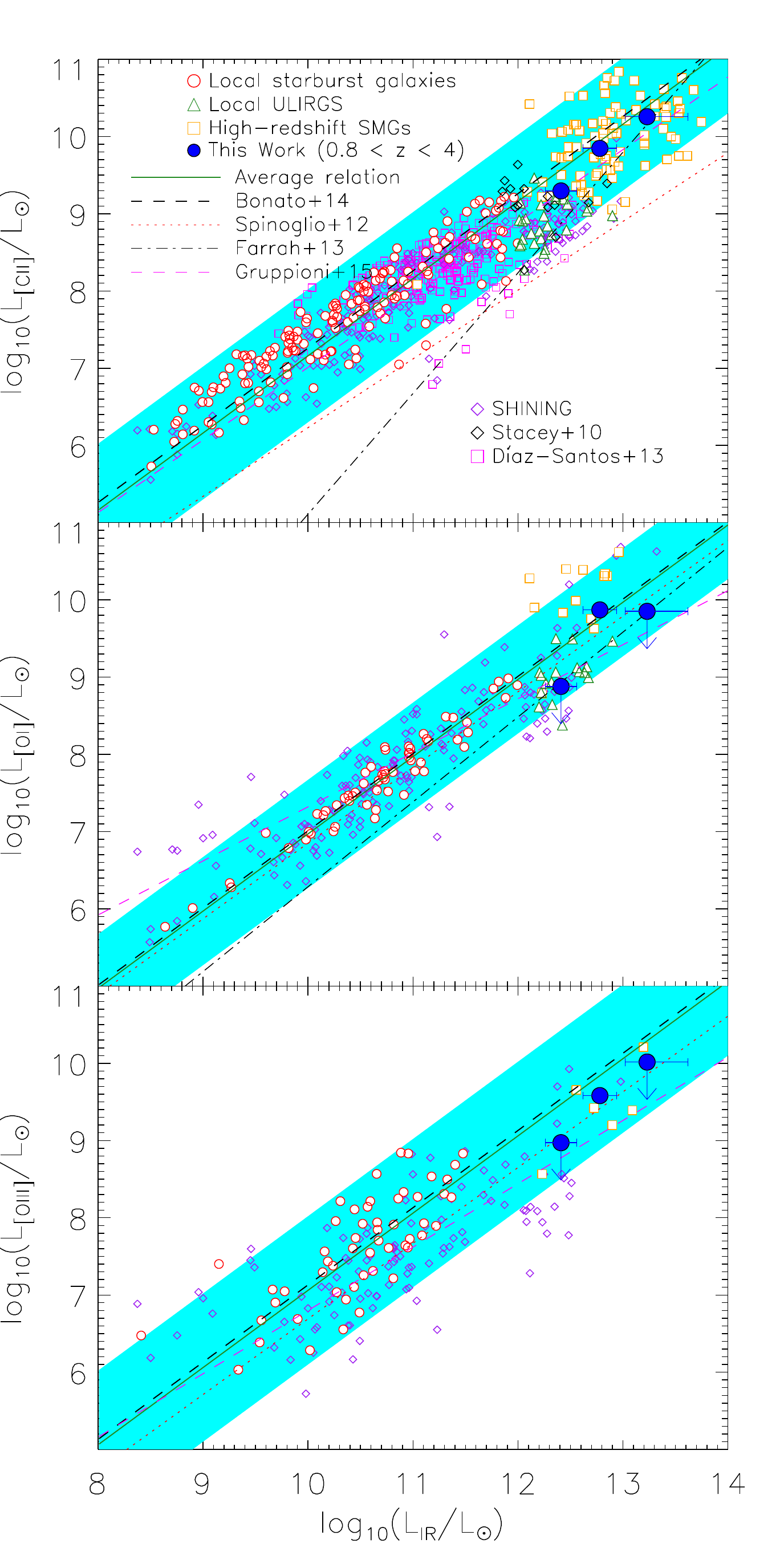}
\caption{Line versus infrared luminosity (rest-frame $8-1000\,\mu$m), L$_{\rm IR}$, of star-forming galaxies for [C\,II], [O\,I], and [O\,III] fine
structure lines at high redshift. Background data are from the literature sources listed in the text. The solid green lines correspond to the average L$_{\rm line}$/L$_{\rm IR}$ ratios (-3.03,
-2.94 and -2.84) for the [O\,I] 63.18\,$\mu$m, [O\,III] 88.36\,$\mu$m, and [C\,II] 157.7\,$\mu$m lines from the literature, respectively.
The reason for the choice of a linear relation is explained in the text. The cyan stripes
correspond to two times the dispersion around the mean relation ($\sigma = 0.35, 0.48$ and $0.43$,
respectively). Also shown, for comparison, are the $L_{\rm line}$/L$_{\rm IR}$ relations found in the literature
(see text).}
\label{data_calibration}
\end{figure}

\begin{figure*}[!th]
\centering
\includegraphics[trim=1cm 0cm 0cm 0cm, scale=0.75]{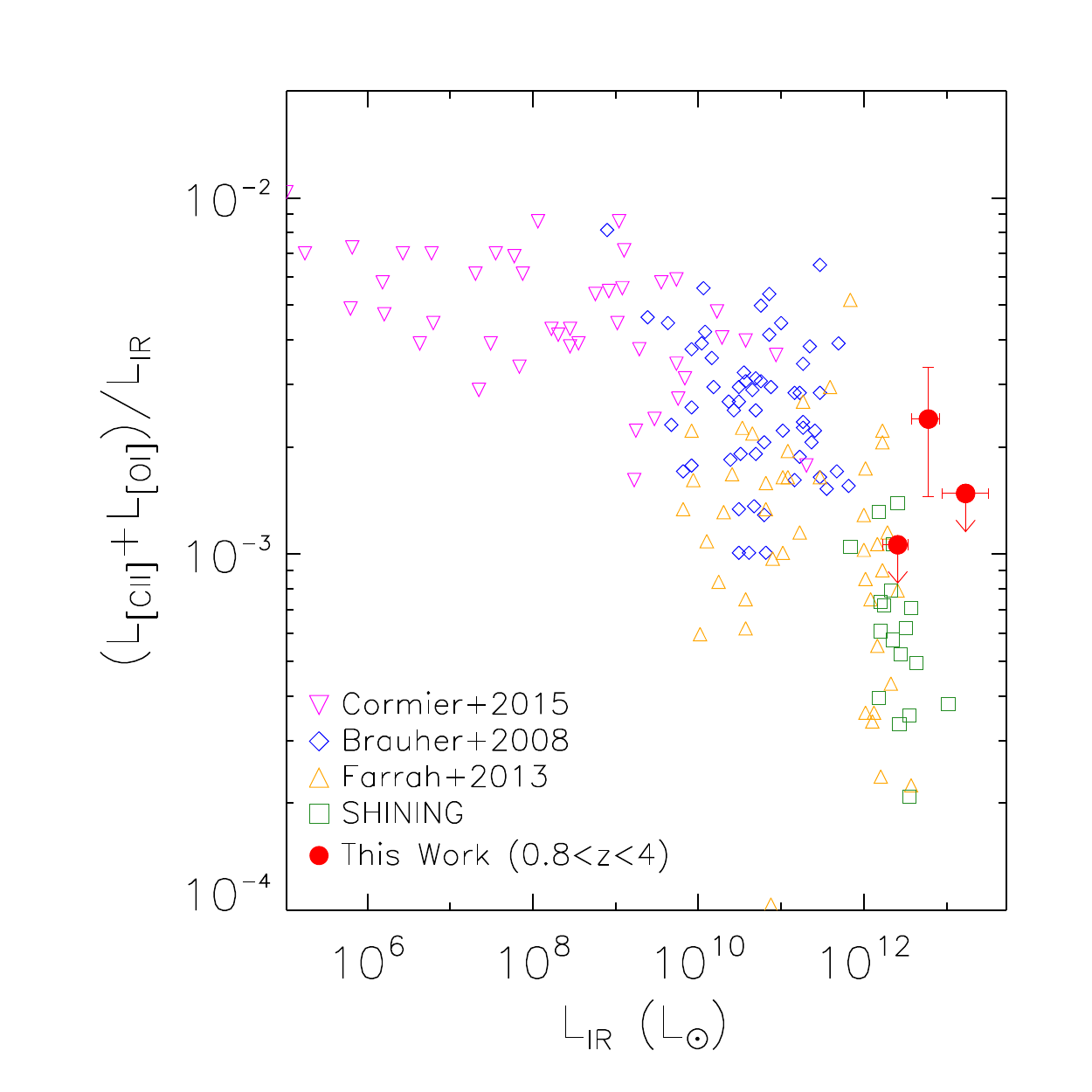}
\includegraphics[trim=1cm 0cm 0cm 0cm, scale=0.75]{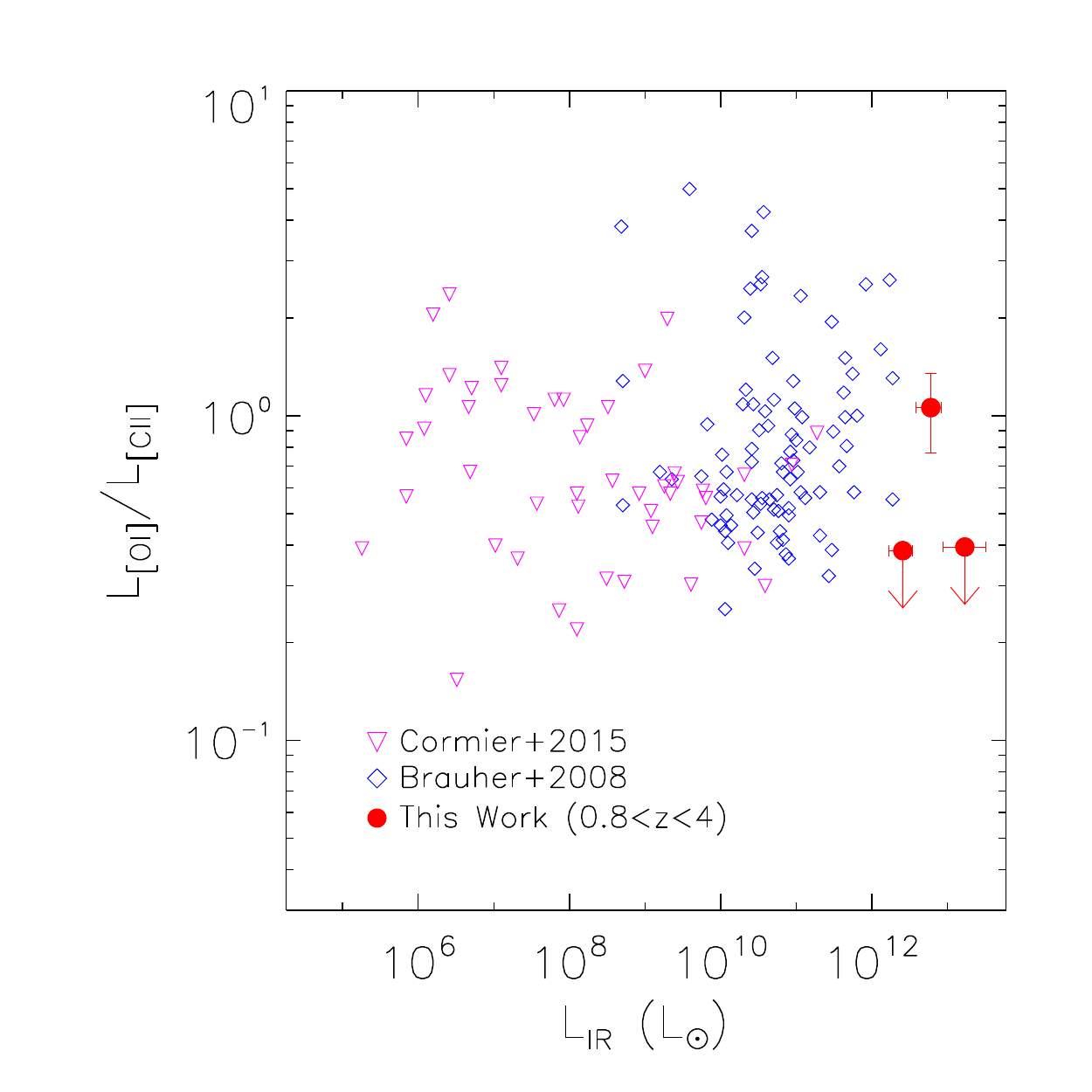}\\
\includegraphics[trim=1cm 0cm 0cm 1cm, scale=0.75]{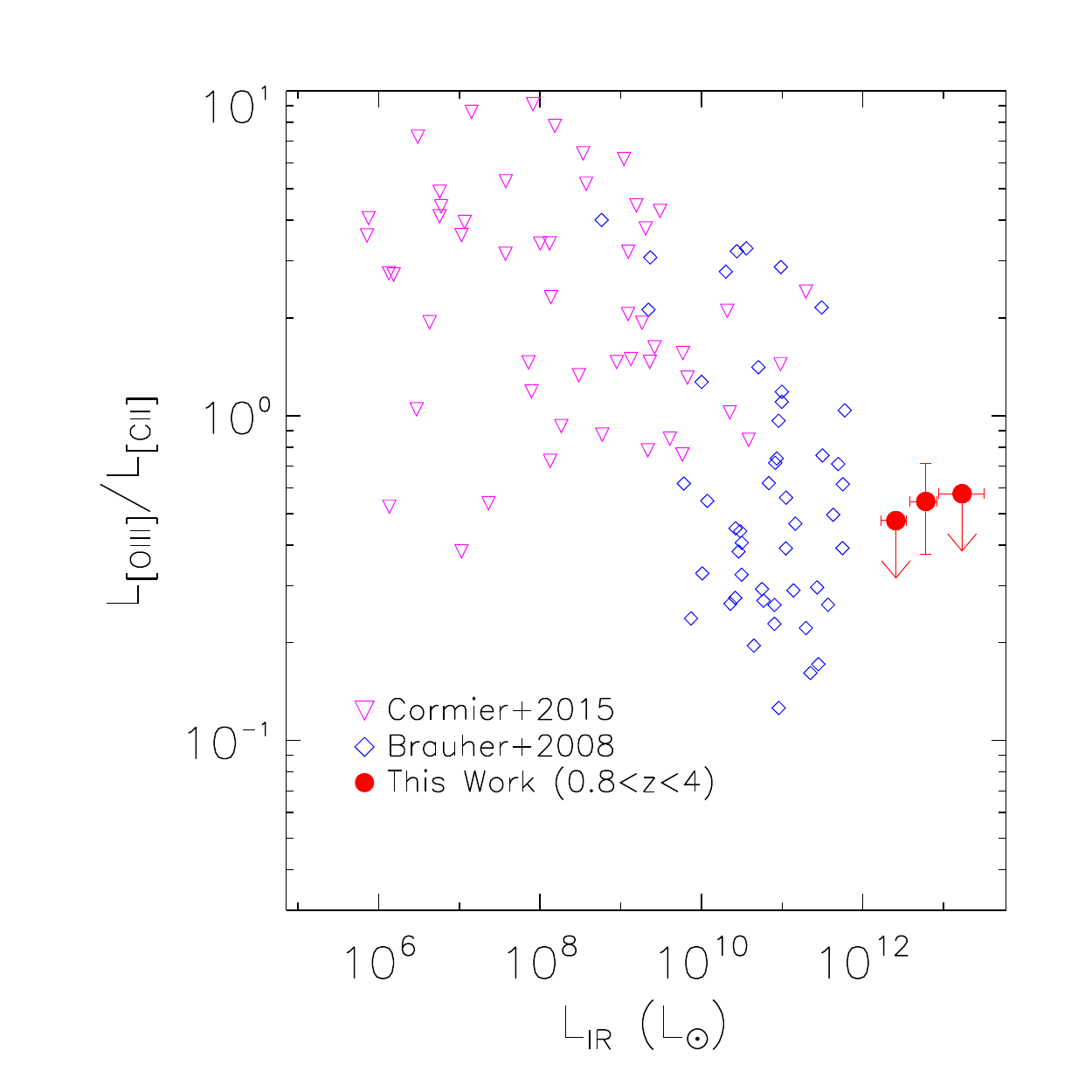}
\includegraphics[trim=1cm 0cm 0cm 1cm, scale=0.75]{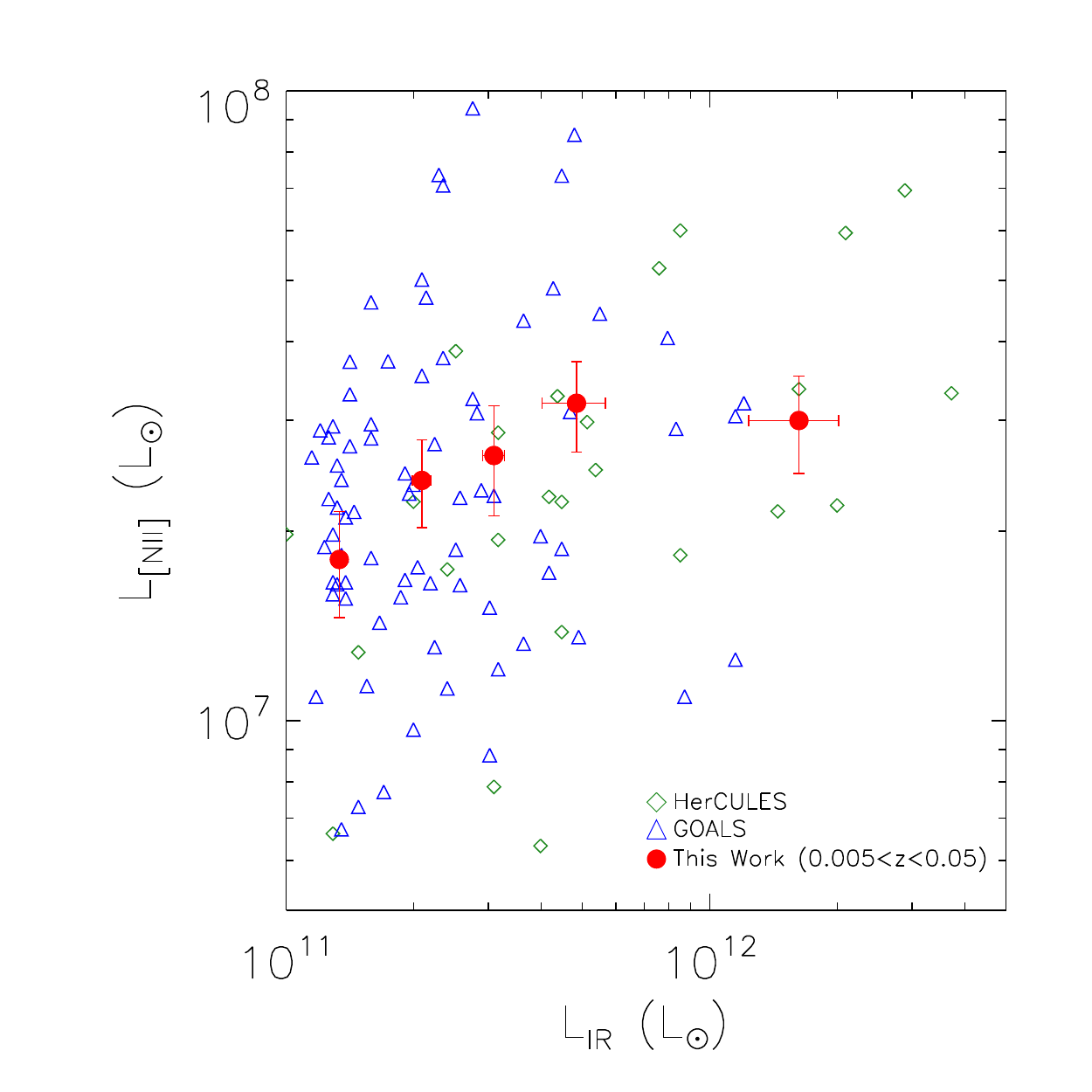}
\label{fig:cii_plus_oi}
\caption{{\it Top Left:} Ratio of ([C\,II]+[O\,I]) luminosity to total infrared luminosity (rest-frame $8-1000\,\mu$m) in three luminosity bins for sources with $0.8 < z < 4$ as a function of total infrared luminosity. The breakdown of the three luminosity bins is as follows: L$_{\rm IR} < 10^{12.5}\,$L$_{\odot}$, $10^{12.5}\,$L$_{\odot} \, < $L$_{\rm IR} < 10^{13}\,$L$_{\odot}$, and L$_{\rm IR} > 10^{13}\,$L$_{\odot}$; however, [O\,I] is only detected in the middle luminosity bin. For comparison, we show data from \citet{Cormier2015,Brauher2008,Farrah2013} and \citet{SHINING2011}. {\it Bottom Left and Top Right:} Line ratios as a function of total infrared luminosity in three luminosity bins for sources with $0.8 < z < 4$. For comparison, we show data from \citet{Cormier2015} and \citet{Brauher2008}. {\it Right:} Line luminosity of the [N\,II] transition in luminosity bins for sources at $0.005 < z < 0.05$. Background data were produced by fitting to the [NII] lines in individual spectra in the HerCULES and GOALS samples.}
\end{figure*}

To obtain the average stacked spectrum in each of the redshift bins or luminosity bins as we discuss later we follow the
stacking procedure  outlined by \citet{Spilker2014}.  It involves
scaling the flux densities in individual spectra in each redshift bin to the flux densities that the source would have were it located at some common redshift (which we take to be the mean redshift in each bin) and then scaling to a common luminosity so that we can present an average spectrum
of the sample. For simplicity, we take the mean redshift and median infrared luminosity in each bin and both scale up and scale down individual galaxy
spectra in both redshift and luminosity to avoid introducing biases in the average stacked spectrum; however we note that the sample does contain biases
associated with initial sample selections in the proposals that were accepted for {\it Herschel}/SPIRE-FTS observations. We discuss how such selections impact
a precise interpretation of the spectra in the discussion. We now outline the process used in the scaling of spectra.

The background-subtracted flux densities of the spectra are scaled to the flux values that they would have at the common redshift, which was taken to be the mean redshift in each of the redshift categories; namely, $z_{\rm com} = 0.02$ for the $0.005<z<0.05$ sources, $z_{\rm com} = 0.1$ for $0.05<z<0.2$ sources, $z_{\rm com} = 0.3$ for $0.2<z<0.5$ sources, $z_{\rm com} = 1.4$ for $0.8<z<2$ sources, and $z_{\rm com} = 2.8$ for $2<z<4$ sources. The choice between median or mean redshift does not significantly affect the overall spectrum or line fluxes.
The flux density and error values (error values are obtained from the error column of the level-2 spectrum products from the {\it Herschel} Science Archive) of each spectrum are multiplied by the scaling factor given in \citet{Spilker2014}:
\begin{equation}
f=\bigg(\frac{D_{\rm L}(z_{\rm src})}{D_{\rm L}(z_{\rm com})}\bigg)^2 \times\bigg(\frac{1 + z_{\rm com}}{1 + z_{\rm src}}\bigg)
\label{scale_factor}
\end{equation}
where $D_{\rm L}$ is the luminosity distance. The flux density and error values of each spectrum are then representative of the flux density and error values that the source would have were it located at $z_{\rm com}$. The frequency axes of the scaled spectra are then converted from observed-frame frequencies to rest-frame frequencies. 

To normalize the spectra, all spectrum flux densities and errors are scaled by a factor such that each source will have the same total infrared luminosity (rest-frame $8-1000 \, \mu$m); namely, L$_{\rm IR}  =  10^{11.35} \,$L$_{\odot}$, $10^{12.33}\,$L$_{\odot}$, $10^{11.89}\,$L$_{\odot}$, $10^{12.53}\,$L$_{\odot}$ and $10^{12.84}\,$L$_{\odot}$ in each of the five bins, respectively. In the two highest redshift bins, we calculate a total infrared luminosity by fitting a single-temperature, optically-thin, modified blackbody (i.e. greybody with $S({\nu})\propto\nu^{\beta}B_{\nu}(T)$ where $B_{\nu}(T)$ is the Planck function) spectral energy
distribution (SED) (commonly used in the literature, e.g. \citealp{Casey2012, Bussmann2013}) to the available photometry in the infrared from {\it Herschel} and public IRSA data. For this we use the publicly available code developed by \citet{Casey2012} assuming a fixed emissivity ($\beta = 1.5$) (e.g. \citealp{Bussmann2013}). The resulting infrared luminosities are presented in Table \ref{table:all_targets}, along with lensing magnification factors and references. Luminosities in the tables are corrected for lensing magnification (where applicable), and we ignore the uncertainty in magnification from existing lens models. Sources without a magnification of factor $\mu$ are not affected by gravitational lensing.

After the spectra are scaled to a common IR luminosity, a second-order polynomial is then fit to the continuum of each source and is subsequently subtracted from each source spectrum. Instrumental noise impacts the continuum subtraction and leads to residuals in the continuum-subtracted spectrum. These residuals in return impact the detection of faint lines.
A number of objects have multiple FTS spectra, taken at multiple time intervals as part of the same program or
observations conducted in different programs. 
Multiples of the same object are combined into a single average spectrum by calculating the mean flux density at each frequency for each of the repeats. This mean spectrum is what is used in the stacking procedure.
After the spectra are calibrated and scaled, the flux values at each frequency in the rest frame of the spectra are stacked using an inverse variance weighting scheme with the inverse of the square of the flux errors as weights. In the $0.005 < z < 0.05$ stack, a minority of the sources (though still a significant subset of the total) have high signal-to-noise ratios and thus dominate over the other sources when using the inverse variance weighting scheme. To avoid this bias without throwing out sources, we stack the $0.005 < z < 0.05$ bin by calculating the mean stack without inverse variance weighting. The unweighted mean stack is shown in Figure \ref{fig:z0-005}. The inverse variance weighted stack for this redshift bin is presented in the Appendix for comparison.

The noise level of the stacked spectrum in each of the five redshift bins is estimated using a jackknife technique in which we remove one source from the sample and then stack. The removed source is replaced, and this process is repeated for each source in the sample. The jackknife error in the mean of the flux densities at each frequency from the jackknifed stacks is taken to be the 1$\sigma$ noise level in the overall stacked spectrum in each redshift bin. The red curves in the upper panels of Figures \ref{fig:z0-005} - \ref{fig:z2-4} are found by smoothing the jackknife error curve.


\section{Stacking Results}

The stacked spectra in each of the five redshift bins are shown in Figures \ref{fig:z0-005} - \ref{fig:z2-4}, while in Figure \ref{fig:lowz_lirbins} we show the mean stacks (no inverse-variance weighting) for the $0.005 < z < 0.05$ bin by sub-dividing the sample into five luminosity bins given by $10^{11.0}\,$L$_{\odot}<$ L$_{\rm IR}<10^{11.2}\,$L$_{\odot}$, $10^{11.2}\,$L$_{\odot}<$ L$_{\rm IR}<10^{11.4}\,$L$_{\odot}$, $10^{11.4}\,$L$_{\odot}<$ L$_{\rm IR}<10^{11.6}\,$L$_{\odot}$, $10^{11.6}\,$L$_{\odot}<$ L$_{\rm IR}<10^{12.0}\,$L$_{\odot}$, and L$_{\rm IR}>10^{12.0}\,$L$_{\odot}$. For the purposes of this study and for PDR model interpretations, we concentrate on lines that are detected at a 
signal-to-noise ratio greater than 3.5. The stacks do reveal detections with a signal-to-noise ratios at the level of 2.5 to 3; we will return to those lines in future papers.

The natural line shape of the SPIRE FTS is a sinc profile \citep{Swinyard2014}. A sinc profile is typically used to fit unresolved spectral lines. However, a sinc profile may be too thin to fully capture the width of broad partially-resolved extragalactic spectral lines, in which case a sinc-Gauss (sinc convolved with a Gaussian) can provide a better fit\setcounter{footnote}{0}\footnote{\url{http://herschel.esac.esa.int/hcss-doc-15.0/index.jsp\#spire_drg:_start}}. For spectral lines with the same intrinsic line width, the sinc-Gauss fit gives a higher flux measurement than the sinc fit; the ratio of sinc-Gauss to sinc flux increases as a function of increasing spectral line frequency. For broad line-widths, the sinc-Gauss fit contains significantly more flux than the pure sinc fit. Because the stacked SPIRE/FTS spectra contain a variety of widths for each spectral line and because the width of each line is altered when scaling the frequency axis of the spectra to the common-redshift frame, the sinc profile appeared to under-fit all of the spectral lines in the stacked spectra, so a sinc-Gauss profile was used for flux extraction. See Figures \ref{fig:z0-005_post} - \ref{fig:high_z_post}. The width of the sinc component of the fit was fixed at the native SPIRE FTS resolution of 1.184\,GHz, and the width of the Gaussian component was allowed to vary. The integral of the fitted sinc-Gauss profile was taken to be the measured flux. The fluxes from the fits are presented in Tables \ref{table:linefluxes1} - \ref{table:linefluxes3}. In the case of an undetected line (i.e., the feature has less than 3.5$\sigma$ significance), we place an upper limit on its flux by injecting an artificial line with velocity width 300 km s$^{-1}$ (a typical velocity width for these lines; e.g., \citealt{Magdis2014}) into the stack at the expected frequency and varying the amplitude of this line until it is measured with 2$\sigma$ significance. The flux of this artificial line is taken to be the upper limit on the flux of the undetected line.

The error on the fluxes includes a contribution from the uncertainty in the fits to the spectral lines as well as a 6$\%$ uncertainty from the absolute calibration of the FTS. The error due to the fit is estimated by measuring the ``bin-to-bin'' spectral noise of the residual spectrum in the region around the line of interest (see SPIRE Data Reduction Guide). The residual spectrum is divided into bins with widths of 30 GHz, and the standard deviation of the flux densities within each bin is taken to be the noise level in that bin. Additionally, we incorporate a 15$\%$ uncertainty for corrections to the spectra for (semi)-extended sources \citep{Rosenberg2015} in the lowest redshift stack. This 15\% uncertainty is not included for sources with $z > 0.05$, as these are all point sources (as verified by inspection).

We now discuss our stacking results for the five redshift bins; for simplicity we define low-redshift as $0.005 < z< 0.2$, intermediate as $0.2 < z< 0.5$ and high-redshift as $0.8 < z<4$; both low and high-redshift have two additional redshift bins. Within these bins we also consider luminosity bins when adequate statistics allow us to further divide
the samples.

\subsection{Low-redshift stacks}

Figures \ref{fig:z0-005} and \ref{fig:z005-02} show the stacked FTS spectra and corresponding uncertainty along with major atomic and molecular emission and absorption lines for the $0.005<z<0.05$ and $0.05<z<0.2$ bins respectively. With the large number of galaxy samples, the far-IR spectrum of lowest redshift bin results in a highly reliable average spectrum showing a number of ISM atomic and molecular emission lines. In particular we detect all the CO lines with $J_{\rm upper}>5$ out to the high excitation line of $\rm CO(13-12)$. This allows us to construct the CO spectral line energy distribution (SLED) and to explore the ISM excitation state in DSFGs in comparison with other starbursts and that of normal star-forming galaxies (see Section 5). We further detect multiple H$_2$O emission lines in these stacks which arise from the very dense regions in starbursts. The strength of the rotational water lines rivals that of the CO transition lines.  We additionally detect the [C\,I] (1-0) at 609\,$\mu$m and [C\,I] (2-1) at 370\,$\mu$m along with [N\,II] at 205\,$\mu$m in both redshift bins. We will use these measured line intensity ratios in Section 5 to construct photodissociation region models of the ISM and to study the density and ionizing photon intensities. We note here that the [C\,I] line ratios are very sensitive to the ISM conditions and would therefore not always agree with more simplistic models of the the ISM. We will discuss these further in Section 5. For comparison to Figure \ref{fig:z0-005}, which is stacked using an unweighted mean, Figure \ref{fig:z005-02_inv} shows the $0.005 < z < 0.05$ sources stacked with an inverse variance weighting. A few absorption lines also appear in the low-redshift stack. Despite Arp 220 \citep{Rangwala2011} being the only individual source with strong absorption features, many of the absorption features are still present in the stack due to the high signal-to-noise ratio of Arp 220 in conjunction with an inverse variance weighting scheme for stacking. The SPIRE FTS spectrum of Arp 220 has been studied in detail in \citet{Rangwala2011} and is characterized by strong absorption features in water and related molecular ions OH$^+$ and H$_2$O$^+$ interpreted as a massive molecular outflow.

The best-fit profiles of the detected lines in the low-redshift stacks are shown in Figures \ref{fig:z0-005_post} and \ref{fig:z005-02_post} for the $0.005<z<0.05$ and $0.05<z<0.2$ redshift bins, respectively. Fluxes in $\rm{W\,m}^{-2}$ are obtained by integrating the best-fit line profiles. Table \ref{table:linefluxes1} summarizes these line fluxes as well as velocity-integrated fluxes from the sinc-Gauss fits for detections with $\rm S/N > 3.5$ in these stacks.

As discussed above, we further stack the lowest redshift bin ($0.005<z<0.05$) in five infrared luminosity bins. Figure \ref{fig:lowz_lirbins} shows the stacked FTS spectra each of these luminosity bins. See the caption in Figure \ref{fig:lowz_lirbins} for the redshift and luminosity breakdown of the sample. By comparing these stacks we can look at the effects of infrared luminosity on emission line strengths. It appears from these stacked spectra that the high-$J$ CO lines are comparable in each of the luminosity bins. We explore the variation in the [N\,II] line in the discussion. Fluxes for the lines in each luminosity bin are tabulated in Figure \ref{table:linefluxes2}.

\subsection{Intermediate-redshift stacks}

We show the intermediate-redshift ($0.2<z<0.5$) stack in Figure \ref{fig:z02-5}. Due to the limited number of galaxies observed with SPIRE/FTS in this redshift range, we only detect a bright [C\,II] line with our threshold signal-to-noise ratio of 3.5. The [C\,II] 158\,$\mu$m fine structure line is a main ISM cooling line and is the most pronounced ISM emission line detectable at high redshifts, when it moves into mm bands, revealing valuable information on the state of the ISM. We further discuss these points in Section 5.
Figure \ref{fig:int_z_post} shows the best-fit profile to the [C\,II] line in the intermediate redshift. The measured fluxes from this profile are reported in Table \ref{table:linefluxes1}. The average  [C\,II] flux from the stack is lower than the measurements reported in \citet{Magdis2014} for individual sources (note that our $0.2 < z < 0.5$ is comprised almost entirely of the sources from \citet{Magdis2014}, the exception being the source IRAS 00397-1312). Stacking without IRAS 00397-1312 leads to similar results. We attribute the deviation of the stack [C\,II] flux toward lower values to the scalings we apply when shifting spectra to a common redshift and common luminosity during the stacking process.

\subsection{High-redshift stacks}

The high redshift ($0.8<z<2$ and $2<z<4$) FTS stacks are shown in Figures \ref{fig:z08-2} and \ref{fig:z2-4} consisting of 36 total individual spectra for sources in Table \ref{table:all_targets}. The stack at $0.8<z<2$ also suffers from a limited number of galaxies observed with the FTS. At $0.8<z<2$, [C\,II] 158\,$\mu$m and [O\,III] 88\,$\mu$m appear. We detect [C\,II] at 158\,$\mu$m, [O\,III] at 88\,$\mu$m and [O\,I] at 63\,$\mu$m atomic emission lines with $\rm S/N>3.5$ in the stacked spectra at $2<z<4$. The relative line ratios of these main atomic fine structure cooling lines will be used to construct the photodissociation region model of the ISM of DSFGs at these extreme redshifts to investigate the molecular density and radiation intensity. 

To study the strengths of spectral lines at different luminosities, all sources with $z>0.8$ were combined into a single sample and then divided into three luminosity bins with roughly the same number of sources in each bin. The average luminosities in the three bins are $10^{12.41}\,$L$_{\odot}$, $10^{12.77}\,$L$_{\odot}$, and $10^{13.24}\,$L$_{\odot}$. See Tables \ref{table:linefluxes3} and \ref{table:highz_flux} for the precise breakdown of the sample and measured fluxes. Each of the subsamples is separately stacked, and the line fluxes are measured as a function of far-infrared luminosity. Figure \ref{fig:high_z_post} shows the best-fit line profiles to the three main detected emission lines in the three infrared luminosity bins. The ISM emission lines are more pronounced with increasing infrared luminosity. This agrees with results of individual detected atomic emission lines at high redshifts \citep{Magdis2014, Riechers2014} although deviations from a main sequence are often observed depending on the physics of the ISM in the form of emission line deficits \citep{Stacey2010}. These are further discussed in the next section. 

\section{Discussion}

The ISM atomic and molecular line emissions observed in the stacked spectra of DSFGs can be used to characterize the physical condition of the gas and radiation in the ISM across a wide redshift range. This involves investigating the CO and water molecular line transitions and the atomic line diagnostic ratios with respect to the underlying galaxy infrared luminosity for comparison to other populations and modeling of those line ratios to characterize the ISM.  

\subsection{The CO SLED}

The CO molecular line emission intensity depends on the conditions in the ISM. Whereas the lower-$J$ CO emission traces the more extended cold molecular ISM, the high-$J$ emissions are observational evidence of ISM in more compact starburst clumps (e.g., \citealt{Swinbank2011}). In fact, observations of the relative strengths of the various CO lines have been attributed to a multi-phase ISM with different spatial extension and temperatures \citep{Kamenetzky2016}. The CO spectral line energy distribution (SLED), plotted as the relative intensity of the CO emission lines as a function of the rotational quantum number, $J$, hence reveals valuable information on the ISM conditions (e.g., \citealt{Lu2014}. 

Figure \ref{fig:co_lumbins} shows the high-$J$ CO SLED of the DSFGs for stacks in the two low redshift bins of $0.005<z<0.05$ and $0.05<z<0.2$. Here we are limited to the $J_{\rm upper}>5$ CO SLED covered by the SPIRE/FTS in the redshift range probed. A combined {\it Herschel}/SPIRE and PACS stacked spectra of DSFGs and corresponding full CO SLED will be presented in Wilson et al. in prep. The CO SLED is normalized to CO (5-4) line flux density and plotted as a function of ${J_{\rm upper}}$. The background colored regions in \ref{fig:co_lumbins} are from \citet{Rosenberg2015} in which they determined a range of CO flux ratios for three classes of galaxies from the HerCULES sample: star-forming objects, starbursts and Seyferts, and ULIRGs and QSOs. The $0.005<z<0.05$ sample is consistent with the starbursts and Seyfert regions whereas line measurements from stacked spectra in $0.05<z<0.2$ redshift bin are more consistent with ULIRGs and QSO regions. Both measurements are higher than the expected region for normal star-forming galaxies which indicates a heightened excitation state in DSFGs specifically at the high-$J$ lines linked to stronger radiation from starbursts and/or QSO activity. 

Increased star-formation activity in galaxies is often accompanied by an increase in the molecular gas reservoirs. This is studied locally as a direct correlation between the observed infrared luminosity and CO molecular gas emission in individual LIRGs and ULIRGs \citep{Kennicutt2012}. To further investigate this correlation, we looked at the CO SLED in our low-$z$ ($0.005<z<0.05$) sample in bins of infrared luminosity (Figure \ref{fig:lowz_lirbins}). Figure \ref{fig:co_lumbins} further shows the CO SLED for the the different luminosity bins. The stronger radiation present in the higher luminosity bin sample, as traced by the total infrared luminosity, is responsible for the increase in the CO line intensities. In the high luminosity bin sample, the excitation of the high-$J$ lines could also partially be driven by AGN activity given the larger fraction of QSO host galaxies in the most IR luminous sources (e.g., \citealt{Rosenberg2015}).

\subsection{ISM Emission Lines}

\subsubsection{Atomic and Molecular Line Ratios}

We detect several $\rm H_2O$ emission lines in the two lowest redshift bins of $0.005<z<0.05$ and $0.05<z<0.2$. Fluxes from detected water rotational lines are plotted in Figure \ref{fig:co_lumbins}, along with data from fits made to individual spectra from the sample that exhibited strong water line emission. These include well-known sources such as Arp 220 at $z = 0.0181$ \citep{Rangwala2011} and Mrk 231 at $z = 0.0422$ \citep{Vanderwerf2010,Gonzalez2010}. $\rm H_2O$ lines are normally produced in the warm and most dense regions of starbursts \citep{Danielson2011} and may indicate infrared pumping by AGN \citep{GonzalezAlfonso2010,Bradford2011}. Figure \ref{fig:co_lumbins} also shows the different water emission lines and the ISM temperatures required for their production. As we see from the figure, at the highest temperature end the emission is more pronounced in galaxies in the $0.05<z<0.2$ redshift range. These systems tend to have a higher median infrared luminosity (Figure \ref{fig:lumhist}) and hence hotter ISM temperatures which are believed to drive the high temperature water emissions \citep{Takahashi1983}. Figure \ref{fig:co_lumbins} also shows the dependence of the water emission lines on the infrared luminosity for three of our five luminosity bins in the $0.005<z<0.05$ sample with the strongest H$_2$O detections. Using a sample of local {\it Herschel} FTS/SPIRE spectra with individual detections, \citet{Yang2013} showed a close to linear relation between the strength of water lines and that of L$_{\rm IR}$. We observe a similar relation in our stacked binned water spectra of DSFGs across all different transitions with higher water emission line intensities in the more IR-luminous sample.

The first two neutral [C\,I] transitions ([C\,I] (1-0) at 609\,$\mu$m and [C\,I] (2-1) at 370\,$\mu$m) are detected in both low-$z$ stacks (see Figures \ref{fig:z0-005} and \ref{fig:z005-02}). We look at the [C\,I] line ratios in terms of gas density and kinetic temperature using the non-LTE radiative transfer code RADEX\footnote{\url{http://home.strw.leidenuniv.nl/~moldata/radex.html}} \citep{vanderTak2007}. To construct the RADEX models, we use the collisional rate coefficients by \citet{Schroder1991} and use the same range of ISM physical conditions reported in \citet{Pereira2013} (with $T=10-1000\,{\rm K}$, $n_{\rm H_2}=10-10^8\,{\rm cm^{-3}}$ and $N_{\rm C}/\Delta v=10^{12}-10^{18}\,{\rm cm^{-2}/(km\,s^{-1})}$). Figure \ref{temp_ci_lumbins} shows the expected kinetic temperature and molecular hydrogen density derived by RADEX for the observed [C\,I] ratios in the low-$z$ stacks for the different infrared luminosity bins with contours showing the different models. The [C\,I] emission is observed to originate from the colder ISM traced by  CO (1-0) rather than the warm molecular gas component traced by the high-$J$ CO lines \citep{Pereira2013} and in fact the temperature is well constrained from these diagrams for high gas densities. 

The fine structure emission line relative strengths are important diagnostics of the physical conditions in the ISM. Here we focus on the three main atomic lines detected at $z>0.8$ ([C\,II] at 158\,$\mu$m, [O\,I] at 63\,$\mu$m and [O\,III] at 88\,$\mu$m) and study their relative strengths as well as their strength in comparison to the infrared luminosity of the galaxy. We break all sources with $ z > 0.8 $ into three smaller bins based on total infrared luminosity. Table \ref{table:highz_flux} lists the infrared luminosity bins used. The [C\,II] line is detected in each subset of the high-redshift stack whereas [O\,I] and [O\,III] are only detected in the $10^{12.5}$ L$_{\odot}$ $<$ $10^{13}$ L$_{\odot}$ infrared luminosity bin.

Figure \ref{data_calibration} shows the relation between emission line luminosity and total infrared luminosity. Total infrared luminosity is integrated in the rest-frame wavelength range $8-1000\,\mu$m. Luminosities in different wavelength ranges in the literature have been converted to L$_{\rm IR}$ using the mean factors derived from Table 7 of \citet{Brisbin2015}:

\begin{subequations}\label{grp}
\begin{align}
&{\rm log}(\rm{L}_{\rm IR}) = {\rm log}(\rm{L}(42.5\,\mu{\rm m} - 122.5\,\mu{\rm m})) + 0.30 \\
&{\rm log}(\rm{L}_{\rm IR}) = {\rm log}(\rm{L}(40\,\mu{\rm m} - 500\,\mu{\rm m})) + 0.145 \\
&{\rm log}(\rm{L}_{\rm IR}) = {\rm log}(\rm{L}(30\,\mu{\rm m} - 1000\,\mu{\rm m})) + 0.09
\end{align}
\end{subequations}

For the [C\,II] 158\,$\mu$m line we used data from a compilation by \citet{Bonato2014}; references therein, \citet{Georgethesis}, \citet{Brisbin2015}, \citet{Oteo2016}, \citet{Gullberg2015}, \citet{Schaerer2015}, \citet{Yun2015}, \citet{Magdis2014}, \citet{Farrah2013}, \citet{Stacey2010}, \citet{Diaz2013}, and a compilation of data from SHINING \citep{SHINING2011}. For the [O\,I] 63\,$\mu$m line we used data from compilation by \citet{Bonato2014}; references therein, \citet{Ferkinhoff2014}, \citet{Brisbin2015}, \citet{Farrah2013}, and SHINING \citep{SHINING2011}. For the [O\,III] 88\,$\mu$m line we used data from a compilation by \citet{Bonato2014}; references therein, \citet{Georgethesis}, and SHINING \citep{SHINING2011}.
As in \citet{Bonato2014}, we excluded all objects for which there is evidence for a substantial AGN contribution.
The line and continuum measurements of strongly lensed galaxies given by \citet{Georgethesis} were corrected using the gravitational magnifications, $\mu$, estimated by \citet{Ferkinhoff2014} while those by \citet{Gullberg2015} were corrected using the magnification estimates from
\citet{Hezaveh2013} and \citet{Spilker2016} available for 17 out of the 20 sources. For the
other three sources we used the median value of $\mu_{\rm med}$ = 7.4.
The solid green lines in Figure 15 correspond to the average L$_{\rm line}$/L$_{\rm IR}$ ratios of -3.03,
-2.94 and -2.84 for the [O\,I] 63\,$\mu$m, [O\,III] 88\,$\mu$m and [C\,II] 158\,$\mu$m lines from the literature, respectively. 
The [CII] line luminosity-to-IR luminosity ratio is at least an order of magnitude higher than the typical value of 10$^{-4}$ quoted in the literature for
local nuclear starburst ULIRGS and high-z QSOs.

Since the data come from heterogeneous samples, a least square fitting is susceptible to selection
effects that may bias the results. To address this issue, \citet{Bonato2014} have carried
out an extensive set of simulations of the expected emission line intensities as a function of
infrared luminosity for different properties (density, metallicity, filling factor) of the emitting gas, different
ages of the stellar populations and a range of dust obscuration. For a set of lines, including
those considered in this paper the simulations were consistent with a direct proportionality
between L$_{\rm line}$ and L$_{\rm IR}$. Based on this result, we have adopted a linear relation.
The other lines show L$_{\rm line}-$L$_{\rm IR}$ relations found in the literature, namely:
\begin{subequations}\label{grp}
\begin{align}
&{\rm log}(\rm{L}{\rm [O\,I]}\,63\,\mu{\rm m}) = {\rm log}(\rm{L}_{\rm IR}) - 2.99,\\
&{\rm log}(\rm{L}{\rm [O\,III]}\,88\,\mu{\rm m}) = {\rm log}(\rm{L}_{\rm IR}) - 2.87, \\
&{\rm log}(\rm{L}{\rm [C\,II]}\,158\,\mu{\rm m}) = {\rm log}(\rm{L}_{\rm IR}) - 2.74,
\end{align}
\end{subequations}
from \citet{Bonato2014},
\begin{subequations}\label{grp}
\begin{align}
&{\rm log}(\rm{L}{\rm [O\,I]}\,63\,\mu{\rm m}) = 0.98\times{\rm log}(\rm{L}_{\rm IR}) - 2.95,\\
&{\rm log}(\rm{L}{\rm [O\,III]}\,88\,\mu{\rm m}) = 0.98\times{\rm log}(\rm{L}_{\rm IR}) - 3.11, \\
&{\rm log}(\rm{L}{\rm [C\,II]}\,158\,\mu{\rm m}) = 0.89\times{\rm log}(\rm{L}_{\rm IR}) - 2.67,
\end{align}
\end{subequations}
from \citet{Spinoglio2014},
\begin{subequations}\label{grp}
\begin{align}
&{\rm log}(\rm{L}{\rm [O\,I]}\,63\,\mu{\rm m}) = 0.70\times{\rm log}(\rm{L}_{\rm IR}) + 0.32,\\
&{\rm log}(\rm{L}{\rm [O\,III]}\,88\,\mu{\rm m}) = 0.82\times{\rm log}(\rm{L}_{\rm IR}) - 1.40, \\
&{\rm log}(\rm{L}{\rm [C\,II]}\,158\,\mu{\rm m}) = 0.94\times{\rm log}(\rm{L}_{\rm IR}) - 2.39,
\end{align}
\end{subequations}
from \citet{Gruppioni2016}, and 
\begin{subequations}\label{grp}
\begin{align}
&{\rm log}(\rm{L}{\rm [O\,I]}\,63\,\mu{\rm m}) = 1.10\times{\rm log}(\rm{L}_{\rm IR}) - 4.70,\\
&{\rm log}(\rm{L}{\rm [C\,II]}\,158\,\mu{\rm m}) = 1.56\times{\rm log}(\rm{L}_{\rm IR}) - 10.52,
\end{align}
\end{subequations}
from \citet{Farrah2013}, respectively.

In the high-$z$ bin at $z > 1$, we find that [O\,III] and [O\,I] detections are limited to only one of the three luminosity bins. The ISM emission lines show a deficit (i.e. deviating from a one to one relation) compared to the infrared luminosity. This in particular is more pronounced in our stacked high-$z$ DSFG sample compared to that of local starbursts and is similar to what is observed in local ULIRGs. This deficit further points towards an increase in the atomic ISM lines optical depth in these very dusty environments. There is no clear trend in the measured lines with the infrared luminosities, given the measured uncertainties, however there is some evidence pointing towards a further decrease with increasing IR luminosity. Figure 16 shows the [O\,I]/[C\,II] line ratio for the stacks of DSFGs compared to \citet{Brauher2008} and \citet{Cormier2015}. Although both lines trace neutral gas, they have different excitation energies (with the [O\,I] being higher). Given the uncertainties, we don't see a significant trend in this line ratio with the infrared luminosity. 

Due to the wavelength coverage of SPIRE/FTS, we are unable to study the [N\,II] 205\,$\mu$m line in the high-$z$ bin. Instead, we concentrate on the luminosity dependence of the [N\,II] 205\,$\mu$m line in the low-$z$ bin. This [N\,II] ISM emission cooling line is usually optically thin, suffering less dust attenuation compared to optical lines and hence is a strong star-formation rate indicator \citep{Zhao2013,Herrera2016,Hughes2016, Zhao2016}. The [N\,II] line luminosity in fact shows a tight correlation with SFR for various samples of ULIRGs \citep{Zhao2013}. Given the ionization potential of [N\,II] at 14.53\,eV, this line is also a good tracer of the warm ionized ISM regions \citep{Zhao2016}. Figure 16 shows the [N\,II]  emission for our low-$z$ stack ($0.005<z<0.05$) as a function of infrared luminosity for the five luminosity bins outlined in Figure 8. The [N\,II] line luminosity probes the same range as observed for other samples of ULIRGs and consistently increases with infrared luminosity (a proxy for star-formation) \citep{Zhao2013}. The [N\,II]/L$_{\rm IR}$ ratio is $\sim 10^{-5}$ compared to the [C\,II]/L$_{\rm IR}$ at $\sim 10^{-3}$ \citep{Diaz2013, Ota2014,Herrera2015, Rosenberg2015}.

\clearpage

\begin{table}

  \rotatebox{90}{%
    \begin{minipage}{9in}
      \begin{center}
  \fontsize{7}{10}\selectfont
  
\caption{Fluxes of observed spectral lines in each of the redshift bins.}

\begin{tabular}{| c c | c c | c c | c c | c c | c c |}   \hline\hline

\\ [0.1ex]

  &       & \multicolumn{2}{c|}{$0.005 < z < 0.05$} & \multicolumn{2}{c|}{$0.05 < z < 0.2$} & \multicolumn{2}{c|}{$0.2 < z < 0.5$} & \multicolumn{2}{c|}{$0.8 < z < 2$} & \multicolumn{2}{c|}{$2 < z < 4$}\\

  \\
  \hline \hline
Line  & Rest Freq. & Flux & Flux & Flux & Flux & Flux & Flux & Flux & Flux & Flux & Flux \\ [0.5ex]  
  & [$\rm{GHz}$]  & [$10^{-18}$ $\rm{W}\rm{m^{-2}}$] & [$\rm{Jy}\,\rm{km}\,\rm{s^{-1}}$]& [$10^{-18}$ $\rm{W}\rm{m^{-2}}$] & [$\rm{Jy}\,\rm{km}\,\rm{s^{-1}}$]& [$10^{-18}$ $\rm{W}\rm{m^{-2}}$] & [$\rm{Jy}\,\rm{km}\,\rm{s^{-1}}$]& [$10^{-18}$ $\rm{W}\rm{m^{-2}}$] & [$\rm{Jy}\,\rm{km}\,\rm{s^{-1}}$]& [$10^{-18}$ $\rm{W}\rm{m^{-2}}$] & [$\rm{Jy}\,\rm{km}\,\rm{s^{-1}}$]\\ [0.5ex] 
\hline \hline
&\\
CO (5-4)               & 576.268  & 15 $\pm$ 3     & 790 $\pm$ 130  & 2.8 $\pm$ 0.4    & 160 $\pm$ 30  &               -          &        -       &       -        &      -          &         -        &     -    \\
CO (6-5)               & 691.473  & 14 $\pm$ 3     & 620 $\pm$ 100  & 3.8 $\pm$ 0.4    & 180 $\pm$ 20  &          $<$ 0.40        &      $<$ 23    &       -        &      -          &         -        &     -    \\
CO (7-6)               & 806.653  & 12 $\pm$ 2     & 440 $\pm$ 80   & 4.7 $\pm$ 0.4    & 190 $\pm$ 20  &          $<$ 0.38        &      $<$ 19    &       -        &      -          &         -        &     -    \\
CO (8-7)               & 921.800  & 11 $\pm$ 2     & 360 $\pm$ 60   & 4.5 $\pm$ 0.4    & 160 $\pm$ 20  &          $<$ 0.24        &      $<$ 10    &       -        &      -          &         -        &     -    \\
CO (9-8)               & 1036.914 & 9.7 $\pm$ 1.7  & 280 $\pm$ 50   & 4.0 $\pm$ 0.5    & 130 $\pm$ 20  &          $<$ 0.21        &      $<$ 7.7   &    $<$ 0.48    &      $<$ 33     &         -        &     -    \\
CO (10-9)              & 1151.985 & 9.6 $\pm$ 1.7  & 250 $\pm$ 50   & 5.7 $\pm$ 0.6    & 160 $\pm$ 20  &          $<$ 0.32        &      $<$ 11    &    $<$ 0.34    &      $<$ 21     &         -        &     -    \\
CO (11-10)             & 1267.016 & 4.9  $\pm$ 1.0 & 120 $\pm$ 30   & 3.9 $\pm$ 0.4    & 100 $\pm$ 20  &          $<$ 0.50        &      $<$ 16    &    $<$ 0.21    &      $<$ 12     &         -        &     -    \\
CO (12-11)             & 1381.997 & 5.4  $\pm$ 1.1 & 120 $\pm$ 30   & 3.5 $\pm$ 0.5    & 84 $\pm$ 10   &          $<$ 0.34        &      $<$ 9.5   &    $<$ 0.26    &      $<$ 14     &         -        &     -    \\
CO (13-12)             & 1496.926 & 2.3  $\pm$ 0.6 & 54 $\pm$ 13    & 2.7 $\pm$ 0.5    & 60 $\pm$ 9    &          $<$ 0.37        &      $<$ 9.7   &    $<$ 0.33    &      $<$ 16     &     $<$ 0.38     &  $<$ 29  \\
$\rm{H_2O}$ 211-202    & 752.032  & 1.9  $\pm$ 0.4 & 78 $\pm$ 17    & 1.1 $\pm$ 0.3    & 49 $\pm$ 9    &          $<$ 0.49        &      $<$ 26    &        -       &       -         &         -        &     -    \\
$\rm{H_2O}$ 202-111    & 987.927  & 5.5  $\pm$ 1.2 & 170 $\pm$ 40   & 2.3 $\pm$ 0.3    & 78 $\pm$ 9    &          $<$ 0.30        &      $<$ 12    &     $<$ 0.50   &      $<$ 37     &         -        &     -    \\
$\rm{H_2O}$ 312-303    & 1097.365 & 2.7  $\pm$ 0.7 & 75 $\pm$ 19    & 2.3 $\pm$ 0.3    & 70 $\pm$ 9    &          $<$ 0.23        &      $<$ 8.2   &     $<$ 0.43   &      $<$ 29     &         -        &     -    \\
$\rm{H_2O}$ 312-221    & 1153.128 &  -             &  -             &  -               & -             &               -          &        -       &        -       &       -         &         -        &     -    \\
$\rm{H_2O}$ 321-312    & 1162.910 & 2.7 $\pm$ 0.7  & 72 $\pm$ 18    & 2.9 $\pm$ 0.3    & 82 $\pm$ 9    &          $<$ 0.32        &      $<$ 11    &     $<$ 0.31   &      $<$ 19     &         -        &     -    \\
$\rm{H_2O}$ 422-413    & 1207.638 &    $<$ 1.2     &   $<$ 30       & 1.6 $\pm$ 0.5    & 44 $\pm$ 12   &          $<$ 0.42        &      $<$ 14    &     $<$ 0.25   &      $<$ 15     &         -        &     -    \\
$\rm{H_2O}$ 220-211    & 1228.789 & 3.9  $\pm$ 1.0 & 96 $\pm$ 23    & 1.6 $\pm$ 0.4    & 43 $\pm$ 11   &          $<$ 0.50        &      $<$ 16    &     $<$ 0.24   &      $<$ 14     &         -        &     -    \\
$\rm{H_2O}$ 523-514    & 1410.615 & $<$ 1.4        &  $<$ 30        & 1.8 $\pm$ 0.4    & 41 $\pm$ 9    &          $<$ 0.35        &       $<$ 9.7  &     $<$ 0.36   &      $<$ 19     &         -        &     -    \\
$[\rm C\,I] \, (1-0)$  & 492.161  & 9.2  $\pm$ 4.1 & 570 $\pm$ 250  & 2.5 $\pm$ 0.8    & 170 $\pm$ 50  &           -              &        -       &       -        &      -          &         -        &     -    \\
$[\rm C\,I] \,(2-1)$   & 809.340  & 15  $\pm$ 3    & 570 $\pm$ 100  & 3.0 $\pm$ 0.3    & 120 $\pm$ 10  &          $<$ 0.39        &      $<$ 18    &       -        &      -          &         -        &     -    \\
$[\rm N\,II]$          & 1461.132 & 96 $\pm$ 16    & 2000 $\pm$ 400 & 5.4 $\pm$ 0.5    & 120 $\pm$ 10  &          $<$ 0.39        &      $<$ 11    &     $<$ 0.14   &      $<$ 6.9    &    $<$ 0.52      & $<$ 41     \\
$[\rm C\,II]$          & 1901.128 &      -         &        -       &         -        &       -       &      4.0 $\pm$ 0.4       & 83 $\pm$ 7     & 1.3 $\pm$ 0.2  & 51 $\pm$ 5      & 0.22 $\pm$ 0.04  & 13 $\pm$ 2 \\
$[\rm N\,II]$          & 2461.250 &       -        &        -       &         -        &       -       &              -           &        -       &    $<$ 0.17    &    $<$ 4.8      &    $<$ 0.048     &  $<$ 2.2    \\
$[\rm O\,III]$         & 3393.006 &      -         &        -       &         -        &       -       &              -           &        -       & 1.1 $\pm$ 0.3  & 23 $\pm$ 6      & 0.14 $\pm$ 0.03  & 4.5 $\pm$ 1.0\\
$[\rm O\,I]$           & 4744.678 &      -         &        -       &         -        &       -       &              -           &        -       &      -         &      -          & 0.14 $\pm$ 0.05  & 3.5 $\pm$ 1.1\\

&  \\[0.5ex]
 
\hline\hline \end{tabular}
\tablecomments{$\rm CO (10-9)$ is contaminated by emission from $\rm H_2O \, 312-221$, so we quote only the combined flux for the two emission lines in the $\rm CO (10-9)$ row. In the five redshift bins ($0.005 < z < 0.05$, $0.05 < z < 0.2$, $0.2 < z < 0.5$, $0.8 < z < 2$, and $2 < z < 4$), the  mean redshifts are $z = 0.02$, $z = 0.1$, $z = 0.3$, $z = 1.4$, $z = 2.8$, respectively, and the median IR luminosities are $10^{11.35}$ L$_{\odot}$, $10^{12.33}$ L$_{\odot}$, $10^{11.89}$ L$_{\odot}$, $10^{12.53}$ L$_{\odot}$, and $10^{12.84}$ L$_{\odot}$, respectively.}

\label{table:linefluxes1}
      \end{center}
    \end{minipage}

  }

\end{table}

\clearpage

\clearpage

\begin{table}

  \rotatebox{90}{%
    \begin{minipage}{9in}
      \begin{center}
  \fontsize{7}{10}\selectfont
  
\caption{ Measured fluxes of observed spectral lines from sources with $0.005 < z < 0.05$ in five luminosity bins.}

\begin{tabular}{| c c | c c | c c | c c | c c | c c |}   \hline \hline

  \\
  &       & \multicolumn{2}{c|}{$10^{11.0} \,$L$_{\odot} \, < $ L $< 10^{11.2} \,$L$_{\odot}$} & \multicolumn{2}{c|}{$10^{11.2} \,$L$_{\odot} \, < $ L $< 10^{11.4} \,$L$_{\odot}$} & \multicolumn{2}{c|}{$10^{11.4} \,$L$_{\odot} \, < $ L $< 10^{11.6} \,$L$_{\odot}$} & \multicolumn{2}{c|}{$10^{11.6} \,$L$_{\odot} \, < $ L $< 10^{12.0} \,$L$_{\odot}$} & \multicolumn{2}{c|}{ L $> 10^{12.0} \,$L$_{\odot}$}\\
  \\
  \hline \hline
\\ [0.1ex]
Line  & Rest Freq. & Flux & Flux & Flux & Flux & Flux & Flux & Flux & Flux & Flux & Flux \\ [0.5ex]  
  & [$\rm{GHz}$]  & [$10^{-18}$ $\rm{W}\rm{m^{-2}}$] & [$\rm{Jy}\,\rm{km}\,\rm{s^{-1}}$]& [$10^{-18}$ $\rm{W}\rm{m^{-2}}$] & [$\rm{Jy}\,\rm{km}\,\rm{s^{-1}}$]& [$10^{-18}$ $\rm{W}\rm{m^{-2}}$] & [$\rm{Jy}\,\rm{km}\,\rm{s^{-1}}$]& [$10^{-18}$ $\rm{W}\rm{m^{-2}}$] & [$\rm{Jy}\,\rm{km}\,\rm{s^{-1}}$]& [$10^{-18}$ $\rm{W}\rm{m^{-2}}$] & [$\rm{Jy}\,\rm{km}\,\rm{s^{-1}}$]\\ [0.5ex] 
\hline \hline
& \\

&\\
CO (5-4)               & 576.268  & 22 $\pm$ 4       & 1200 $\pm$ 200 & 17 $\pm$ 3    & 880 $\pm$ 150  & 16 $\pm$ 3    & 840 $\pm$ 150  & 20 $\pm$ 4    & 1100 $\pm$ 200 & 18 $\pm$ 4    & 980 $\pm$ 170   \\
CO (6-5)               & 691.473  & 16 $\pm$ 3       & 720 $\pm$ 120  & 16 $\pm$ 3    & 710 $\pm$ 120  & 18 $\pm$ 3    & 820 $\pm$ 150  & 20 $\pm$ 4    & 910 $\pm$ 200  & 22 $\pm$ 4    & 1000 $\pm$ 200  \\
CO (7-6)               & 806.653  & 13 $\pm$ 3       & 480 $\pm$ 80   & 12 $\pm$ 3    & 470 $\pm$ 80   & 15 $\pm$ 3    & 580 $\pm$ 100  & 20 $\pm$ 4    & 760 $\pm$ 130  & 24 $\pm$ 4    & 910 $\pm$ 150   \\
CO (8-7)               & 921.800  & 10 $\pm$ 2       & 330 $\pm$ 60   & 11 $\pm$ 2    & 370 $\pm$ 70   & 15 $\pm$ 3    & 500 $\pm$ 90   & 19 $\pm$ 3    & 630 $\pm$ 110  & 27 $\pm$ 5    & 930 $\pm$ 160   \\
CO (9-8)               & 1036.914 & 8.5 $\pm$ 2.0    & 250 $\pm$ 60   & 7.7 $\pm$ 1.7 & 230 $\pm$ 50   & 14 $\pm$ 3    & 410 $\pm$ 80   & 16 $\pm$ 3    & 490 $\pm$ 90   & 24 $\pm$ 5    & 730 $\pm$ 130   \\
CO (10-9)              & 1151.985 & 8.5 $\pm$ 1.9    & 230 $\pm$ 50   & 10 $\pm$ 2    & 260 $\pm$ 50   & 14 $\pm$ 4    & 380 $\pm$ 90   & 17 $\pm$ 3    & 460 $\pm$ 80   & 34 $\pm$ 6    & 930 $\pm$ 160   \\
CO (11-10)             & 1267.016 & $<$ 7.0          & $<$ 170        & 3.4 $\pm$ 1.2 & 82 $\pm$ 27    & 12 $\pm$ 4    & 290 $\pm$ 100  & 10 $\pm$ 2    & 250 $\pm$ 50   & 21 $\pm$ 4    & 520 $\pm$ 90    \\
CO (12-11)             & 1381.997 & $<$ 5.0          & $<$ 110        & 4.6 $\pm$ 1.5 & 100 $\pm$ 30   & 6.4 $\pm$ 1.9 & 140 $\pm$ 40   & 11 $\pm$ 2    & 250 $\pm$ 50   & 14 $\pm$ 3    & 320 $\pm$ 60    \\
CO (13-12)             & 1496.926 & $<$ 3.9          & $<$ 80         & $<$ 4.8       &  $<$ 97        & $<$ 9.3       & $<$ 190        & 11 $\pm$ 3    & 220 $\pm$ 50   & 15 $\pm$ 3    & 310 $\pm$ 60    \\
$\rm{H_2O}$ 211-202    & 752.032  & $<$ 1.5          & $<$ 59         & 2.4 $\pm$ 0.6 & 97 $\pm$ 25    & 5.2 $\pm$ 1.4 & 210 $\pm$ 60   & 3.0 $\pm$ 0.6 & 120 $\pm$ 30   & 9.3 $\pm$ 1.7 & 390 $\pm$ 70    \\
$\rm{H_2O}$ 202-111    & 987.927  & $<$ 3.2          & $<$ 99         & 5.4 $\pm$ 1.2 & 170 $\pm$ 40   & $<$ 6.1       & $<$ 190        & 4.8 $\pm$ 1.1 & 150 $\pm$ 40   & 18 $\pm$ 4    & 580 $\pm$ 110   \\
$\rm{H_2O}$ 312-303    & 1097.365 & $<$ 6.1          & $<$ 170        & $<$ 3.2       & $<$ 88         & $<$ 5.9       & $<$ 170        & $<$ 4.8       & $<$ 140        & 12 $\pm$ 3    & 350 $\pm$ 70    \\
$\rm{H_2O}$ 312-221    & 1153.128 &  -               &    -           &     -         &        -       &      -        &       -        &        -      &         -      &        -      &        -        \\
$\rm{H_2O}$ 321-312    & 1162.910 & $<$ 2.7          & $<$ 69         & 3.5 $\pm$ 1.1 & 93 $\pm$ 28    & $<$ 5.0       & $<$ 140        & 3.8 $\pm$ 1.1 & 100 $\pm$ 30   & 19 $\pm$ 4    & 520 $\pm$ 90    \\
$\rm{H_2O}$ 422-413    & 1207.638 & $<$ 2.7          & $<$ 67         & $<$ 2.4       & $<$ 60         & $<$ 2.7       & $<$ 68         & $<$ 3.4       & $<$ 87         & 8.6 $\pm$ 1.9 & 220 $\pm$ 50    \\
$\rm{H_2O}$ 220-211    & 1228.789 & $<$ 4.6          & $<$ 120        & 4.6 $\pm$ 1.5 & 110 $\pm$ 37   & 6.1 $\pm$ 1.9 & 150 $\pm$ 50   & 3.0 $\pm$ 0.9 & 75 $\pm$ 22    & 16 $\pm$ 3    & 400 $\pm$ 80    \\
$\rm{H_2O}$ 523-514    & 1410.615 & $<$ 3.0          & $<$ 65         & $<$ 3.6       & $<$ 77         & $<$ 2.8       & $<$ 61         & $<$ 1.8       & $<$ 40         & 7.5 $\pm$ 1.9 & 170 $\pm$ 40    \\
$[\rm C\,I] \, (1-0)$  & 492.161  & 14 $\pm$ 5       & 850 $\pm$ 250  & 11 $\pm$ 3    & 680 $\pm$ 140  & 10 $\pm$ 3    & 640 $\pm$ 150  & 9.6 $\pm$ 2.3 & 600 $\pm$ 140  & 8.8 $\pm$ 2.7 & 560 $\pm$ 170   \\
$[\rm C\,I] \,(2-1)$   & 809.340  & 21 $\pm$ 4       & 790 $\pm$ 130  & 19 $\pm$ 4    & 700 $\pm$ 120  & 20 $\pm$ 4    & 750 $\pm$ 130  & 17 $\pm$  3   & 640 $\pm$ 110  & 16 $\pm$ 3    & 610 $\pm$ 110   \\
$[\rm N\,II]$          & 1461.132 & 160 $\pm$ 30     & 3300 $\pm$ 600 & 130 $\pm$ 20  & 2600 $\pm$ 500 & 100 $\pm$ 20  & 2100 $\pm$ 400 & 73 $\pm$ 12   & 1500 $\pm$ 300 & 34 $\pm$ 6    & 730 $\pm$ 120   \\
$[\rm C\,II]$          & 1901.128 &   -              &    -           &    -          &        -       &      -        &        -       &       -       &        -       &        -      &        -        \\
$[\rm N\,II]$          & 2461.250 &   -              &    -           &    -          &        -       &      -        &        -       &       -       &        -       &        -      &        -        \\
$[\rm O\,III]$         & 3393.006 &   -              &    -           &    -          &        -       &      -        &        -       &       -       &        -       &        -      &        -        \\
$[\rm O\,I]$           & 4744.678 &   -              &    -           &    -          &        -       &      -        &        -       &       -       &        -       &        -      &        -        \\

&  \\[0.5ex]

\hline\hline \end{tabular} 

\tablecomments{$\rm CO (10-9)$ is contaminated by emission from $\rm H_2O \, 312-221$, so we quote only the combined flux for the two emission lines in the $\rm CO (10-9)$ row. In the luminosity ranges $10^{11.0-11.2}$L$_{\odot}$, $10^{11.2-11.4}$L$_{\odot}$, $10^{11.4-11.6}$L$_{\odot}$, $10^{11.6-12.0}$L$_{\odot}$, and L $ > \, 10^{12}$ L$_{\odot}$, the mean redshifts are $z = 0.015$, $z = 0.018$, $z = 0.021$, $z = 0.027$, and $z = 0.038$, respectively, and the median IR luminosities are $10^{11.12}$ L$_{\odot}$, $10^{11.32}$ L$_{\odot}$, $10^{11.49}$ L$_{\odot}$, $10^{11.69}$ L$_{\odot}$, and $10^{12.21}$ L$_{\odot}$, respectively.}

\label{table:linefluxes2}
\end{center}
    \end{minipage}
    }

\end{table}

\clearpage

\clearpage

\begin{table}

    \rotatebox{90}{%
      \begin{minipage}{9in}
          \begin{center}
  \fontsize{7}{10}\selectfont
  
\caption{Measured fluxes of observed spectral lines from sources with $0.8 < z < 4$ in three luminosity bins.}

\begin{tabular}{| c c | c c | c c | c c |}    \hline \hline

  \\
  &       & \multicolumn{2}{c|}{$10^{11.5} \,$L$_{\odot} \, < $ L $< 10^{12.5} \,$L$_{\odot}$} & \multicolumn{2}{c|}{$10^{12.5} \,$L$_{\odot} \, < $ L $< 10^{13.0} \,$L$_{\odot}$} & \multicolumn{2}{c|}{$10^{13.0} \,$L$_{\odot} \, < $ L $< 10^{14.5} \,$L$_{\odot}$} \\
  \\
 \hline\hline\\ [0.1ex]   
Line  & Rest Freq. & Flux & Flux & Flux & Flux & Flux & Flux \\ [0.5ex]  
  & [$\rm{GHz}$]  & [$10^{-18}$ $\rm{W}\rm{m^{-2}}$] & [$\rm{Jy}\,\rm{km}\,\rm{s^{-1}}$]& [$10^{-18}$ $\rm{W}\rm{m^{-2}}$] & [$\rm{Jy}\,\rm{km}\,\rm{s^{-1}}$]& [$10^{-18}$ $\rm{W}\rm{m^{-2}}$] & [$\rm{Jy}\,\rm{km}\,\rm{s^{-1}}$]\\ [0.5ex] 
\hline \hline
& \\

&\\
CO (5-4)               & 576.268  &          -            &          -        &          -        &          -       &          -      &          -    \\
CO (6-5)               & 691.473  &          -            &          -        &          -        &          -       &          -      &          -    \\
CO (7-6)               & 806.653  &          -            &          -        &          -        &          -       &          -      &          -    \\
CO (8-7)               & 921.800  &          -            &          -        &          -        &          -       &          -      &          -    \\
CO (9-8)               & 1036.914 &      $<$ 1.5          &      $<$ 130      &          -        &          -       &          -      &          -    \\
CO (10-9)              & 1151.985 &      $<$ 1.1          &      $<$ 89       &     $<$ 0.51      &     $<$ 46       &          -      &          -    \\
CO (11-10)             & 1267.016 &      $<$ 0.66         &      $<$ 50       &     $<$ 0.21      &     $<$ 17       &          -      &          -    \\
CO (12-11)             & 1381.997 &      $<$ 0.18         &      $<$ 12       &     $<$ 0.20      &      $<$ 15      &          -      &          -    \\
CO (13-12)             & 1496.926 &      $<$ 0.11         &      $<$ 6.8      &     $<$ 0.16      &      $<$ 11      &          -      &          -    \\
$\rm{H_2O}$ 211-202    & 752.032  &          -            &         -         &           -       &        -         &         -       &        -      \\
$\rm{H_2O}$ 202-111    & 987.927  &          -            &         -         &           -       &        -         &         -       &        -      \\
$\rm{H_2O}$ 312-303    & 1097.365 &       $<$ 0.96        &      $<$ 84       &     $<$ 0.53      &      $<$ 49      &          -      &          -    \\
$\rm{H_2O}$ 312-221    & 1153.128 &           -           &         -         &           -       &        -         &         -       &        -      \\
$\rm{H_2O}$ 321-312    & 1162.910 &       $<$ 0.99        &      $<$ 82       &     $<$ 0.51      &      $<$ 45      &          -      &          -    \\
$\rm{H_2O}$ 422-413    & 1207.638 &       $<$ 0.92        &      $<$ 73       &     $<$ 0.31      &      $<$ 26      &          -      &          -    \\
$\rm{H_2O}$ 220-211    & 1228.789 &       $<$ 0.91        &      $<$ 71       &     $<$ 0.24      &      $<$ 20      &          -      &          -    \\
$\rm{H_2O}$ 523-514    & 1410.615 &       $<$ 0.14        &      $<$ 9.4      &     $<$ 0.18      &      $<$ 13      &          -      &          -    \\
$[\rm C\,I] \, (1-0)$  & 492.161  &          -            &          -        &          -        &          -       &          -      &          -    \\
$[\rm C\,I] \,(2-1)$   & 809.340  &          -            &          -        &          -        &          -       &          -      &          -    \\
$[\rm N\,II]$          & 1461.132 &       $<$ 0.12        &        $<$ 7.5    &     $<$ 0.18      &       $<$ 13     &                 &          -    \\
$[\rm C\,II]$          & 1901.128 & 0.20 $\pm$ 0.02       & 10 $\pm$ 1        & 0.56 $\pm$ 0.06   & 30 $\pm$ 4       & 0.89 $\pm$ 0.25 & 55 $\pm$ 15   \\
$[\rm N\,II]$          & 2461.250 &       $<$ 0.025       &        $<$ 0.97   &     $<$ 0.066     &       $<$ 2.7    &     $<$ 0.21    &    $<$ 10     \\
$[\rm O\,III]$         & 3393.006 &  $<$ 0.094            & $<$ 2.7           & 0.31 $\pm$ 0.09   & 9.2 $\pm$ 2.5    & $<$ 0.37        & $<$ 13       \\
$[\rm O\,I]$           & 4744.678 &  $<$ 0.076            & $<$ 1.6           & 0.59  $\pm$ 0.15  & 13 $\pm$ 3       & $<$ 0.35        & $<$ 8.5        \\

&  \\[0.5ex]

\hline\hline \end{tabular}
\tablecomments{$\rm CO (10-9)$ is contaminated by emission from $\rm H_2O \, 312-221$, so we quote only the combined flux for the two emission lines in the $\rm CO (10-9)$ row. In the luminosity ranges $10^{11.5-12.5}$L$_{\odot}$, $10^{12.5-13.0}$L$_{\odot}$, $10^{13.0-14.5}$L$_{\odot}$, the mean redshifts are $z = 2.19$, $z = 2.40$, and $z = 2.93$, respectively, and the median IR luminosities are $10^{12.41}$ L$_{\odot}$, $10^{12.77}$ L$_{\odot}$, and $10^{13.24}$ L$_{\odot}$, respectively.}

\label{table:linefluxes3}
\end{center}
      \end{minipage}
      }

\end{table}

\clearpage

\begin{table*}
\caption{Uncorrected line ratios used in PDR modeling for high-redshift sources in three luminosity bins based on lensing-corrected luminosity.}
\begin{center} 
\begin{tabular}{c c c c c c c} 
\hline\hline\\ [0.1ex] 
Range  & Median  & Number of & $[\rm O\,I]/[\rm C\,II]$ & $[\rm C\,II]$/FIR  &  $[\rm O\,I]$/FIR   &($[\rm O\,I]$+$[\rm C\,II]$)/FIR \\ [0.5ex] 
 [$\rm{log_{10}(L_{\odot})}$] & [$\rm{log_{10}(L_{\odot})}$]  & Sources &  &  ($\times\rm{10^{-4}}$) &    ($\times\rm{10^{-4}}$)  & ($\times \rm{10^{-4}}$)\\ [0.5ex] 
\hline 
11.5 - 12.5 & 12.41 $\pm$0.12 & 11  &   $<$ 0.38 [36]    & 7.8$\pm$2.3 [1]   & $<$ 3.0 [4]           &  $<$ 11 [1.8]         \\

12.5 - 13.0 & 12.77 $\pm$0.17 & 15  & 1.1$\pm$0.3 [36]   & 12$\pm$5 [1]      &  13 $\pm$ 6 [4]       &  24$\pm$11 [2.6]      \\

13.0 - 14.5 & 13.24 $\pm$0.32 & 10  & $<$ 0.40 [36]      & 11$\pm$9 [1]      & $<$ 4.1 [4]           & $<$ 15 [1.8]          \\
& & \\[0.5ex] 
\hline\hline\\ [0.1ex] 
\end{tabular}
\end{center}

\tablecomments{The median luminosities in each bin are $10^{12.41}\,$L$_{\odot}$, $10^{12.77}\,$L$_{\odot}$, and $10^{13.24}\,$L$_{\odot}$, and the mean redshifts are 2.19, 2.40, and 2.93. These ratios are uncorrected for [O\,I] optical thickness, filling factors, and non-PDR [C\,II] emission, or for a plane-parallel PDR model FIR. The total correction factor (i.e., ([A]/[B])$_{\rm corrected}$/([A]/[B])$_{\rm uncorrected}$) for each ratio is given in brackets. The plots in Figure \ref{n_go} do take these correction factors into account.}
\label{table:highz_flux}
\end{table*}

\begin{table*}
\caption{Uncorrected line ratios used in the PDR modeling of the observed lines in the $0.005<z<0.05$ and $0.05<z<0.2$ redshift bins.}
\begin{center} 
\begin{tabular}{c c c c c c c c c} 
  \hline\hline\\ [0.1ex]

  Range  & Median  & Number of & $\frac{[\rm C\,I](2-1)}{[\rm C\,I](1-0)}$ & $\frac{[\rm C\,I] (1-0)}{\rm{CO (7-6)}}$  &   $\frac{[\rm C\,I] (2-1)}{\rm{CO (7-6)}}$   & $\frac{[\rm C\,I] (2-1)}{\rm{FIR}}$   &$\frac{[\rm C\,I] (1-0)}{\rm{FIR}}$   &$\frac{\rm{CO} (7-6)}{\rm{FIR}}$   \\ [0.5ex] 

  & [$\rm{log_{10}(L_{\odot})}$]  & Sources &  &   &   & ($\times \rm{10^{-5}}$)& ($\times \rm{10^{-5}}$)& ($\times \rm{10^{-5}}$)\\ [0.5ex] 
\hline 
$0.005<z<0.05$  & 11.35 $\pm$1.03  & 115 & 1.6$\pm$0.8 [1] & 0.77$\pm$0.37 [1]  & 1.3$\pm$0.4 [1] & 1.6$\pm$3.7 [0.5]& 0.97$\pm$2.29 [0.5]& 1.3 $\pm$ 2.9 [0.5] \\

$0.05<z<0.2$    & 12.33$\pm$0.23   & 34  & 1.2$\pm$0.4 [1] & 0.53$\pm$0.18 [1]& 0.63$\pm$0.09 [1]& 0.93$\pm$0.51 [0.5] & 0.78$\pm$0.48 [0.5]& 1.5$\pm$0.8 [0.5]\\

& & \\[0.5ex] 
\hline\hline\\ [0.1ex] 
\end{tabular}
\end{center}
\tablecomments{The median luminosities of sources in these bins are L$_{\rm IR}  =  10^{11.35} \,$L$_{\odot}$ and $10^{12.33}\,$ L$_{\odot}$, and the mean redshifts are $z=0.02$ and $z=0.1$, respectively. These ratios do not account for the corrections given in the text. The total correction factor (i.e., ([A]/[B])$_{\rm corrected}$/([A]/[B])$_{\rm uncorrected}$) for each ratio is given in brackets, where applicable. The large uncertainties reported in the $0.005 < z < 0.05$ bin stem from the large standard deviation of source FIR luminosities.}
\label{table:lowz_pdr_ratios}
\end{table*}

\subsubsection{PDR Modeling}

The average gas number density and radiation field strength in the interstellar medium can be inferred using 
photodissociation regions (PDR) models. About 1\% of far-ultraviolet (FUV) photons from young stars collide with neutral gas in the interstellar medium and strip electrons off of small dust grains and polycyclic aromatic hydrocarbons via the photoelectric effect. These electrons transfer some of their kinetic energy to the gas, heating it. The gas is subsequently cooled by the emission of the far-infrared lines that we observe. The remaining fraction of the UV light is reprocessed in the infrared by large dust grains via thermal continuum emission \citep{Hollenbach1999}. Understanding the balance between the input radiation source and the underlying atomic and molecular cooling mechanisms is essential in constraining the physical properties of the ISM.

We use the online PDR Toolbox\footnote{\url{http://dustem.astro.umd.edu/pdrt/}} \citep{Poundwolfire2008,Kaufman2006} to infer the average conditions in the interstellar medium that correspond to the measured fluxes of both the stacked low ($0.005<z<0.05$ and $0.05<z<0.2$) and high-redshift ($0.8<z<4$) spectra. The PDR toolbox uses the ratios between the fluxes of fine structure lines and of the FIR continuum to constrain the PDR gas density and strength of the incident FUV radiation (given in units of the Habing field, $1.6\times10^-3\,\rm{erg}\,\rm{cm}^{-2}\,\rm{s}^{-1}$). At low redshifts, the PDR models take into account the lines [C\,I] (1-0), [C\,I] (2-1), CO (7-6), and the FIR continuum; at high redshits, the models use [C\,II] 158 $\mu$m, [O\,I] 63 $\mu$m, and the FIR continuum. We do not attempt PDR models of the intermediate redshift sample as we only detect the [C\,II] line in that redshift bin which would not allow us to constrain the parameters characterizing the ISM (in particular constraining the radiation field-gas density parameter space).  

As previously discussed, all sources with $ z > 0.8 $ are divided into three smaller bins based on total infrared luminosity. The [C\,II] line is detected in each subset of the high-redshift stack. In the high-redshift stacks, we observed emission from singly-ionized carbon ([C\,II] at 158\,$\mu$m) as well as some weak emission from neutral oxygen ([O\,I] at 63\,$\mu$m). We perform PDR modeling for only one of three luminosity bins. In this bin (12.5  L$_{\odot} \, < $ L $<$ 13.0 L$_{\odot}$), the [C\,II] and [O\,I] detections were the strongest, while in the other two bins, the detections were either too weak or nonexistent.

Before applying measured line ratios to the PDR toolbox, we must make a number of corrections to the measured fluxes. First, the PDR models of \citet{Kaufman1999} and \citet{Kaufman2006} assume a single, plane-parallel, face-on PDR. However, if there are multiple clouds in the beam or if the clouds are in the active regions of galaxies, there can be emission from the front and back sides of the clouds, requiring the total infrared flux to be cut in half in order to be consistent with the models (e.g., \citealt{Kaufman1999,DeLooze2016}). Second, [O\,I] can be optically thick and suffers from self-absorption, so the measured [O\,I] is assumed to be only half of the true [O\,I] flux; i.e., we multiply the measured [O\,I] flux by two (e.g., \citealt{DeLooze2016,Contursi2013}). [C\,II] is assumed to be optically thin, so no correction is applied. Similarly, no correction is applied for [C\,I] and CO at low redshifts. Third, the different line species considered will have different beam filling factors for the SPIRE beam. We follow the method used in \citet{Wardlow2017} and apply a correction to only the [O\,I]/[C\,II] ratio using a relative filling factor for M82 from the literature. Since the large SPIRE beam size prevents measurement of the relative filling factors, the [O\,I]/[C\,II] ratio is corrected by a factor of 1/0.112, which is the measured relative filling factor for [O\,I] and [C\,II] in M82 \citep{Stacey1991, Lord1996, Kaufman1999, Contursi2013}. \citet{Wardlow2017} note that the M82 correction factor is large, so the corrected [O\,I]/[C\,II] ratio represents an approximate upper bound. Lastly, it is possible that a significant fraction of the [C\,II] flux can come from ionized gas in the ISM and not purely from the neutral gas in PDRs (e.g., \citealt{Abel2006,Contursi2013}). As a limiting case, we assume that 50\% of the [C\,II] emission comes from ionized regions. This correction factor is equivalent to the correction for ionized gas emission used in \citet{Wardlow2017} and is consistent with the results of \citet{Abel2006}, who finds that the ionized gas component makes up between 10-50\% of [C\,II] emission.

To summarize: a factor of 0.5 is applied to the FIR flux to account for the plane-parallel model of the PDR Toolbox, a factor of 2 is applied to the [O\,I] flux to account for optical thickness, a factor of 0.5 is applied to the [C\,II] flux to account for ionized gas emission, and lastly, a correction factor of 1/0.112 is applied to the [O\,I]/[C\,II] ratio to account for relative filling factors. We do not apply any corrections to the [C\,I] (1-0), [C\,I] (2-1), or CO (7-6) fluxes used in the PDR modeling of the lower-redshift stacks. These correction factors can significantly alter the flux ratios; for example, the ratio ([O\,I]/[C\,II])$_{\rm corrected}$ = 36$\times$([O\,I]/[C\,II])$_{\rm uncorrected}$. Tables \ref{table:highz_flux} and \ref{table:lowz_pdr_ratios} contain the uncorrected line ratios with the total correction factor for each ratio given in brackets.

\begin{figure*}[!th]
  \centering
  \begin{minipage}{\columnwidth}
    \includegraphics[width=\columnwidth,trim=3cm 0cm 0cm 1cm, scale=0.7]{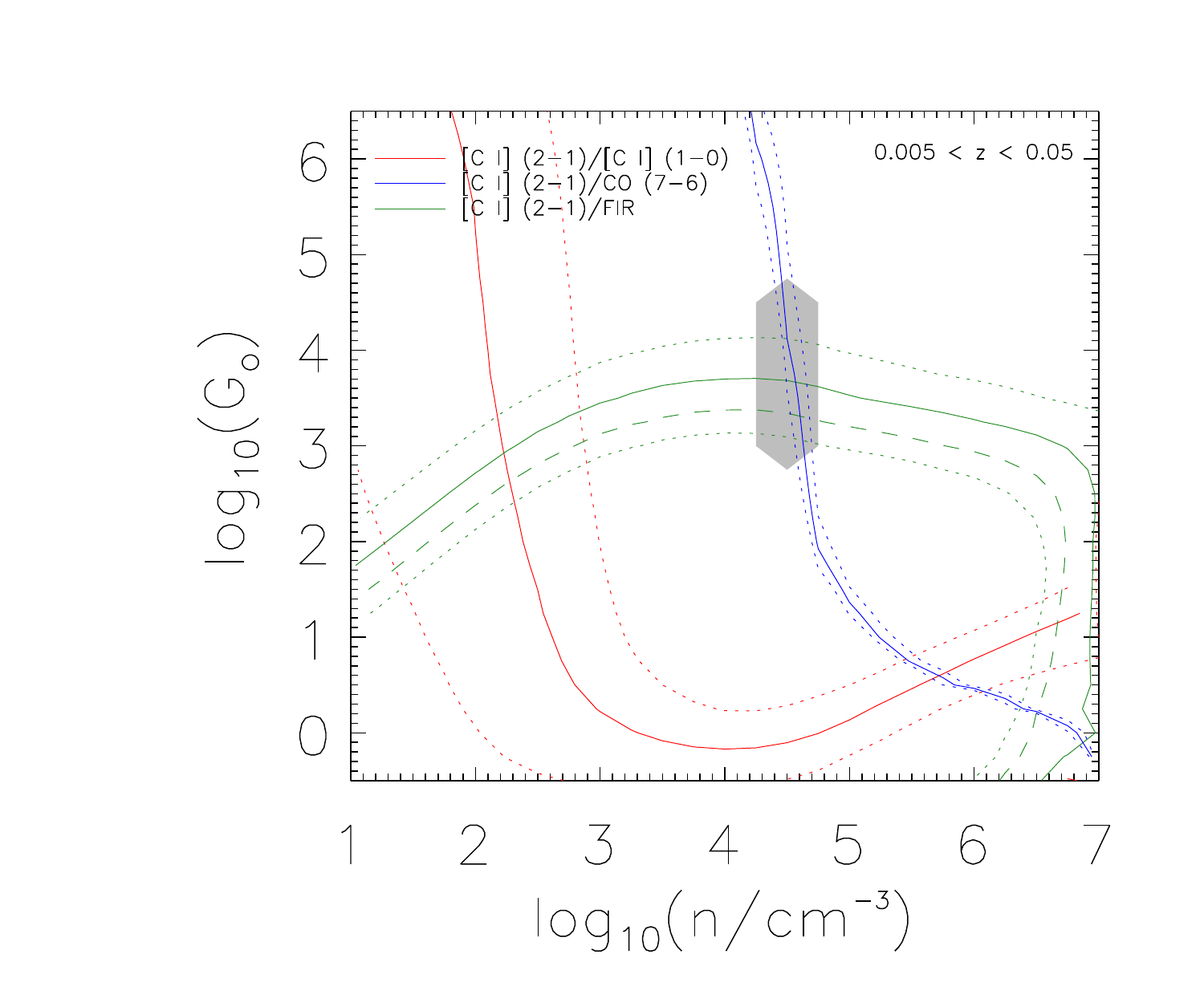}
  \end{minipage}
  \hfill
  \begin{minipage}{\columnwidth}
    \includegraphics[width=\columnwidth,trim=3cm 0cm 0cm 1cm, scale=0.7]{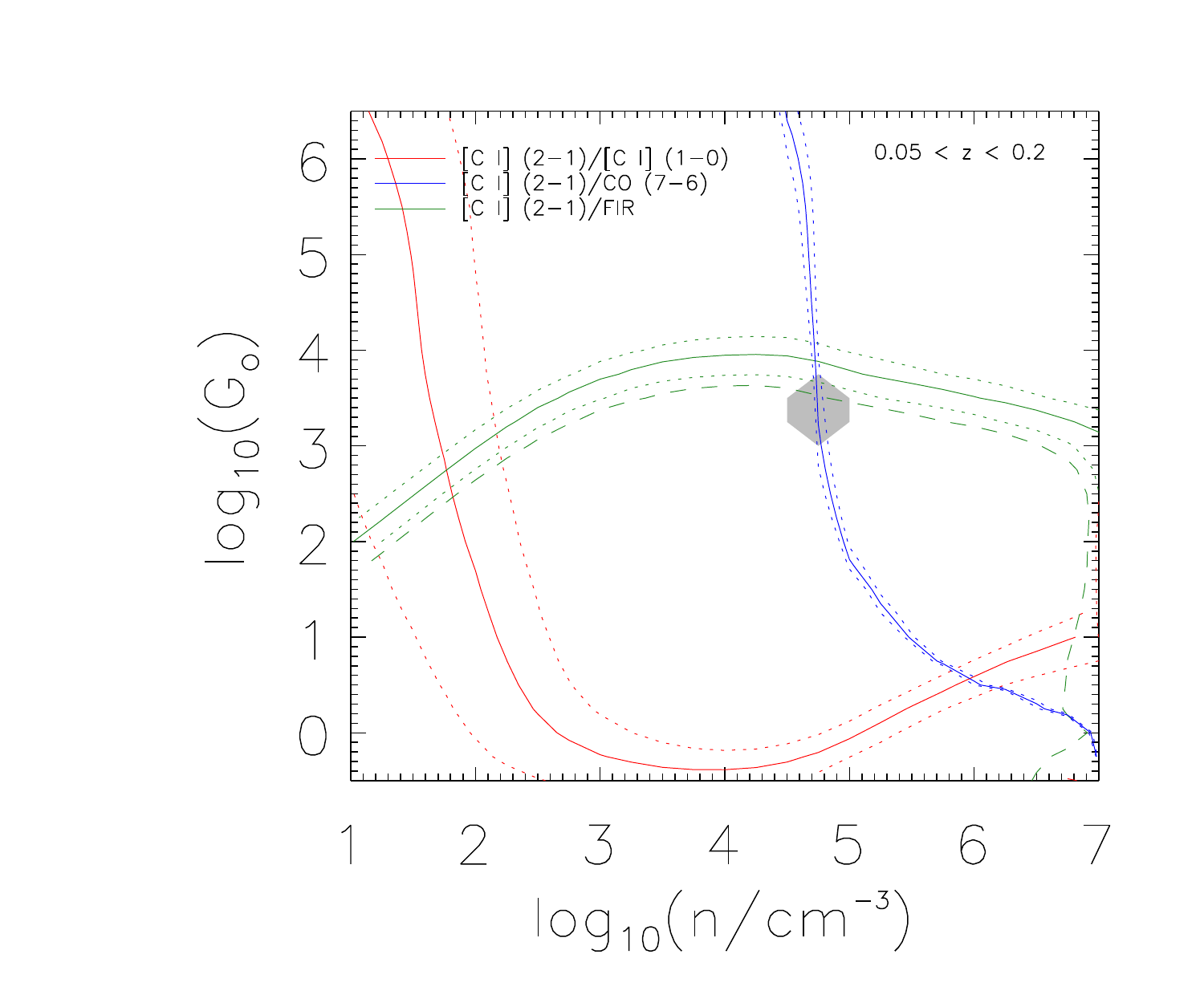}
  \end{minipage}

  \caption{PDR modeling of observed fluxes in $0.005 < z < 0.05$ bin (left) and $0.05 < z < 0.2$ (right). The solid lines are constraint contours determined from modeling, and the dotted lines are the 1$\sigma$ uncertainties. The dashed lines indicate the changes in line flux ratios when the FIR correction (see text) is applied. The gray regions indicate the most likely values of $n$ and $G_0$ determined from a likelihood analysis using the corrected flux values of FIR. Table \ref{table:lowz_pdr_ratios} lists the flux values for these two redshift bins before FIR corrections were applied. The line fluxes are in units of $\rm{W m^{-2}}$, and the L$_{\rm IR}$ is the far-infrared flux, where the wavelength range that defines L$_{\rm IR}$ is converted to 30-1000 $\mu$m \citep{Farrah2013}.}
    \label{n_go_ci}
\end{figure*}

\begin{figure*}[!th]
  \centering
  \begin{minipage}{\columnwidth}
    \includegraphics[width=\columnwidth,trim=3cm 0cm 0cm 1cm, scale=0.7]{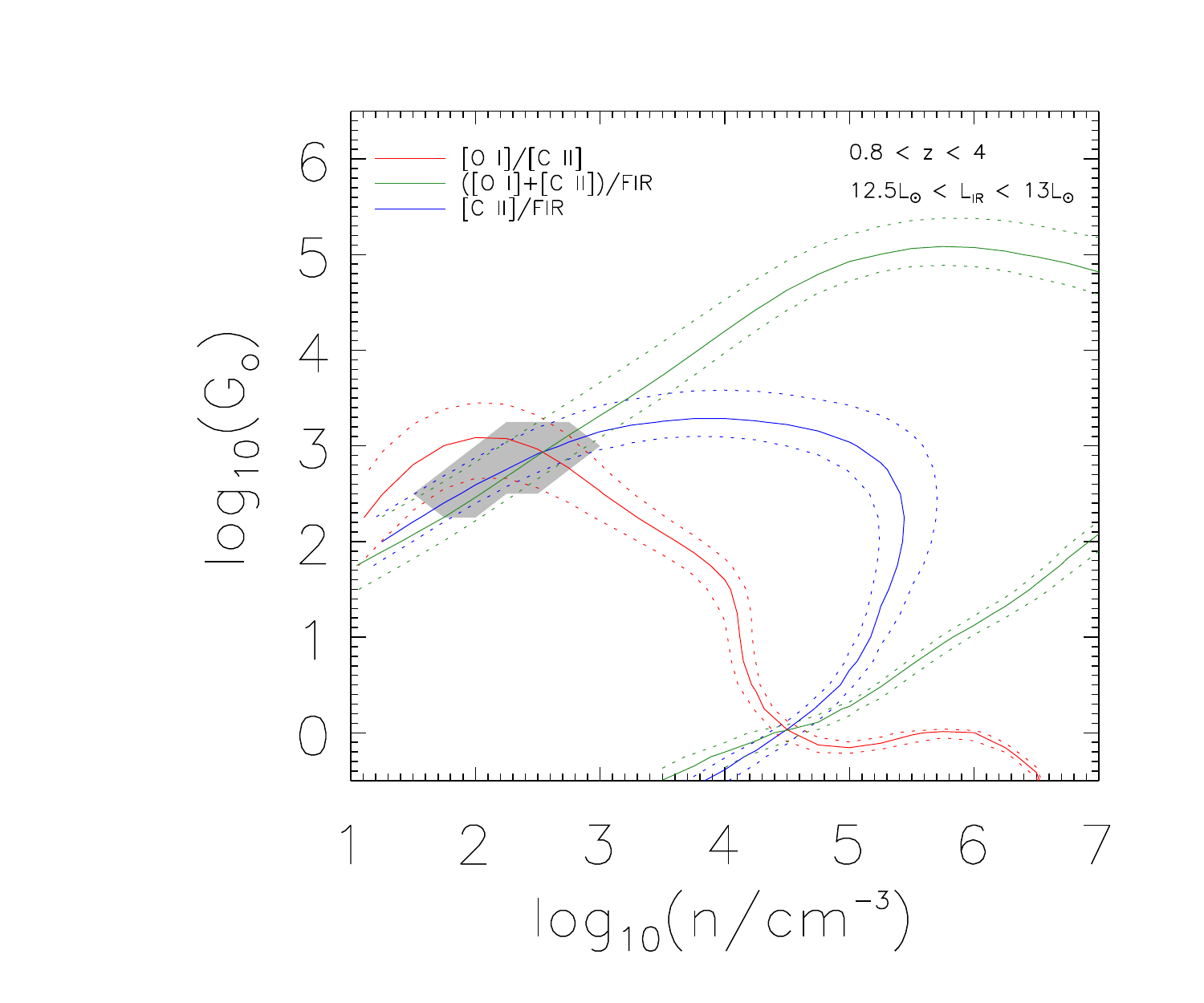}
  \end{minipage}
  \hfill
  \begin{minipage}{\columnwidth}
    \includegraphics[width=\columnwidth,trim=3cm 0cm 0cm 1cm, scale=0.7]{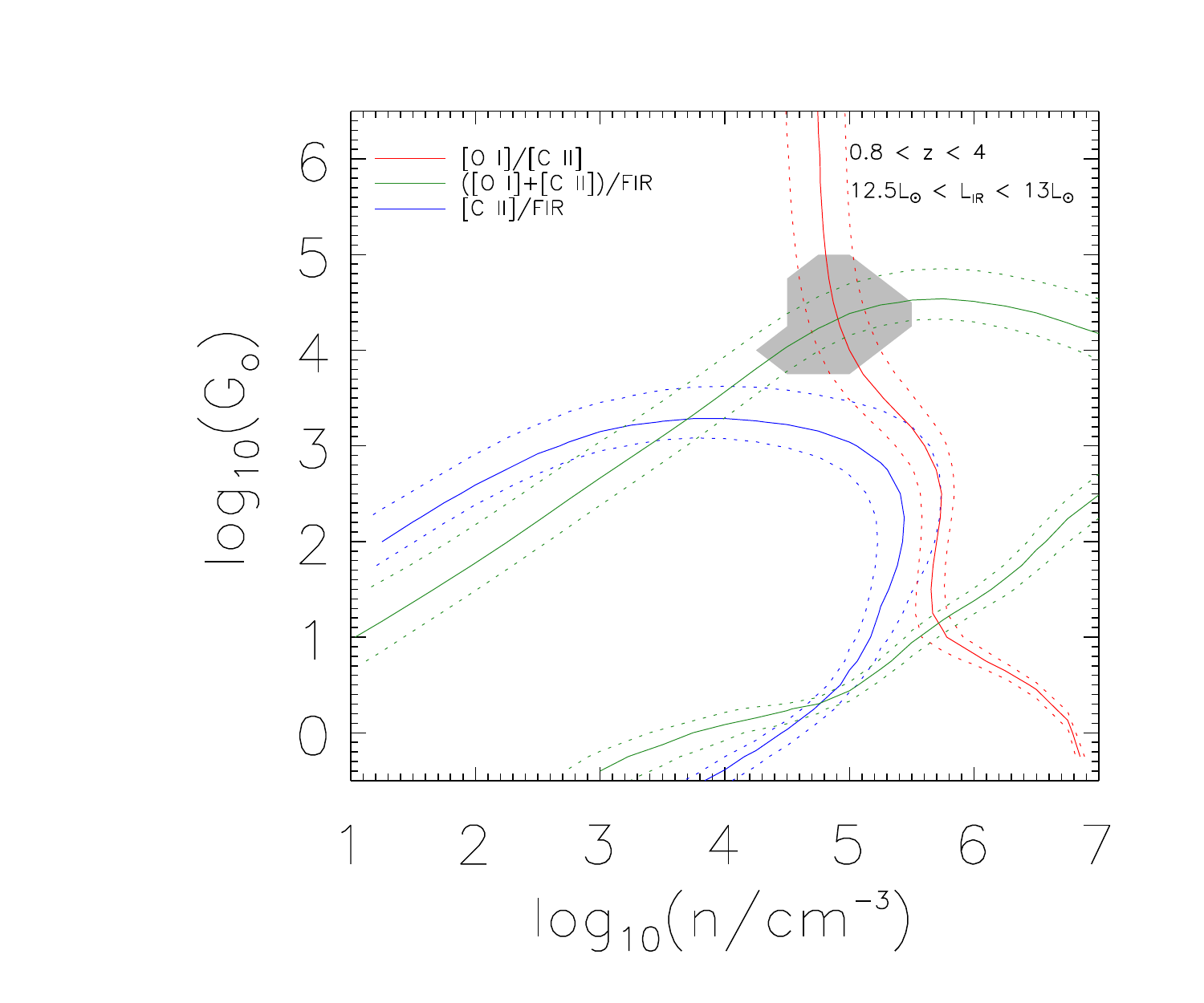}
  \end{minipage}

  \caption{\textit{Left:} PDR modeling of observed fluxes for sources with $0.8 < z < 4$ in the luminosity bin $10^{12.5}\,L_{\odot} < L_{\rm IR} < 10^{13}\,L_{\odot}$. No correction factors (see text) are applied to the line and line-FIR ratios in this plot. The gray regions indicates the most likely values of $n$ and $G_0$ determined from a likelihood analysis. The uncorrected ratios used for PDR modeling are given in Table \ref{table:highz_flux}. The line fluxes are in units of $\rm{W m^{-2}}$, and the FIR is the far-infrared flux, where the wavelength range that defines L$_{\rm IR}$ is converted to 30-1000\,$\mu$m \citep{Farrah2013}. Though sources in this redshift range are split into three bins based on total infrared luminosity in the text (L$_{\rm IR} < 10^{12.5}\,$L$_{\odot}$, $10^{12.5}\,$L$_{\odot} \, <\, $L$_{\rm IR} < 10^{13}\,$L$_{\odot}$, and L$_{\rm IR} > 10^{13}\,$L$_{\odot}$), the lack of [O\,I] detections in the first and third bins mean that PDR models for only the second bin are presented. \textit{Right:} Same PDR model as on the left but with the correction factors discussed in the text taken into account. The most variation appears in the [O\,I]/[C\,II] ratio, which shifts the intersection region from log($n$) $\sim$ 2.5 and log($G_0$) $\sim$ 2.5 to log($n$) $\sim$ 5 and log($G_0$) $\sim$ 4.}
    \label{n_go}
\end{figure*}

Naturally, these corrections introduce a large amount of uncertainty into our estimated line ratios. To demonstrate the effects that these corrections have on the results, we include contours from uncorrected and corrected line ratios in Figures \ref{n_go_ci} and \ref{n_go}. In Figure \ref{n_go_ci} (low redshifts), the only flux correction carried out is the correction to the FIR flux. This correction is indicated by the dashed line in each of the plots. In Figure \ref{n_go}, the lefthand-side plot displays the constraints on gas density and radiation field intensity (n,$\,$G$_0$) for high-redshift sources in the luminosity bin 12.5 L$_{\odot}$ $<$ L $<$ 13.0 L$_{\odot}$ determined from the uncorrected line ratios. The righthand-side plot shows the same contours but with the aforementioned correction factors taken into account. Clearly, the corrections can shift the intersection locus (the gray regions) to very different parts of n-G$_0$ parameter space. However, the correction factors should be treated with caution and represent limiting cases. The most variation is observed in the [O\,I]/[C\,II] ratio (shown in red), so the [O\,I]/[C\,II] contours on the lefthand and righthand plots in Figure \ref{n_go} represent the two extreme locations that this contour can occupy. The uncorrected line ratios are summarized in Tables \ref{table:highz_flux} and \ref{table:lowz_pdr_ratios}. These tables include line ratios that are not included in Figures \ref{n_go_ci} and \ref{n_go} (for example, Table \ref{table:highz_flux} contains the ratio [O\,I]/FIR, which does not appear in Figure \ref{n_go}). The figures contain only the independent ratios; the tables contain more (though not all independent ratios) for completeness.


\begin{figure*}[!th]
\centering
\includegraphics[trim=3cm 0cm 0cm 1cm, scale=1]{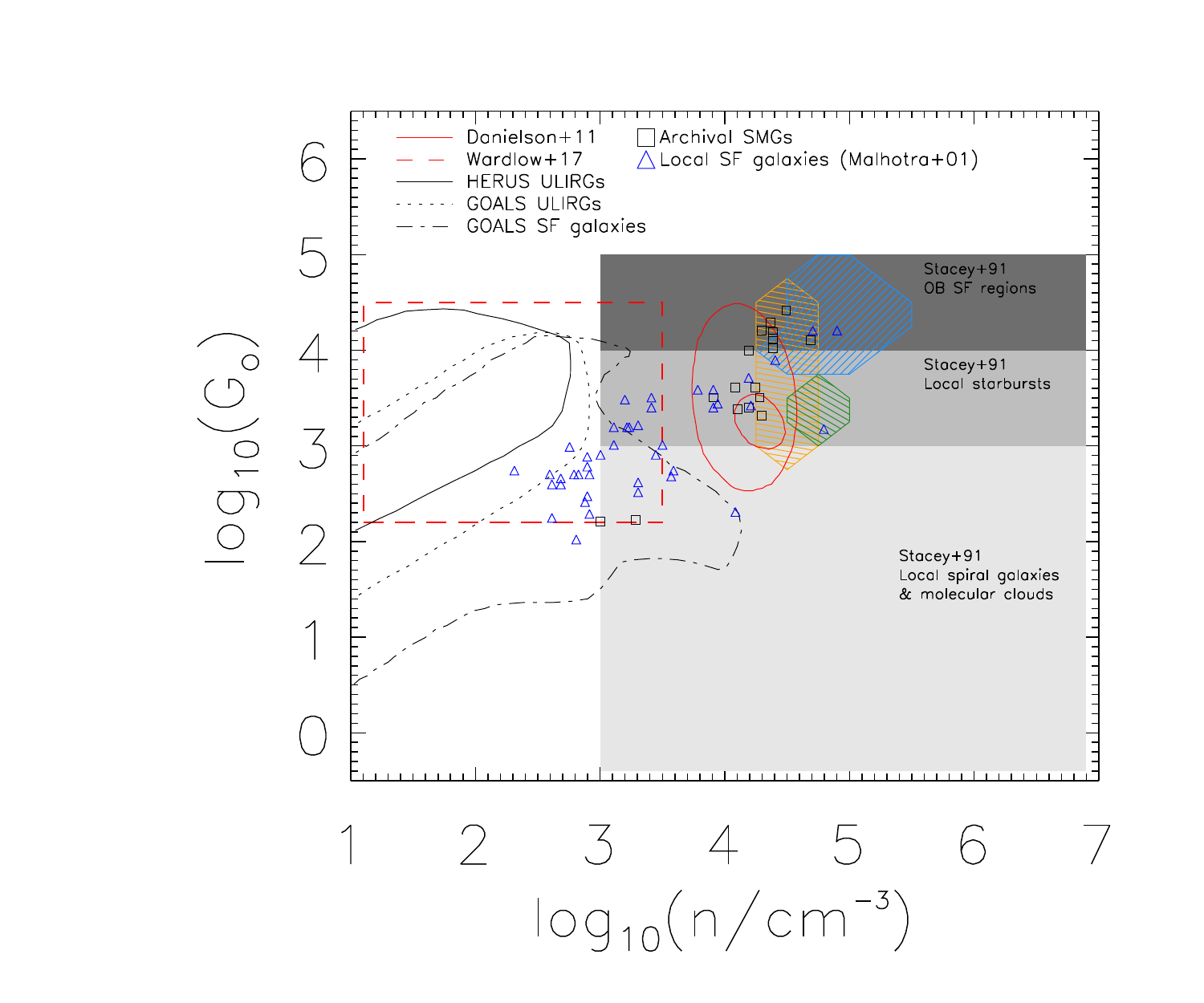}
\caption{The results of PDR modeling compared to results from the literature. The light blue region represents the derived n-G$_0$ for sources with $0.8 < z < 4$ and $12.5 < L/L_{\odot} < 13.0$. The orange and green regions represent the derived quantities for $0.005 < z < 0.05$ and $0.05 < z < 0.2$ subsamples, respectively. The regions shown here take into the account the correction factors discussed in the text. For comparison, the conditions for local spiral galaxies, molecular clouds, local starbursts, and galactic OB star-forming regions from \citet{Stacey1991} are shown, as well as data points for local star-forming galaxies from \citet{Malhotra2001} and for SMGs come from \citet{Wardlow2017, Sturm2010,Cox2011,Danielson2011,Valtchanov2011,Alaghband-Zadeh2013,Huynh2014}, and \citet{Rawle2014}.}
\label{fig:patches}
\end{figure*}

The gray shaded regions in Figures \ref{n_go_ci} and \ref{n_go} represent the most likely values of n and G$_0$ given the measured line flux ratios. To generate these regions, we perform a likelihood analysis using a method adapted from \cite{Ward2003}. The density n and radiation field strength G$_0$ are taken as free parameters. For measured line ratios $\vec{R}$ with errors $\vec{\sigma}$, we take a Gaussian form for the probability distribution; namely,

  \begin{equation}
    \mathrm{P(}\vec{\mathrm{R}}\,|\,\mathrm{n,G_0,}\, \vec{\sigma}) = \prod\limits_{i=1}^{\mathrm{N}} \frac{1}{\sqrt{2\pi}\sigma_i} \exp{\bigg\{-\frac{1}{2}\bigg[\frac{\mathrm{R}_i - \mathrm{M}_i}{\sigma_i}\bigg]^2\bigg\}}
    \end{equation}

  where the R$_i$ are the measured line ratios (i.e., [O\,I]/[C\,II], [C\,II]/FIR, etc.), N is the number of independent line ratios, and the M$_i$ are the theoretical line ratio plots from the PDR toolbox. A grid of discrete points in n, G$_0$-space ranging from $1 < $ log$_{10}($n$) < 7$ and $-0.5 < \, $log$_{10}($G$_0) \, < \, 6.5$ is constructed. To compute the most likely values of n and G$_0$, we use Bayes' theorem:

  \begin{equation}
    \mathrm{P(n,G_0}\,|\,\vec{\mathrm{R}},\vec{\sigma}) = \frac{\mathrm{P(n,G_0)\,P(}\vec{\mathrm{R}}\,|\,\mathrm{n,G_0},\vec{\sigma})}{\sum\limits_{\mathrm{n,G_0}} \, \mathrm{P(n,G_0)\,P(}\vec{\mathrm{R}}\,|\,\mathrm{n,G_0},\vec{\sigma})}
    \end{equation}

  The prior probability density function, P(n,G$_0$), is set equal to 1 for all points in the grid with G$_0>10^{2}$. Points with G$_0<10^{2}$ are given a prior probability of 0. The reason for this choice of prior stems from the argument that, given the intrinsic luminosities of our sources ($\sim 10^{11.5-13.5} $L$_{\odot}$), low values of G$_0$ (which include, for example, the value of G$_0$ at the line convergence in the high-$z$ PDR plot at $\rm {log(n/cm^{-3}) \sim 4.5}$ and $\rm{log(G_0) \sim  0.2}$) would correspond to galaxies with sizes on the order of hundreds of kpc or greater \citep{Wardlow2017}. Such sizes are expected to be unphysical, as typical measurements put galaxy sizes with these luminosities at ~$0.5-10 \,$kpc (see \citealt{Wardlow2017} and references therein). P(n, G$_0\,|\,\vec{\rm R},\vec{\sigma})$ gives the probability for each point in the n-G$_0$ grid that that point represents the actual conditions in the PDR, given the measured flux ratios. The gray regions in Figures \ref{n_go} and \ref{n_go_ci} are 68.2\% confidence regions. The relative likelihoods of each of the points in the grid are sorted from highest to lowest, and the cumulative sum for each grid point (the likelihood associated with that grid point summed with the likelihoods of the points preceding it in the high-to-low ordering) is computed. Grid points with a cumulative sum less than 0.682 represent the most likely values of density n and UV radiation intensity G$_0$, given the measured fluxes, with a total combined likelihood of 68.2\%. These points constitute the gray regions.

 The data constrain the interstellar gas density to be in the range $\rm{log(n/cm^{-3}) \sim 4.5  -  5.5}$ for both low-$z$ and high-$z$, where these values are estimated from the PDR models with correction factors taken into account. The FUV radiation is constrained to be in the range of $\rm{log(G_0) \sim  3 - 4}$ and $\rm{log(G_0) \sim  3 - 5}$ for low-$z$ and high-$z$, respectively.

The [C\,I] (2-1)/[C\,I] (1-0) line ratio is observed to deviate from the region of maximum likelihood on the $\rm G_0$-density diagram (Figure \ref{n_go_ci}). The region of maximum likelihood is shaded in gray in the figure. In fact this ratio is very sensitive to the conditions in the ISM, such that a modest change in the radiation strength or density would shift the line towards the expected locus \citep{Danielson2011}. The PDR models also constrain the assumption for the production of [C\,I] to that of a thin layer on the surface of far-UV heated molecular ISM whereas several studies \citep{Papadopoulos2004} point to the coexistence of neutral [C\,I] along CO in the same volume. These assumptions could also result in the deviations observed in the PDR models. 

Figure \ref{fig:patches} summarizes our main results of the PDR modeling based on the low and high redshift ISM emission lines from the stacked FTS spectra. We compare these measurements with that of local star-forming galaxies \citep{Malhotra2001}, local starbursts \citep{Stacey1991} and archival SMGs. We see from Figure \ref{fig:patches} that local DSFGs are on average subject to stronger UV radiation than that of local star-forming galaxies and are more consistent with local starbursts. Our measured density and radiation field strengths are further in agreement with results reported in \citet{Danielson2011} for a single DSFG at $z\sim2$. Given the uncertainty in filling factors and in the fraction of non-PDR [C\,II] emission, the [O\,I]/[C\,II] ratio contour in Figure \ref{n_go} may shift downward and to the left toward smaller density and radiation field strength where it would be more consistent with the results in \citet{Wardlow2017} for {\it Herschel}/PACS stacked spectra of DSFGs.


\section{Summary}

\begin{itemize}
\item We have stacked a diverse sample of \textit{Herschel} dusty, star-forming galaxies from redshifts $0.005<z<4$ and with total infrared luminosities from from LIRG levels up to luminosities in excess of $10^{13}\,$L$_{\odot}$. The sample is heterogeneous, consisting of starbursts, QSOs, and AGN, among other galaxy types. With this large sample, we presented a stacked statistical analysis of the archival spectra in redshift and luminosity bins.

\item We present the CO and H$_2$O spectral line energy distributions for the stacked spectra.
  
\item Radiative transfer modeling with RADEX places constraints on the gas density and temperature based on [C\,I] (2-1) 370 $\mu$m and [C\,I] (1-0) 609 $\mu$m measurements.

\item We use PDR modeling in conjunction with measured average fluxes to constrain the interstellar gas density to be in the range $\rm{log(n/cm^{-3}) \sim 4.5  -  5.5}$ for stacks at low and high redshifts. The FUV radiation is constrained to be in the range of $\rm{log(G_0) \sim  3 - 4}$ and $\rm{log(G_0) \sim  3 - 5}$, for low redshifts and high redshifts, respectively. Large uncertainties are present, especially due to effects such as contributions to the [C\,II] line flux due to non-PDR emission for which we can only estimate the correction factors to the observed line fluxes. Such uncertainties may lead to further discrepancies between the gas conditions at high- and low-redshifts, which may be understood in terms of nuclear starbursts of local DSFGs and luminous and ultra-luminous infrared galaxies compared to $\sim$ 10 kpc-scale massive starbursts of high-$z$ DSFGs. 
\end{itemize}


\section*{Acknowledgments} 
The authors thank an anonymous referee for his/her helpful comments and suggestions. The authors also thank Rodrigo Herrera-Camus, Eckhard Sturm, Javier Gracia-Carpio, and SHINING for sharing a compilation of [C\,II], [O\,III], and [O\,I] line measurements as well as FIR data to which we compare our results. We wish to thank Paul Van der Werf for the very useful suggestions and recommendations. Support for this paper was provided in part by NSF grant AST-1313319, NASA grant NNX16AF38G, GAANN P200A150121, HST-GO-13718, HST-GO-14083, and NSF Award \#1633631. JLW is supported by a European Union COFUND/Durham Junior Research Fellowship under EU grant agreement number 609412, and acknowledges additional support from STFC (ST/L00075X/1). GDZ acknowledges support from the ASI/INAF agreement n.~2014-024-R.1. The \textit{Herschel} spacecraft was designed, built, tested, and launched under a contract to ESA managed by the \textit{Herschel}/Planck Project team by an industrial consortium under the overall responsibility of the prime contractor Thales Alenia Space (Cannes), and including Astrium (Friedrichshafen) responsible for the payload module and for system testing at spacecraft level, Thales Alenia Space (Turin) responsible for the service module, and Astrium (Toulouse) responsible for the telescope, with in excess of a hundred subcontractors. SPIRE has been developed by a consortium of institutes led by Cardiff University (UK) and including Univ. Lethbridge (Canada); NAOC (China); CEA, LAM (France); IFSI, Univ. Padua (Italy); IAC (Spain); Stockholm Observatory (Sweden); Imperial College London, RAL, UCL-MSSL, UKATC, Univ. Sussex (UK); and Caltech, JPL, NHSC, Univ. Colorado (USA). This development has been supported by national funding agencies: CSA (Canada); NAOC (China); CEA, CNES, CNRS (France); ASI (Italy); MCINN (Spain); SNSB (Sweden); STFC, UKSA (UK); and NASA (USA). HIPE is a joint development by the Herschel Science Ground Segment Consortium, consisting of ESA, the NASA Herschel Science Center, and the HIFI, PACS and SPIRE consortia. This research has made use of the NASA/IPAC Extragalactic Database (NED) which is operated by the Jet Propulsion Laboratory, California Institute of Technology, under contract with the National Aeronautics and Space Administration.


\bibliographystyle{apj}
\bibliography{ftsbib}

\newpage
\appendix

\begin{figure*}
\centering
\includegraphics[trim=0cm 0cm 0cm 0cm, scale=0.7]{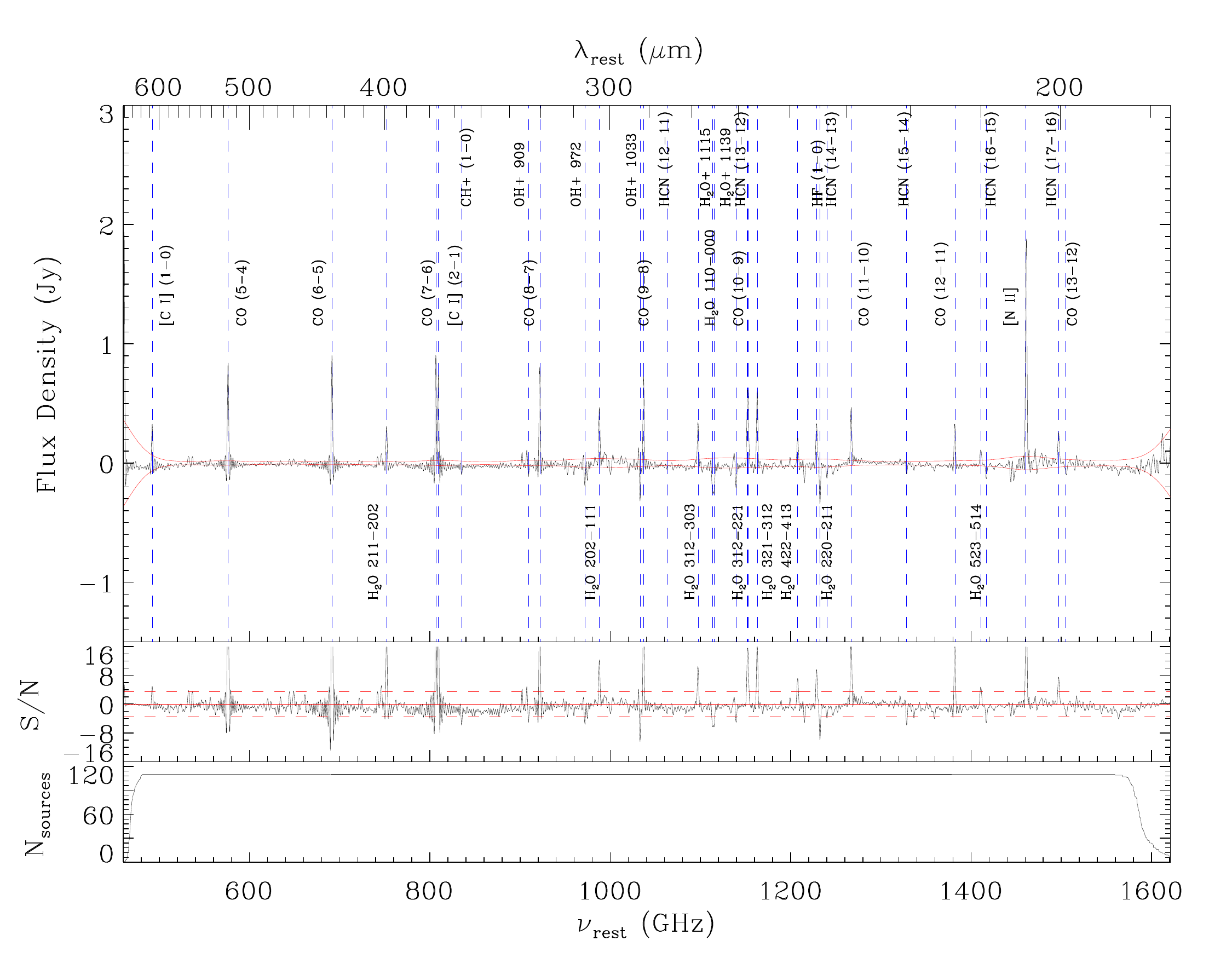}
\caption{\textit{Top}: Stacked SPIRE/FTS spectrum of archival sources with $0.005 < z < 0.05$ when stacked according to an inverse variance weighting scheme, unlike Figure \ref{fig:z0-005} which is an unweighted mean stack. We present this stack to show how sources such as Arp 220 that were measured with high signal-to-noise can dominate the stack if an inverse variance weighting scheme is used. In particular, strong absorption features from Arp 220 are still identifiable even after stacking. Fluxes from the emission lines in this figure can differ from the fluxes from Figure \ref{fig:z0-005} by as little as 10\% or up to a few hundred percent. Large differences in flux are apparent in the H$_2$O lines, which are significant in the Arp 220 spectrum, but which are reduced in significance when domination of the stack by sources like Arp 220 is removed. Overlaid is the $1\sigma$ jackknifed noise level in red and dashed vertical lines showing the locations of main molecular emission lines. \textit{Middle}: Signal-to-noise ratio. The horizontal dashed line indicates $\rm S/N = 3.5$, and the solid red line indicates $\rm S/N = 0$. Lines with $\rm S/N > 3.5$ were considered detected.  \textit{Bottom}: The number of sources that contribute to the stack at each wavelength.}
\label{fig:z005-02_inv}
\end{figure*}

Figure \ref{fig:z005-02_inv} shows the stack at $0.005 < z < 0.05$ resulting from an inverse variance weighting scheme. In the main text, an unweighted average is used for this redshift bin. In Figure \ref{fig:z005-02_inv}, sources with low signal-to-noise, such as Arp 220, dominate the stack. Notable in this stack are the absorption features, which are present primarily in Arp 220 and survive the stacking process. Tables \ref{table:obsids} and \ref{table:all_targets} enumerate the sources and source properties used in this work. 


\begin{longtable*}{p{5cm} p{5cm} p{5cm} p{2cm} } 
\caption{All sources, along with their respective integration times.} \\
\label{table:obsids}  \\
\hline\hline\\ [0.1ex] 
Target & Obs. ID & Program & Int. Time [s] \\ [0.5ex] 
\hline\hline\\
& &\\[0.5ex] 

Mrk 231            & 1342187893 & SDP\textunderscore pvanderw\textunderscore 3 & 6601 \\
                   & 1342210493 & SDP\textunderscore pvanderw\textunderscore 3 & 13722 \\
Arp 220            & 1342190674 & KPGT\textunderscore wilso01\textunderscore 1 & 9772 \\
IRAS F17207-0014   & 1342192829 & SDP\textunderscore pvanderw\textunderscore 3 & 6334 \\
IRAS F18293-3413   & 1342192830 & SDP\textunderscore pvanderw\textunderscore 3 & 5279 \\
NGC 1614           & 1342192831 & SDP\textunderscore pvanderw\textunderscore 3 & 6334 \\
IRAS F05189-2524   & 1342192832 & SDP\textunderscore pvanderw\textunderscore 3 & 16360 \\
                   & 1342192833 & SDP\textunderscore pvanderw\textunderscore 3 & 16360 \\
IC 4687            & 1342192993 & SDP\textunderscore pvanderw\textunderscore 3 & 13986 \\
SDP.81             & 1342197467 & GT1\textunderscore ivaltcha\textunderscore 1 & 13194 \\
SDP.130            & 1342197469 & GT1\textunderscore ivaltcha\textunderscore 1 & 13194 \\
NGC 7552           & 1342198428 & SDP\textunderscore pvanderw\textunderscore 3 & 1717 \\
Arp 299            & 1342199248 & SDP\textunderscore pvanderw\textunderscore 3 & 4620 \\
                   & 1342199249 & SDP\textunderscore pvanderw\textunderscore 3 & 4620 \\
NGC 7469           & 1342199252 & SDP\textunderscore pvanderw\textunderscore 3 & 11875 \\
NGC 34             & 1342199253 & SDP\textunderscore pvanderw\textunderscore 3 & 14249 \\
NGC 3256           & 1342201201 & SDP\textunderscore pvanderw\textunderscore 3 & 4883 \\
ESO 173-G015       & 1342202268 & SDP\textunderscore pvanderw\textunderscore 3 & 1717 \\
NGC 1365           & 1342204020 & SDP\textunderscore pvanderw\textunderscore 3 & 3168 \\
                   & 1342204021 & SDP\textunderscore pvanderw\textunderscore 3 & 5279 \\
Mrk 273            & 1342209850 & SDP\textunderscore pvanderw\textunderscore 3 & 13062 \\
Arp 193            & 1342209853 & SDP\textunderscore pvanderw\textunderscore 3 & 14249 \\
ESO 320-G030       & 1342210861 & SDP\textunderscore pvanderw\textunderscore 3 & 5675 \\
IC 1623            & 1342212314 & SDP\textunderscore pvanderw\textunderscore 3 & 12799 \\
Mrk 331            & 1342212316 & SDP\textunderscore pvanderw\textunderscore 3 & 13061 \\
NGC 7771           & 1342212317 & SDP\textunderscore pvanderw\textunderscore 3 & 14249 \\
IRAS 13120-5453    & 1342212342 & SDP\textunderscore pvanderw\textunderscore 3 & 3828 \\
NGC 5135           & 1342212344 & SDP\textunderscore pvanderw\textunderscore 3 & 14250 \\
CGCG 049-057       & 1342212346 & SDP\textunderscore pvanderw\textunderscore 3 & 14250 \\
NGC 6052           & 1342212347 & OT1\textunderscore nlu\textunderscore 1      & 2641 \\
MCG +12-02-001     & 1342213377 & SDP\textunderscore pvanderw\textunderscore 3 & 1385 \\
MCG-03-04-014      & 1342213442 & OT1\textunderscore nlu\textunderscore 1      & 5279 \\
CGCG 436-030       & 1342213443 & OT1\textunderscore nlu\textunderscore 1      & 5280 \\
NGC 6240           & 1342214831 & SDP\textunderscore pvanderw\textunderscore 3 & 12798 \\
ESO 286-G035       & 1342216901 & OT1\textunderscore nlu\textunderscore 1      & 2640 \\
NGC 2623           & 1342219553 & SDP\textunderscore pvanderw\textunderscore 3 & 12007 \\
SMMJ2135-0102      & 1342219562 & OT1\textunderscore rivison\textunderscore 1  & 13194 \\
GOODS-N07          & 1342219575 & OT1\textunderscore apope\textunderscore 1    & 9237 \\
CGCG 448-020       & 1342221679 & OT1\textunderscore nlu\textunderscore 1      & 2641 \\
MCG+04-48-002      & 1342221682 & OT1\textunderscore nlu\textunderscore 1      & 1321 \\
UGC 12150          & 1342221699 & OT1\textunderscore nlu\textunderscore 1      & 2640 \\
IC 5298            & 1342221700 & OT1\textunderscore nlu\textunderscore 1      & 2640 \\
NGC 7679           & 1342221701 & OT1\textunderscore nlu\textunderscore 1      & 2640 \\
NGC 7592           & 1342221702 & OT1\textunderscore nlu\textunderscore 1      & 2641 \\
NGC 0232           & 1342221707 & OT1\textunderscore nlu\textunderscore 1      & 2641 \\
ESO 244-G012       & 1342221708 & OT1\textunderscore nlu\textunderscore 1      & 2641 \\
NGC 3221           & 1342221714 & OT1\textunderscore nlu\textunderscore 1      & 1322 \\
NGC 6286           & 1342221715 & OT1\textunderscore nlu\textunderscore 1      & 1322 \\
NGC 6621           & 1342221716 & OT1\textunderscore nlu\textunderscore 1      & 2641 \\
IRAS 03158+4227    & 1342224764 & OT1\textunderscore dfarrah\textunderscore 1  & 9237  \\
NGC 0695           & 1342224767 & OT1\textunderscore nlu\textunderscore 1      & 5279 \\
IRAS 23365+3604    & 1342224768 & OT1\textunderscore dfarrah\textunderscore 1  & 5279 \\
IRAS 14378-3651    & 1342227456 & OT1\textunderscore dfarrah\textunderscore 1  & 5939 \\
UGC 03094          & 1342227522 & OT1\textunderscore nlu\textunderscore 1      & 2641 \\
IRAS 04271+3849    & 1342227786 & OT1\textunderscore nlu\textunderscore 1      & 2641 \\
NGC 1961           & 1342228708 & OT1\textunderscore nlu\textunderscore 1      & 1321 \\
MCG+02-20-003      & 1342228728 & OT1\textunderscore nlu\textunderscore 1      & 2641 \\
NGC 2342           & 1342228729 & OT1\textunderscore nlu\textunderscore 1      & 2641 \\
IRAS 05223+1908    & 1342228738 & OT1\textunderscore nlu\textunderscore 1      & 2641 \\
UGC 03608          & 1342228744 & OT1\textunderscore nlu\textunderscore 1      & 2641 \\
IRAS 05442+1732    & 1342230413 & OT1\textunderscore nlu\textunderscore 1      & 1322 \\
MCG+08-11-002      & 1342230414 & OT1\textunderscore nlu\textunderscore 1      & 1322 \\
UGC 03351          & 1342230415 & OT1\textunderscore nlu\textunderscore 1      & 1322 \\
IRAS F17138-1017   & 1342230418 & OT1\textunderscore nlu\textunderscore 1      & 1321 \\
ESO 099-G004       & 1342230419 & OT1\textunderscore nlu\textunderscore 1      & 5279 \\
IRAS 06035-7102    & 1342230420 & OT1\textunderscore dfarrah\textunderscore 1  & 7258 \\
IRAS 08311-2459    & 1342230421 & OT1\textunderscore dfarrah\textunderscore 1  & 7258 \\
IRAS 06206-6315    & 1342231038 & OT1\textunderscore dfarrah\textunderscore 1  & 9237 \\
IRAS 19254-6315    & 1342231039 & OT1\textunderscore dfarrah\textunderscore 1  & 7258 \\
ESO 069-IG006      & 1342231040 & OT1\textunderscore nlu\textunderscore 1      & 7918 \\
NGC 6156           & 1342231041 & OT1\textunderscore nlu\textunderscore 1      & 1322 \\
ESO 138-G027       & 1342231042 & OT1\textunderscore nlu\textunderscore 1      & 2641 \\
IRAS 17578-0400    & 1342231047 & OT1\textunderscore nlu\textunderscore 1      & 1322 \\
IRAS 20087-0308    & 1342231049 & OT1\textunderscore dfarrah\textunderscore 1  & 9237 \\
NGC 6926           & 1342231050 & OT1\textunderscore nlu\textunderscore 1      & 2640 \\
UGC 11041          & 1342231061 & OT1\textunderscore nlu\textunderscore 1      & 2640 \\
IRAS 09022-3615    & 1342231063 & OT1\textunderscore nlu\textunderscore 1      & 7917 \\
NGC 4194           & 1342231069 & OT1\textunderscore nlu\textunderscore 1      & 1321 \\
NGC 2388           & 1342231071 & OT1\textunderscore nlu\textunderscore 1      & 1321 \\
UGC 03410          & 1342231072 & OT1\textunderscore nlu\textunderscore 1      & 1321 \\
IRAS 19297-0406    & 1342231078 & OT1\textunderscore dfarrah\textunderscore 1  & 5279  \\
NGC 2369           & 1342231083 & OT1\textunderscore nlu\textunderscore 1      & 1322 \\
ESO 255-IG007      & 1342231084 & OT1\textunderscore nlu\textunderscore 1      & 5279 \\
NGC 3110           & 1342231971 & OT1\textunderscore nlu\textunderscore 1      & 1321 \\
IRAS 08355-4944    & 1342231975 & OT1\textunderscore nlu\textunderscore 1      & 2641 \\
G09v1.40           & 1342231977 & OT1\textunderscore rivison\textunderscore 1  & 13194 \\
IRAS 08572+3915    & 1342231978 & OT1\textunderscore dfarrah\textunderscore 1  & 5279 \\
HerMES-Lock01      & 1342231980 & OT2\textunderscore rivison\textunderscore 2  & 13195 \\
G09v1.326          & 1342231985 & OT1\textunderscore rivison\textunderscore 1  & 13195 \\
SDP.9              & 1342231986 & OT1\textunderscore rivison\textunderscore 1  & 13195 \\
G09v1.97           & 1342231988 & OT1\textunderscore rivison\textunderscore 1  & 13194 \\
SPT0538-50         & 1342231989 & OT1\textunderscore dmarrone\textunderscore 1 & 13194 \\
ESO 339-G011       & 1342231990 & OT1\textunderscore nlu\textunderscore 1      & 2640 \\
NGC 6701           & 1342231994 & OT1\textunderscore nlu\textunderscore 1      & 1321 \\
NGC 5010           & 1342236996 & OT1\textunderscore nlu\textunderscore 1      & 1321 \\
VV 340             & 1342238241 & OT1\textunderscore nlu\textunderscore 1      & 5279 \\
UGC 545            & 1342238246 & SDP\textunderscore pvanderw\textunderscore 3 & 15042 \\
IRAS 16090-0139    & 1342238699 & OT1\textunderscore dfarrah\textunderscore 1  & 10556 \\
G15v2.235          & 1342238700 & OT1\textunderscore rivison\textunderscore 1  & 13194 \\
G15v2.19           & 1342238701 & OT1\textunderscore rivison\textunderscore 1  & 13194 \\
IRAS 03521+0028    & 1342238704 & OT1\textunderscore dfarrah\textunderscore 1  & 11875 \\
HXMM02             & 1342238706 & OT2\textunderscore rivison\textunderscore 2  & 13195 \\
Mrk 1014           & 1342238707 & SDP\textunderscore pvanderw\textunderscore 3 & 13495  \\
HBootes03          & 1342238709 & OT2\textunderscore rivison\textunderscore 2  & 13195 \\
Mrk 478            & 1342238710 & SDP\textunderscore pvanderw\textunderscore 3 & 5279 \\
IRAS 15250+3609    & 1342238711 & OT1\textunderscore dfarrah\textunderscore 1  & 5279 \\
VV 705             & 1342238712 & OT1\textunderscore nlu\textunderscore 1      & 5279 \\
NGC 0958           & 1342239339 & OT1\textunderscore nlu\textunderscore 1      & 2641 \\
UGC 02238          & 1342239340 & OT1\textunderscore nlu\textunderscore 1      & 2641 \\
UGC 02369          & 1342239341 & OT1\textunderscore nlu\textunderscore 1      & 5279 \\
NGC 0877           & 1342239342 & OT1\textunderscore nlu\textunderscore 1      & 1322 \\
IRAS F01417+1651   & 1342239343 & OT1\textunderscore nlu\textunderscore 1      & 2641 \\
UGC 02608          & 1342239356 & OT1\textunderscore nlu\textunderscore 1      & 2641 \\
NGC 0828           & 1342239357 & OT1\textunderscore nlu\textunderscore 1      & 1321 \\
NGC 0317B          & 1342239358 & OT1\textunderscore nlu\textunderscore 1      & 2641 \\
IC 4734            & 1342240013 & OT1\textunderscore nlu\textunderscore 1      & 1322 \\
NGC 5990           & 1342240016 & OT1\textunderscore nlu\textunderscore 1      & 1323 \\
UGC 02982          & 1342240021 & OT1\textunderscore nlu\textunderscore 1      & 1322 \\
UGC 01845          & 1342240022 & OT1\textunderscore nlu\textunderscore 1      & 1322 \\
NGC 1572           & 1342242588 & OT1\textunderscore nlu\textunderscore 1      & 2640 \\
MCG-05-12-006      & 1342242589 & OT1\textunderscore nlu\textunderscore 1      & 2640 \\
ESO 420-G013       & 1342242590 & OT1\textunderscore nlu\textunderscore 1      & 1321 \\
PG 1613+658        & 1342242593 & SDP\textunderscore pvanderw\textunderscore 3 & 13194 \\
FLS02              & 1342242594 & OT2\textunderscore drigopou\textunderscore 3 & 13194 \\
                   & 1342259071 & OT2\textunderscore drigopou\textunderscore 3 & 13195 \\
IRAS 20414-1651    & 1342243623 & OT1\textunderscore dfarrah\textunderscore 1  & 9237 \\
IRAS 22491-1808    & 1342245082 & OT1\textunderscore dfarrah\textunderscore 1  & 7258 \\
IRAS 01003-2238    & 1342246256 & OT1\textunderscore dfarrah\textunderscore 1  & 11875 \\
IRAS 20100-4156    & 1342245106 & OT1\textunderscore dfarrah\textunderscore 1  & 7258 \\
ESO 286-IG019      & 1342245107 & OT1\textunderscore nlu\textunderscore 1      & 5279 \\
ESO 467-G027       & 1342245108 & OT1\textunderscore nlu\textunderscore 1      & 2641 \\
IC 5179            & 1342245109 & OT1\textunderscore nlu\textunderscore 1      & 1322 \\
ESO 148-IG002      & 1342245110 & OT1\textunderscore nlu\textunderscore 1      & 6994 \\
NGC 7674           & 1342245858 & OT1\textunderscore nlu\textunderscore 1      & 5279 \\
IRAS 00397-1312    & 1342246257 & OT1\textunderscore dfarrah\textunderscore 1  & 13194 \\
IRAS 00188-0856    & 1342246259 & OT1\textunderscore dfarrah\textunderscore 1  & 11875 \\
SWIRE05            & 1342246268 & OT2\textunderscore drigopou\textunderscore 3 & 13194 \\
IRAS 23230-6926    & 1342246276 & OT1\textunderscore dfarrah\textunderscore 1  & 10555 \\
IRAS 23253-5415    & 1342246277 & OT1\textunderscore dfarrah\textunderscore 1  & 11875 \\
ESO 350-IG038      & 1342246978 & OT1\textunderscore nlu\textunderscore 1      & 2641 \\
DAN03              & 1342246979 & OT2\textunderscore drigopou\textunderscore 3 & 13194 \\
IRAS F10565+2448   & 1342247096 & OT1\textunderscore nlu\textunderscore 1      & 5279 \\
BOOTES01           & 1342247113 & OT2\textunderscore drigopou\textunderscore 3 & 13194 \\
IRAS 10378+1109    & 1342247118 & OT1\textunderscore dfarrah\textunderscore 1  & 11875 \\
UGC 08739          & 1342247123 & OT1\textunderscore nlu\textunderscore 1      & 2641 \\
NGC 5653           & 1342247565 & OT1\textunderscore nlu\textunderscore 1      & 1322 \\
NGC 5104           & 1342247566 & OT1\textunderscore nlu\textunderscore 1      & 2641 \\
MCG-02-33-098      & 1342247567 & OT1\textunderscore nlu\textunderscore 1      & 2641 \\
ESO 353-G020       & 1342247615 & OT1\textunderscore nlu\textunderscore 1      & 2641 \\
MCG-02-01-051      & 1342247617 & OT1\textunderscore nlu\textunderscore 1      & 2641 \\
NGC 0023           & 1342247622 & OT1\textunderscore nlu\textunderscore 1      & 1322 \\
G12v2.43           & 1342247744 & OT1\textunderscore rivison\textunderscore 1  & 13194 \\
G12v2.30           & 1342247758 & OT1\textunderscore rivison\textunderscore 1  & 13195 \\
G12v2.257          & 1342247759 & OT1\textunderscore rivison\textunderscore 1  & 13195 \\
IRAS 11095-0238    & 1342247760 & OT1\textunderscore dfarrah\textunderscore 1  & 10556 \\
IRAS 12071-0444    & 1342248239 & OT1\textunderscore dfarrah\textunderscore 1  & 11875 \\
NGP-NC.v1.143      & 1342248412 & OT1\textunderscore rivison\textunderscore 1  & 13194 \\
NGP-NA.v1.56       & 1342248416 & OT1\textunderscore rivison\textunderscore 1  & 13194 \\
NGC 5734           & 1342248417 & OT1\textunderscore nlu\textunderscore 1      & 1321 \\
ESO 507-G070       & 1342248421 & OT1\textunderscore nlu\textunderscore 1      & 2641 \\
MCG-03-34-064      & 1342249041 & OT1\textunderscore nlu\textunderscore 1      & 2640 \\
IC 4280            & 1342249042 & OT1\textunderscore nlu\textunderscore 1      & 2640 \\
ESO 264-G057       & 1342249043 & OT1\textunderscore nlu\textunderscore 1      & 2640 \\
ESO 264-G036       & 1342249044 & OT1\textunderscore nlu\textunderscore 1      & 2640 \\
IRAS 15462-0450    & 1342249045 & OT1\textunderscore dfarrah\textunderscore 1  & 11875  \\
NGC 5936           & 1342249046 & OT1\textunderscore nlu\textunderscore 1      & 1321 \\
Mrk 463            & 1342249047 & OT1\textunderscore dfarrah\textunderscore 1  & 11875 \\
NGP-NB.v1.78       & 1342249063 & OT1\textunderscore rivison\textunderscore 1  & 13194 \\
NGP-NB.v1.43       & 1342249064 & OT1\textunderscore rivison\textunderscore 1  & 13194 \\
NGP-NA.v1.144      & 1342249066 & OT1\textunderscore rivison\textunderscore 1  & 13195 \\
IRAS 14348-1447    & 1342249457 & OT1\textunderscore dfarrah\textunderscore 1  & 5939 \\
ESO 221-IG010      & 1342249461 & OT1\textunderscore nlu\textunderscore 1      & 1322 \\
IRAS 12116-5615    & 1342249462 & OT1\textunderscore nlu\textunderscore 1      & 5279 \\
IC 4518AB          & 1342250514 & OT1\textunderscore nlu\textunderscore 1      & 2641 \\
CGCG 052-037       & 1342251284 & OT1\textunderscore nlu\textunderscore 1      & 5279 \\
IRAS F16399-0937   & 1342251334 & OT1\textunderscore nlu\textunderscore 1      & 5279 \\
IRAS F16516-0948   & 1342251335 & OT1\textunderscore nlu\textunderscore 1      & 5279 \\
SWIRE04            & 1342253658 & OT2\textunderscore drigopou\textunderscore 3 & 13194 \\
IRAS 07598+6508    & 1342253659 & OT1\textunderscore dfarrah\textunderscore 1  & 11875 \\
MM J18423+5938     & 1342253672 & OT2\textunderscore maravena\textunderscore 2 & 15833 \\
                   & 1342255798 & OT2\textunderscore maravena\textunderscore 2 & 17152 \\ 
                   & 1342255810 & OT2\textunderscore maravena\textunderscore 2 & 17152 \\
                   & 1342255811 & OT2\textunderscore maravena\textunderscore 2 & 17152 \\
                   & 1342255812 & OT2\textunderscore maravena\textunderscore 2 & 17152 \\
                   & 1342256357 & OT2\textunderscore maravena\textunderscore 2 & 17152 \\
MG 0751+2716       & 1342253966 & OT1\textunderscore mbradfor\textunderscore 1 & 21110 \\
APM 08279+5255     & 1342253967 & OT1\textunderscore mbradfor\textunderscore 1 & 21110 \\
SWIRE01            & 1342254034 & OT2\textunderscore drigopou\textunderscore 3 & 13194 \\
SWIRE03            & 1342254035 & OT2\textunderscore drigopou\textunderscore 3 & 13194 \\
SWIRE02            & 1342255265 & OT2\textunderscore drigopou\textunderscore 3 & 13195 \\
SDP.17             & 1342255280 & OT1\textunderscore rivison\textunderscore 1  & 13194 \\
SDP.11             & 1342255281 & OT1\textunderscore rivison\textunderscore 1  & 13194 \\
G09v1.124          & 1342255282 & OT1\textunderscore rivison\textunderscore 1  & 13194 \\
IRAS FSC10214+4724 & 1342255799 & OT1\textunderscore mbradfor\textunderscore 1 & 21110 \\
GOODS-N26          & 1342256083 & OT1\textunderscore apope\textunderscore 1    & 9236 \\
GOODS-N19          & 1342256358 & OT1\textunderscore apope\textunderscore 1    & 9236 \\
GOODS-NC1          & 1342256359 & OT1\textunderscore apope\textunderscore 1    & 9236 \\
NGC 7591           & 1342257346 & OT1\textunderscore nlu\textunderscore 1      & 2640 \\
BOOTES03           & 1342257936 & OT2\textunderscore drigopou\textunderscore 3 & 13195 \\
HXMM01             & 1342258698 & OT2\textunderscore drigopou\textunderscore 3 & 13194 \\
FLS01              & 1342258701 & OT2\textunderscore drigopou\textunderscore 3 & 13194 \\
BOOTES02           & 1342259073 & OT2\textunderscore drigopou\textunderscore 3 & 13195 \\
Cloverleaf         & 1342259573 & OT1\textunderscore mbradfor\textunderscore 1 & 21110 \\
CDFS01             & 1342259582 & OT2\textunderscore drigopou\textunderscore 3 & 13195 \\
CDFS02             & 1342259583 & OT2\textunderscore drigopou\textunderscore 3 & 13194 \\
CDFS04             & 1342259584 & OT2\textunderscore drigopou\textunderscore 3 & 13195 \\
SMMJ02399-0136     & 1342262900 & OT2\textunderscore cferkinh\textunderscore 1 & 17811 \\

\end{longtable*}


\begin{longtable}{p{3cm} p{2cm} p{2cm} p{2cm} p{2cm} p{2cm} p{3cm}} 
\label{table:all_targets} \\
\caption{Properties of sources included in the stacks.} \\
\hline\hline\\ [0.1ex] 
Target & R.A.  & DEC. & $z_{spec}$  & $\mu$ & L$_{\rm IR}$ & Refs.  \\ [0.5ex] 
\hline\hline\\
& & & & & & \\[0.5ex] 

NGC 7552            & 23:16:10.80 & -42:35:04.65 & 0.0054   & - & 11.11 & R15,G14,A09\\
NGC 1365            & 03:33:36.61 & -36:08:18.05 & 0.00546  & - & 11.00 & R15,A09 \\
NGC 4194            & 12:14:09.78 & +54:31:34.36 & 0.008342 & - & 11.10 & A09,L17 \\
NGC 3256            & 10:27:51.18 & -43:54:14.21 & 0.0094   & - & 11.64 & R15,G14,A09\\
ESO 173-G015        & 13:27:23.73 & -57:29:22.96 & 0.0097   & - & 11.38 & R15,G14,A09\\
NGC 5010            & 13:12:26.52 & -15:47:51.75 & 0.009924 & - & 11.50 & A09,L17 \\
ESO 221-IG010       & 13:50:56.87 & -49:03:18.66 & 0.010337 & - & 11.22 & A09,L17 \\
Arp 299             & 11:28:33.31 & +58:33:44.89 & 0.0104   & - & 11.93 & R15,NED,A09\\
ESO 320-G030        & 11:53:11.63 & -39:07:49.34 & 0.0108   & - & 11.17 & R15,G14,A09\\
NGC 2369            & 07:16:37.96 & -62:20:35.71 & 0.010807 & - & 11.16 & A09,L17 \\
NGC 6156            & 16:34:52.26 & -60:37:06.06 & 0.010885 & - & 11.14 & A09,L17 \\
IC 5179             & 22:16:09.08 & -36:50:36.54 & 0.011415 & - & 11.24 & A09,L17 \\
NGC 5653            & 14:30:09.89 & +31:12:56.97 & 0.011881 & - & 11.13 & A09,L17 \\
ESO 420-G013        & 04:13:49.69 & -32:00:24.34 & 0.011908 & - & 11.07 & A09,L17 \\
NGC 5990            & 15:46:16.41 & +02:24:54.68 & 0.012806 & - & 11.13 & A09,L17 \\
CGCG 049-057        & 15:13:13.18 & +07:13:30.24 & 0.013    & - & 11.35 & R15,A09\\
NGC 0877            & 02:18:00.12 & +14:32:34.34 & 0.013052 & - & 11.10 & A09,L17 \\
UGC 03410           & 06:14:30.27 & +80:27:00.89 & 0.013079 & - & 11.10 & A09,L17 \\
NGC 1961            & 05:42:04.67 & +69:22:42.69 & 0.013122 & - & 11.06 & A09,L17 \\
NGC 6701            & 18:43:12.47 & +60:39:09.42 & 0.013226 & - & 11.12 & A09,L17 \\
NGC 5936            & 15:30:00.76 & +12:59:20.78 & 0.013356 & - & 11.14 & A09,L17 \\
NGC 5135            & 13:25:44.09 & -29:49:59.51 & 0.0137   & - & 11.30 & R15,G14,A09\\
NGC 3221            & 10:22:20.36 & +21:34:21.41 & 0.013709 & - & 11.09 & A09,L17 \\
NGC 5734            & 14:45:09.02 & -20:52:13.17 & 0.013746 & - & 11.15 & A09,L17 \\
NGC 2388            & 07:28:53.51 & +33:49:08.56 & 0.01379  & - & 11.28 & A09,L17 \\
MCG+04-48-002       & 20:28:35.15 & +25:44:03.27 & 0.0139   & - & 11.11 & A09,L17 \\
IRAS 17578-0400     & 18:00:31.78 & -04:00:54.86 & 0.014043 & - & 11.48 & A09,L17 \\
NGC 7771            & 23:51:24.67 & +20:06:40.24 & 0.0143   & - & 11.40 & R15,G14,A09\\
UGC 03351           & 05:45:48.22 & +58:42:05.69 & 0.01486  & - & 11.28 & A09,L17 \\
NGC 0023            & 00:09:53.39 & +25:55:25.98 & 0.015231 & - & 11.12 & A09,L17 \\
UGC 01845           & 02:24:07.87 & +47:58:11.61 & 0.015607 & - & 11.12 & A09,L17 \\
IC 4734             & 18:38:25.65 & -57:29:25.01 & 0.015611 & - & 11.35 & A09,L17 \\
MCG+12-02-001       & 00:54:03.47 & +73:05:10.14 & 0.0157   & - & 11.50 & R15,G14,A09\\
IC 4518AB           & 14:57:41.12 & -43:07:56.00 & 0.015728 & - & 11.23 & A09,L17 \\
NGC 6052            & 16:05:12.93 & +20:32:36.42 & 0.015808 & - & 11.09 & A09,L17 \\
NGC 1614            & 04:33:59.75 & -08:34:44.84 & 0.0159   & - & 11.65 & R15,G14,A09\\
ESO 353-G020        & 01:34:51.30 & -36:08:14.57 & 0.015921 & - & 11.06 & A09,L17 \\
MCG-02-33-098       & 13:02:19.78 & -15:46:03.69 & 0.015921 & - & 11.17 & A09,L17 \\
MCG+02-20-003       & 07:35:43.63 & +11:42:36.02 & 0.016255 & - & 11.13 & A09,L17 \\
UGC 11041           & 17:54:51.76 & +34:46:32.76 & 0.016281 & - & 11.11 & A09,L17 \\
NGC 7469            & 23:03:15.79 & +08:52:28.62 & 0.0163   & - & 11.65 & R15,G14,A09\\
IC 4280             & 13:32:53.31 & -24:12:25.81 & 0.016331 & - & 11.15 & A09,L17 \\
NGC 7591            & 23:18:16.34 & +06:35:09.43 & 0.016531 & - & 11.12 & A09,L17 \\
MCG-03-34-064       & 13:22:24.43 & -16:43:42.76 & 0.016541 & - & 11.28 & A09,L17 \\
UGC 08739           & 13:49:14.26 & +35:15:19.63 & 0.016785 & - & 11.15 & A09,L17 \\
NGC 3110            & 10:04:02.20 & -06:28:28.08 & 0.016858 & - & 11.37 & A09,L17 \\
NGC 7679            & 23:28:46.63 & +03:30:43.34 & 0.017139 & - & 11.11 & A09,L17 \\
ESO 264-G057        & 10:59:01.79 & -43:26:25.61 & 0.017199 & - & 11.14 & A09,L17 \\
IC 4687             & 18:13:39.82 & -57:43:30.54 & 0.0173   & - & 11.62 & R15,G14,A09 \\
IRAS F17138-1017    & 17:16:35.74 & -10:20:41.31 & 0.017335 & - & 11.49 & A09,L17 \\
ESO 286-G035        & 21:04:11.06 & -43:35:30.24 & 0.017361 & - & 11.20 & A09,L17 \\
ESO 467-G027        & 22:14:39.81 & -27:27:50.32 & 0.017401 & - & 11.08 & A09,L17 \\
NGC 2342            & 07:09:18.22 & +20:38:11.26 & 0.017599 & - & 11.31 & A09,L17 \\
UGC 02982           & 04:12:22.51 & +05:32:49.65 & 0.017696 & - & 11.20 & A09,L17 \\
NGC 0828            & 02:10:09.45 & +39:11:25.50 & 0.017926 & - & 11.36 & A09,L17 \\
Arp 220             & 15:34:57.22 & +23:30:12.34 & 0.0181   & - & 12.28 & R11,R15,A09\\
NGC 0317B           & 00:57:40.32 & +43:47:31.94 & 0.018109 & - & 11.19 & A09,L17 \\
IRAS F18293-3413    & 18:32:41.36 & -34:11:25.88 & 0.0182   & - & 11.88 & R15,A09\\
NGC 6286            & 16:58:31.38 & +58:56:15.38 & 0.018349 & - & 11.37 & A09,L17 \\
NGC 2623            & 08:38:24.06 & +25:45:17.00 & 0.0185   & - & 11.60 & R15,G14,A09\\
Mrk 331             & 23:51:26.53 & +20:35:09.12 & 0.0185   & - & 11.50 & R15,G14,A09\\
NGC 5104            & 13:21:22.99 & +00:20:33.32 & 0.018606 & - & 11.27 & A09,L17 \\
IRAS 05442+1732     & 05:47:11.24 & +17:33:47.37 & 0.01862  & - & 11.30 & A09,L17 \\
MCG-05-12-006       & 04:52:04.97 & -32:59:25.66 & 0.018753 & - & 11.17 & A09,L17 \\
IRAS 04271+3849     & 04:30:33.23 & +38:55:48.62 & 0.018813 & - & 11.11 & A09,L17 \\
NGC 0958            & 02:30:42.79 & -02:56:23.96 & 0.01914  & - & 11.20 & A09,L17 \\
MCG+08-11-002       & 05:40:43.73 & +49:41:41.58 & 0.019157 & - & 11.46 & A09,L17 \\
ESO 339-G011        & 19:57:37.37 & -37:56:08.47 & 0.0192   & - & 11.20 & A09,L17 \\
NGC 34              & 00:11:06.54 & -12:06:23.70 & 0.0196   & - & 11.49 & R15,G14,A09\\
NGC 6926            & 20:33:06.02 & -02:01:40.02 & 0.019613 & - & 11.32 & A09,L17 \\
IC 1623             & 01:07:46.72 & -17:30:27.31 & 0.0201   & - & 11.71 & R15,G14,A09\\
NGC 1572            & 04:22:42.75 & -40:36:03.31 & 0.020384 & - & 11.30 & A09,L17 \\
ESO 350-IG038       & 00:36:52.39 & -33:33:17.36 & 0.020598 & - & 11.28 & A09,L17 \\
NGC 6621            & 18:12:55.21 & +68:21:46.48 & 0.020652 & - & 11.29 & A09,L17 \\
ESO 138-G027        & 17:26:43.00 & -59:55:54.34 & 0.020781 & - & 11.41 & A09,L17 \\
ESO 264-G036        & 10:43:07.67 & -46:12:44.83 & 0.021065 & - & 11.32 & A09,L17 \\
UGC 03608           & 06:57:34.59 & +46:24:11.62 & 0.021351 & - & 11.34 & A09,L17 \\
UGC 12150           & 22:41:12.20 & +34:14:53.38 & 0.021391 & - & 11.35 & A09,L17 \\
ESO 507-G070        & 13:02:52.32 & -23:55:17.90 & 0.021702 & - & 11.56 & A09,L17 \\
UGC 02238           & 02:46:17.42 & +13:05:45.17 & 0.021883 & - & 11.33 & A09,L17 \\
NGC 0232            & 00:42:45.76 & -23:33:39.06 & 0.022639 & - & 11.44 & A09,L17 \\
IRAS F16516-0948    & 16:54:23.78 & -09:53:21.05 & 0.022706 & - & 11.32 & A09,L17 \\
ESO 244-G012        & 01:18:08.31 & -44:27:38.75 & 0.022903 & - & 11.38 & A09,L17 \\
Arp 193             & 13:20:35.49 & +34:08:22.52 & 0.0233   & - & 11.73 & R15,G14,A09\\
UGC 02608           & 03:15:01.23 & +42:02:08.76 & 0.023343 & - & 11.41 & A09,L17 \\
NGC 7592            & 23:18:22.22 & -04:24:56.56 & 0.024444 & - & 11.40 & A09,L17 \\
UGC 03094           & 04:35:33.85 & +19:10:18.51 & 0.02471  & - & 11.44 & A09,L17 \\
CGCG 052-037        & 16:30:56.58 & +04:04:58.99 & 0.02449  & - & 11.45 & A09,L17 \\
NGC 6240            & 16:52:59.10 & +02:24:04.07 & 0.0245   & - & 11.93 & R15,G14,A09\\
IRAS 08355-4944     & 08:37:02.00 & -49:54:29.02 & 0.025898 & - & 11.62 & A09,L17 \\
IRAS F16399-0937    & 16:42:40.11 & -09:43:13.41 & 0.027012 & - & 11.63 & A09,L17 \\
IRAS 12116-5615     & 12:14:22.18 & -56:32:32.78 & 0.027102 & - & 11.65 & A09,L17 \\
MCG-02-01-051       & 00:18:50.85 & -10:22:37.73 & 0.027299 & - & 11.48 & A09,L17 \\
IRAS F01417+1651    & 01:44:30.50 & +17:06:08.40 & 0.027399 & - & 11.64 & A09,L17 \\
IC 5298             & 23:16:00.62 & +25:33:22.03 & 0.027422 & - & 11.60 & A09,L17 \\
NGC 7674            & 23:27:56.74 & +08:46:43.49 & 0.028924 & - & 11.56 & A09,L17 \\
ESO 099-G004        & 15:24:57.73 & -63:07:30.44 & 0.029284 & - & 11.74 & A09,L17 \\
IRAS 05223+1908     & 05:25:16.75 & +19:10:49.25 & 0.029577 & - & 11.65 & A09,L17 \\
IRAS 13120-5453     & 13:15:06.42 & -55:09:21.22 & 0.0308   & - & 12.34 & R15,G14\\
UGC 02369           & 02:54:01.75 & +14:58:13.49 & 0.031202 & - & 11.67 & A09,L17 \\
CGCG 436-030        & 01:20:02.47 & +14:21:40.61 & 0.031229 & - & 11.69 & A09,L17 \\
NGC 0695            & 01:51:14.37 & +22:34:55.81 & 0.032472 & - & 11.68 & A09,L17 \\
VV 340              & 14:57:00.78 & +24:37:04.37 & 0.033669 & - & 11.74 & A09,L17 \\
MCG-03-04-014       & 01:10:08.85 & -16:51:11.29 & 0.035144 & - & 11.65 & A09,L17 \\
CGCG 448-020        & 20:57:24.32 & +17:07:37.28 & 0.036098 & - & 11.94 & A09,L17 \\
Mrk 273             & 13:44:42.42 & +55:53:11.78 & 0.0378   & - & 12.21 & R15,G14,A09\\
ESO 255-IG007       & 06:27:21.68 & -47:10:36.83 & 0.03879  & - & 11.90 & A09,L17 \\
VV 705              & 15:18:06.26 & +42:44:43.68 & 0.040191 & - & 11.92 & A09,L17 \\
Mrk 231             & 12:56:14.46 & +56:52:24.92 & 0.0422   & - & 12.57 & R15,G14,A09,V10\\
IRAS F05189-2524    & 05:21:01.29 & -25:21:45.21 & 0.0426   & - & 12.16 & R15,G14,A09\\
IRAS F17207-0014    & 17:23:21.98 & -00:17:00.96 & 0.0428   & - & 12.46 & R15,A09\\
ESO 286-IG019       & 20:58:26.81 & -42:38:59.71 & 0.042996 & - & 12.06 & A09,L17 \\
IRAS F10565+2448    & 10:59:18.17 & +24:32:34.26 & 0.0431   & - & 12.08 & A09,L17 \\
ESO 148-IG002       & 23:15:46.70 & -59:03:14.91 & 0.044601 & - & 12.06 & A09,L17 \\
ESO 069-IG006       & 16:38:11.46 & -68:26:07.94 & 0.046972 & - & 11.98 & A09,L17 \\
Mrk 463             & 13:56:02.86 & +18:22:20.05 & 0.051    & - & 11.79 & M90,F07,F13\\
IRAS 15250+3609     & 15:26:59.56 & +35:58:37.63 & 0.055    & - & 12.00 & M90,F07,I11 \\
IRAS 08572+3915     & 09:00:25.70 & +39:03:54.83 & 0.058    & - & 12.11 & M90,F07,G14\\
UGC 545             & 00:53:34.86 & +12:41:34.87 & 0.0589   & - & 11.93 & S11,E06,NED\\
IRAS 09022-3615     & 09:04:12.87 & -36:27:00.14 & 0.059641 & - & 12.31 & A09,L17 \\
IRAS 19254-6315     & 19:31:20.23 & -72:39:22.89 & 0.063    & - & 12.09 & M90,F07,F13\\
IRAS 23365+3604     & 23:39:01.08 & +36:21:09.28 & 0.064    & - & 12.15 & M90,F07,G14 \\
IRAS 14378-3651     & 14:40:58.89 & -37:04:32.47 & 0.068    & - & 12.23 & HW88,F07,I11\\
IRAS 22491-1808     & 22:51:49.20 & -17:52:23.62 & 0.078    & - & 11.65 & M90,F07,F13\\
IRAS 06035-7102     & 06:02:54.36 & -71:03:08.81 & 0.079    & - & 12.22 & M90,F07 ,F13\\
Mrk 478             & 14:42:07.61 & +35:26:23.22 & 0.0791   & - & 11.52 & E06,S11,NED\\
IRAS 14348-1447     & 14:37:38.29 & -15:00:24.75 & 0.083    & - & 12.42 & HW88,F07,G14\\
IRAS 19297-0406     & 19:32:21.84 & -04:00:02.83 & 0.086    & - & 12.45 & HW88,F07,I11\\
IRAS 20414-1651     & 20:44:18.25 & -16:40:16.02 & 0.087    & - & 12.22 & HW88,F07,F13\\
IRAS 06206-6315     & 06:21:01.47 & -63:17:23.12 & 0.092    & - & 12.23 & M90,F07,F13\\
IRAS 08311-2459     & 08:33:20.65 & -25:09:32.20 & 0.100    & - & 12.50 & HW88,F07 ,F13\\
IRAS 15462-0450     & 15:48:56.83 & -04:59:33.71 & 0.100    & - & 12.24 & M90,F07,F13\\
IRAS 20087-0308     & 20:11:23.73 & -02:59:50.82 & 0.106    & - & 12.42 & HW88,F07,F13\\
IRAS 11095-0238     & 11:12:03.35 & -02:54:23.93 & 0.107    & - & 12.28 & HW88,F07,F13\\
IRAS 23230-6926     & 23:26:03.52 & -69:10:19.11 & 0.107    & - & 12.37 & M90,F07,F13\\
IRAS 01003-2238     & 01:02:49.92 & -22:21:56.47 & 0.118    & - & 12.32 & HW88,F07,F13 \\
IRAS 00188-0856     & 00:21:26.52 & -08:39:26.86 & 0.128    & - & 12.39 & M90,F07 ,F13\\
IRAS 12071-0444     & 12:09:45.13 & -05:01:13.46 & 0.128    & - & 12.41 & M90,F07,F13\\
PG 1613+658         & 16:13:57.15 & +65:43:09.19 & 0.129    & - & 11.88 & NED,E06\\
IRAS 20100-4156     & 20:13:29.68 & -41:47:34.68 & 0.13     & - & 12.67 & HW88,F07,F13\\
IRAS 23253-5415     & 23:28:06.15 & -53:58:30.96 & 0.13     & - & 12.36 & M90,F07,F13\\
IRAS 03158+4227     & 03:19:12.50 & +42:38:28.37 & 0.134    & - & 12.63 & HW88,F07,F13\\
IRAS 16090-0139     & 16:11:40.61 & -01:47:05.86 & 0.134    & - & 12.55 & M90,F07,F13\\
IRAS 10378+1109     & 10:40:29.13 & +10:53:18.36 & 0.136    & - & 12.31 & HW88,F07,F13\\
SWIRE03             & 10:40:43.66 & +59:34:09.66 & 0.148    & - & 12.25 & M90,H03,Y13\\
IRAS 07598+6508     & 08:04:30.36 & +64:59:52.76 & 0.148    & - & 12.50 & M90,F07,F13 \\
IRAS 03521+0028     & 03:54:42.10 & +00:37:00.71 & 0.152    & - & 12.52 & HW88,F07 ,F13\\
Mrk 1014            & 01:59:50.10 & +00:23:39.10 & 0.1631   & - & 12.62 & F13,NED \\
CDFS04              & 03:35:49.16 & -27:49:18.29 & 0.168    & - & 11.80 & E09,D09,M90,M11\\
BOOTES03            & 14:28:49.80 & +34:32:39.81 & 0.219    & - & 11.87 & M14\\
CDFS02              & 03:28:18.03 & -27:43:08.25 & 0.248    & - & 11.82 & M14 \\
BOOTES02            & 14:32:34.88 & +33:28:32.25 & 0.25     & - & 11.91 & M14 \\
SWIRE04             & 10:32:37.47 & +58:08:45.75 & 0.251    & - & 11.80 & M14 \\
IRAS 00397-1312     & 00:42:15.50 & -12:56:03.17 & 0.262    & - & 12.90 & HW88,F07 \\
DAN03               & 00:40:14.68 & -43:20:10.81 & 0.265    & - & 11.59 & M14\\
CDFS01              & 03:29:04.38 & -28:47:52.52 & 0.289    & - & 11.79 & M14 \\
BOOTES01            & 14:36:31.97 & +34:38:29.60 & 0.354    & - & 12.69 & M14 \\
SWIRE02             & 10:51:13.42 & +57:14:25.79 & 0.362    & - & 11.90 & M14 \\
SWIRE05             & 10:35:58.01 & +58:58:46.17 & 0.366    & - & 12.06 & M14 \\
FLS01               & 17:20:17.08 & +59:16:37.47 & 0.417    & - & 11.86 & M14 \\
FLS02               & 17:13:31.69 & +58:58:04.60 & 0.436    & - & 12.41 & M14\\
SWIRE01             & 10:47:53.34 & +58:21:05.99 & 0.887    & - & 12.92 & M14\\
G15v2.19*           & 14:29:35.27 & -00:28:36.23 & 1.027    & 9.7  $\pm$ 0.7  & 12.42 & ME14,W17 \\
GOODS-N26           & 12:36:34.53 & +62:12:39.74 & 1.219    & -               & 12.26 & G14,M12 \\
HBootes03           & 14:28:24.16 & +35:26:19.84 & 1.326    & 3.0  $\pm$ 1.5  & 12.69 & B13 \\
SDP.9               & 09:07:40.22 & -00:41:59.03 & 1.577    & 8.8  $\pm$ 2.2  & 12.58 & B13 \\
NGP-NB.v1.43        & 13:24:27.23 & +28:44:50.37 & 1.68     & 2.8  $\pm$ 0.4  & 12.92 & B13,T16 \\
SDP.11              & 09:10:43.04 & -00:03:22.72 & 1.784    & 10.9 $\pm$ 1.3  & 12.48 & B13 \\
GOODS-N07           & 12:36:21.22 & +62:17:10.32 & 1.9924   & -               & 12.42 & M12\\
GOODS-NC1           & 12:36:00.25 & +62:10:47.13 & 2.0017   & -               & 12.22 & G14 \\
G09v1.40*           & 08:53:59.04 & +01:55:38.05 & 2.0894   & 15.3 $\pm$ 3.5  & 12.45 & B13,W17 \\
G12v2.257*          & 11:58:20.06 & -01:37:52.10 & 2.191    & 13.0 $\pm$ 7.0  & 12.04 & W17 \\
NGP-NA.v1.144*      & 13:36:49.88 & +29:18:00.95 & 2.202    & 4.4  $\pm$ 0.8  & 12.95 & B13,W17\\
IRAS FSC10214+4724  & 10:24:34.58 & +47:09:10.11 & 2.286    & 50              & 12.46 & BL95,RR93,B99 \\
NGP-NA.v1.56*       & 13:44:29.41 & +30:30:36.10 & 2.301    & 11.7 $\pm$ 0.9  & 12.77 & B13,W17 \\
SDP.17              & 09:03:02.83 & -01:41:25.84 & 2.3051   & 4.9  $\pm$ 0.7  & 13.00 & B13\\
HXMM01*             & 02:20:16.53 & -06:01:43.67 & 2.308    & 1.5  $\pm$ 0.3  & 13.30 & B13,W17 \\
SMMJ2135-0102       & 21:35:11.60 & -01:02:52.16 & 2.326    & 37.5 $\pm$ 4.5  & 12.41 & I10b,SW11\\
G09v1.124*          & 08:49:33.31 & +02:14:42.61 & 2.410    & 2.8  $\pm$ 0.2  & 13.14 & B13,W17\\
G15v2.235*          & 14:13:52.18 & -00:00:23.84 & 2.478    & 1.8  $\pm$ 0.3  & 13.26 & B13,W17 \\
GOODS-N19           & 12:37:07.26 & +62:14:07.55 & 2.484    & -               & 12.57 & M12 \\
Cloverleaf          & 14:15:46.26 & +11:29:43.45 & 2.56     & 11              & 12.78 & B92,B99,VS03,U16 \\
G09v1.326*          & 09:18:40.85 & +02:30:47.37 & 2.581    & -               & 12.76 & B13,W17 \\
SDP.130             & 09:13:05.30 & -00:53:42.80 & 2.626    & 2.1  $\pm$ 0.3  & 13.30 & B13 \\
SMMJ02399-0136      & 02:39:51.89 & -01:35:58.56 & 2.795    & 2.38 $\pm$ 0.08 & 12.73 & G14,I10a,T12 \\
SPT0538-50          & 05:38:16.77 & -50:30:53.66 & 2.782    & 21   $\pm$ 4    & 12.54 & BW13 \\
HerMES-Lock01       & 10:57:51.21 & +57:30:27.83 & 2.956    & 9.2  $\pm$ 0.4  & 13.06 & B13 \\
SDP.81              & 09:03:11.50 & +00:39:06.70 & 3.043    & 11.1 $\pm$ 1.1  & 12.57 & B13\\
NGP-NB.v1.78*       & 13:30:08.34 & +24:59:00.07 & 3.111    & 13.0 $\pm$ 1.5  & 12.78 & B13,W17 \\
G12v2.43*           & 11:35:26.16 & -01:46:06.84 & 3.127    & 2.8  $\pm$ 0.4  & 12.70 & B13,W17\\
MG 0751+2716        & 07:51:41.62 & +27:16:32.67 & 3.2      & 16              & 11.77 & A07,BI02,B02,WU09 \\
G12v2.30*           & 11:46:37.96 & -00:11:32.89 & 3.259    & 9.5  $\pm$ 0.6  & 13.06 & B13,W17\\
HXMM02              & 02:18:30.56 & -05:31:31.85 & 3.39     & 4.4  $\pm$ 1.0  & 12.90 & B13 \\
NGP-NC.v1.143       & 12:56:32.65 & +23:36:25.55 & 3.565    & 11.3 $\pm$ 1.7  & 12.91 & B13 \\
G09v1.97            & 08:30:51.12 & +01:32:26.00 & 3.634    & 6.9  $\pm$ 0.6  & 13.20 & B13 \\
APM 08279+5255      & 08:31:41.59 & +52:45:17.81 & 3.91     & 4               & 13.78 & E00,R09,B06,K07,LI11,V11 \\
MM J18423+5938      & 18:42:22.50 & +59:38:29.81 & 3.926    & 12              & 12.13 & L11,D12 \\

\\
\\
\hline
Unused Spectra & & & & \\
\hline
\\
HATLAS 090302-014226$^{\dagger}$        & 09:03:02.95 & -01:42:26.30 & &   \\
SF.V1.88$^{\dagger}$                    & 23:26:23.03 & -34:26:40.34 & &   \\
SF.V1.100$^{\dagger}$                   & 01:24:07.35 & -28:14:35.78 & &   \\
SG.V1.22$^{\dagger}$                    & 01:37:19.98 & -33:19:49.54 & &   \\
SG.V1.29$^{\dagger}$                    & 01:40:31.14 & -32:42:03.51 & &   \\
SG.V1.77$^{\dagger}$                    & 01:48:34.68 & -30:35:32.05 & &   \\
SA.V1.44$^{\dagger}$                    & 22:38:29.06 & -30:41:48.86 & &   \\
SC.V1.128$^{\dagger}$                   & 23:24:19.84 & -32:39:23.71 & &   \\
SD.V1.70$^{\dagger}$                    & 00:09:12.62 & -30:08:09.16 & &   \\
SD.V1.133$^{\dagger}$                   & 00:07:22.37 & -35:20:15.57 & &   \\
SB.V1.143$^{\dagger}$                   & 23:25:55.44 & -30:22:34.89 & &   \\
SGP-B-202$^{\dagger}$                   & 23:26:23.03 & -34:26:40.34 & &   \\
SGP-D-328$^{\dagger}$                   & 00:26:25.11 & -34:17:38.13 & &   \\
SGP-E-165$^{\dagger}$                   & 00:47:36.05 & -27:29:54.03 & &  \\
SGP-A-53$^{\dagger}$                    & 22:25:36.37 & -29:56:46.03 & &   \\
MACS J2043-2144$^{\ddagger}$            & 20:43:14.17 & -21:44:38.80 & & \\
MMJ0107$^{\dagger}$                     & 01:07:02.35 & -73:01:59.80 &  &  \\
SWIRE 07$^{\dagger}$                    & 11:02:05.80 & +57:57:40.75 &  &  \\
XMM 02$^{\dagger}$                      & 02:19:57.26 & -05:23:48.81 &  &  \\
DAN 02$^{\dagger}$                      & 17:20:17.08 & +53:12:43.41 &  &  \\
Locke 01$^{\dagger}$                    & 10:45:30.42 & +58:12:32.86 &  &  \\
HeLMS 44$^{\dagger}$                    & 23:32:55.44 & -03:11:34.34 &  &  \\
HeLMS 45$^{\dagger}$                    & 23:24:39.55 & -04:39:36.18 &  &  \\
SPT 0551-50$^{\dagger\dagger}$                & 05:51:39.41 & -50:58:02.36 & 3.164    & \\
SPT 0512-59$^{\dagger\dagger}$                & 05:12:57.93 & -59:35:41.87 & 2.2331   & \\
NGC 2146$^{\ddagger\ddagger}$                   & 06:18:38.91 & +78:21:25.37 & 0.00298  & \\
NGC 1068$^{\ddagger\ddagger}$                     & 02:42:40.82 & -00:00:47:56 & 0.00379  & \\
NGC 4151$^{\ddagger\ddagger}$                     & 12:10:32.88 & +39:24:18.55 & 0.00332  & \\
NGC 5128$^{\ddagger\ddagger}$                     & 13:25:27.55 & -43:01:09.79 & 0.00183  & \\
M81$^{\ddagger\ddagger}$                          & 09:55:32.78 & +69:03:57.05 & -0.000113& \\
M82$^{\ddagger\ddagger}$                          & 09:55:51.67 & +69:40:48.20 & 0.000677 & \\
NGC 2403$^{\ddagger\ddagger}$                     & 07:36:49.82 & +65:36:12.27 & 0.000445 & \\
NGC 205$^{\ddagger\ddagger}$                      & 00:40:24.03 & +41:41:50.39 & -0.000768& \\
MCG 604$^{\dagger}$                     & 01:34:33.46 & +30:46:47.59 &          &  \\

\hline

\\

\end{longtable}
\footnotesize{Notes: L$_{\rm{IR}}$ values for sources with $z<1$ are taken from the literature. The given values of L$_{\rm{IR}}$ for sources with $z>1$ are computed from fits to continuum photometry and are corrected for lensing magnification. Sources marked with * have PACS spectroscopy and appear in \citet{Wardlow2017}. References: A07: \citet{Alloin2007}, A09: \citet{Armus2009}, B92: \citet{Barvainis_cloverleaf1992}, B99: \citet{Benfordthesis1999}, B02: \citet{Barvainis2002}, B06: \citet{Beelen2006}, B13: \citet{Bussmann2013}, BI02: \citet{Barvainisivison2002}, BL95: \citet{Broadhurstlehar1995}, BW13: \citet{Bothwell2013}, D09: \citet{Dye2009}, D12: \citet{Decarli2012}, E00: \citet{Egami2000}, E06: \citet{Evans2006}, E09: \citet{Eales2009}, F07: \citet{Farrah2007}, F13: \citet{Farrah2013}, G14: \citet{Greve2014}, H03: \citet{Hutchings2003}, HW88: \citet{Helouwalker1988}, I10a \citet{Ivison2010a}, I10b:\citet{Ivison2010}, V10:\citet{Vanderwerf2010}, I11: \citet{Iwasawa2011}, K07: \citet{Krips2007}, L11: \citet{Lestrade2011}, L17:\citet{Lu2017}, LI11: \citet{Lis2011}, M90: \citet{Moshir1990}, M11: \citet{Moncelsi2011}, M12: \citet{Magnelli2012}, M14: \citet{Magdis2014}, ME14: \citet{Messias2014}, NED: NASA/IPAC Extragalactic Database, R09: \citet{Riechers2009}, R11: \citet{Rangwala2011}, R15: \citet{Rosenberg2015}, RR93: \citet{Rowanrobinson1993}, S11: \citet{Sargsyan2011}, S14: \citet{Spilker2014}, SW11: \citet{Swinbank2011}, T12: \citet{Thomson2012}, T16: \citet{Timmons2016}, U16: \citet{Uzgil2016}, V11: \citet{Vanderwerf2011}, VS03 \citet{Venturinisolomon2003}, W17: \citet{Wardlow2017}, WU09: \citet{Wu2009}, Y13: \citet{Yamada2013}.}
\footnotesize{For unused spectra: $^{\dagger}$No spectroscopic redshift and/or magnification factor. $^{\ddagger}$ Multiple objects within beam. $^{\dagger\dagger}$ No magnification factor. $^{\ddagger\ddagger}$Redshift less than 0.005.}

\end{document}